\documentclass[10pt,journal,compsoc]{IEEEtran}
\ifCLASSOPTIONcompsoc
  \usepackage[nocompress]{cite}
\else
  \usepackage{cite}
\fi
\ifCLASSINFOpdf
\else
\fi
\usepackage{amsmath}
\usepackage{algorithmic}

\usepackage{array}
\usepackage{fixltx2e}

\usepackage{stfloats}
\usepackage{url}

\usepackage{subfig}

\usepackage{tikz}
\usepackage{xcolor}
\newcommand*\circled[1]{\tikz[baseline=(char.base)]{
		\node[shape=circle,fill,inner sep=0.5pt] (char) {\textcolor{white}{#1}};}}

\usepackage{mathtools}

\usepackage{color, colortbl}

\definecolor{Grayl}{gray}{0.9}
\definecolor{Grayd}{gray}{0.8}

\usepackage{url}

\usepackage{breakurl}
\usepackage[breaklinks]{hyperref}

\usepackage{cleveref}

\crefformat{section}{\S#2#1#3} 
\crefformat{subsection}{\S#2#1#3}
\crefformat{subsubsection}{\S#2#1#3}
\pdfoutput=1

\begin{document}
%
\title{Efficiently Using Polar Codes in 5G Base Stations to Enhance Rural Connectivity}
%
%
%
%

\author{Aman~Shreshtha
        and~Smruti~R~Sarangi

\IEEEcompsocitemizethanks{\IEEEcompsocthanksitem The authors are with the Department of Computer Science and Engineering, Indian Institute of Technology Delhi, Hauz Khas, New Delhi, India\protect\\
E-mail: amanshreshtha@cse.iitd.ac.in, srsarangi@cse.iitd.ac.in
}
\thanks{Manuscript received Month dd, 2022; revised Month dd, 2022.}}

%
%

\markboth{IEEE Transactions on Mobile Computing,~Vol.~XX, No.~x, Month~2022}%
{Shell \MakeLowercase{\textit{et al.}}: Bare Demo of IEEEtran.cls for Computer Society Journals}
%



\IEEEtitleabstractindextext{%
\begin{abstract}
5G connectivity has become essential to integrate rural communities into the broader digital economy and support critical applications like remote education and remote surgery. A major hindrance to expanding rural broadband coverage, especially in developing countries, is the high cost of installing 5G base stations. Hence, there is a need to reduce the cost of a 5G base station without degrading its performance. Our work proposes a novel approach to efficiently utilize the polar code encoders in a 5G base station. The idea is to use the idle time of the polar encoders during downlink transmission for error correction in the 5G data plane. Polar codes have conventionally been used in the 5G control plane, while LDPC codes are used in the data plane. We perform detailed characterization experiments to show the advantages of using polar codes in the data plane as well. Further, to intelligently distribute the user data packets among the available compute nodes, we propose a set of novel resource allocation algorithms and compare their performance with other algorithms in the literature. Using our proposed optimization techniques, we achieve a 17\% reduction in the cost of a 5G base station. Simultaneously, we are able to improve the performance by 24\% compared to a conventional base station.
\end{abstract}

}

\maketitle

\IEEEdisplaynontitleabstractindextext

%
\IEEEpeerreviewmaketitle

\IEEEraisesectionheading{\section{Introduction}\label{sec:introduction}}

\begin{figure}[!t]
	\centering
	\subfloat[\centering Rural broadband]{{\includegraphics[width=6.7cm]{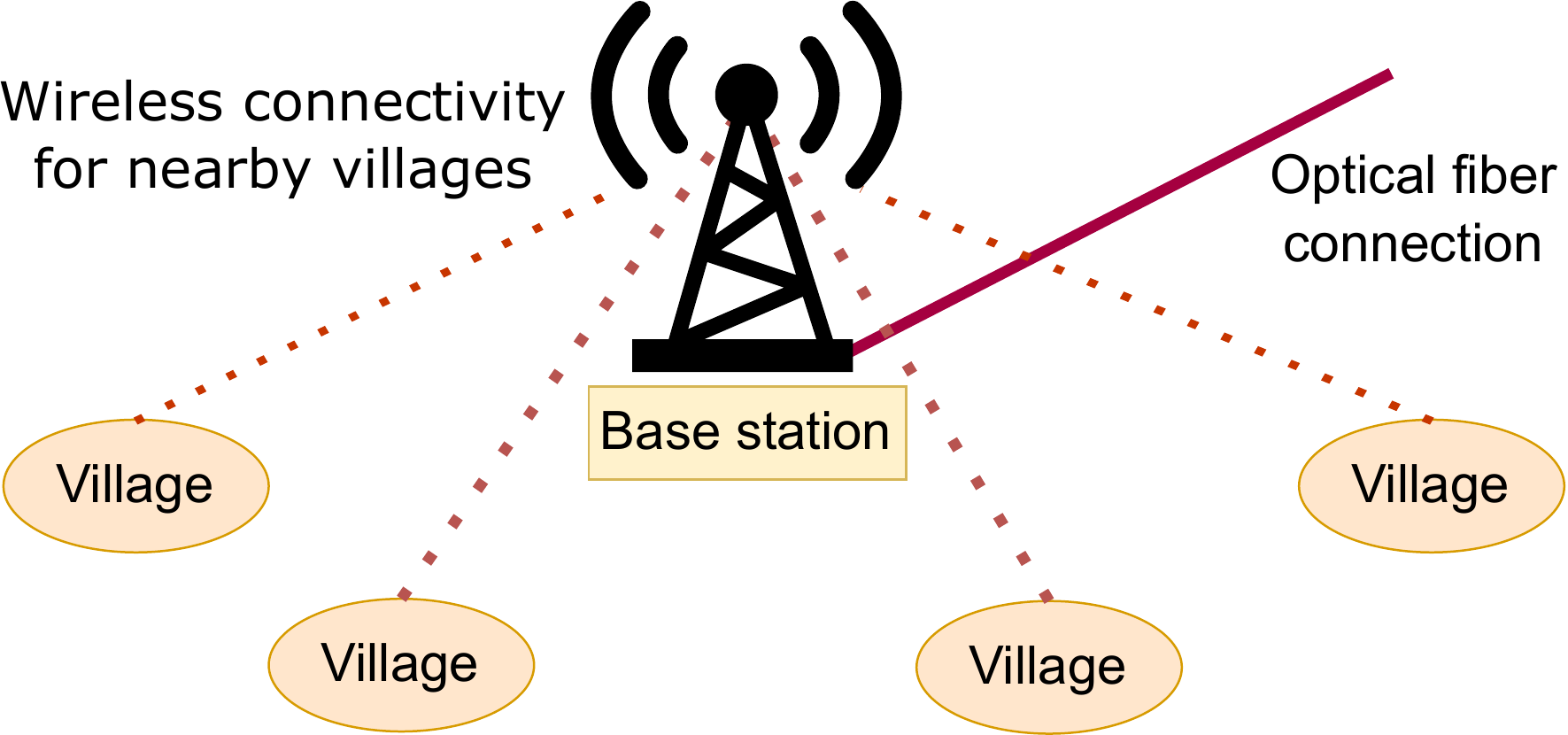} }}%
	\qquad \\
	\subfloat[\centering Edge-computing scenario]{{\includegraphics[width=5cm]{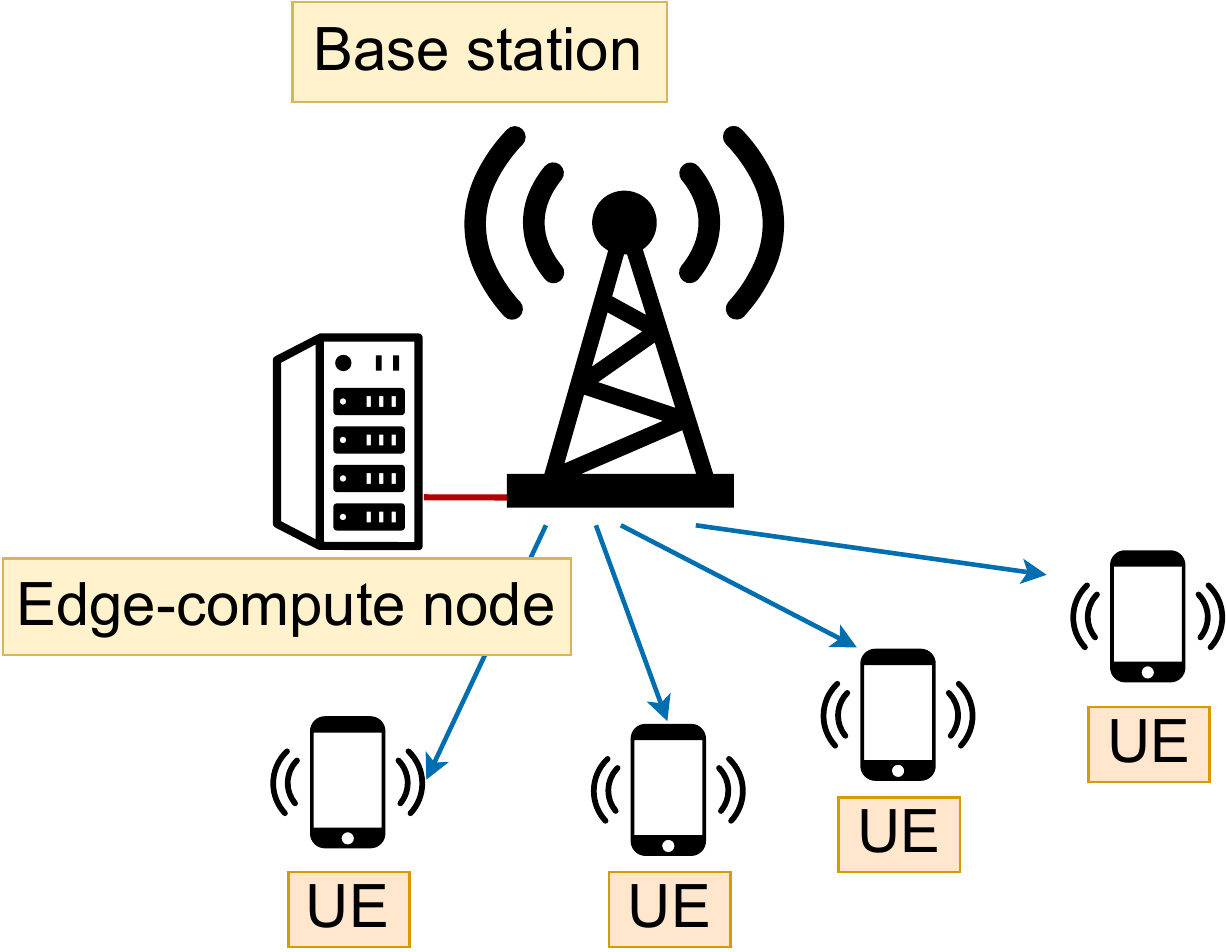} }}%
	\caption{Rural broadband and the edge-computing scenario}%
	\label{fig:intro_combined}%
\end{figure}

\IEEEPARstart{R}{ural}\footnote{Extension of a conference paper: Shreshtha et al.~\cite{polarcode_IPSN}. In the conference paper~\cite{polarcode_IPSN}, we proposed a polar code-based approximate communication system for multimedia web pages. This paper extends the approach proposed in the conference paper to accommodate LDPC codes also. On top of that, we perform a detailed characterization of our proposed approach using the polar and LDPC codes across multiple channel conditions (our previously published conference paper provided limited characterization for only a single channel condition, and that too for only polar codes). We make a paradigmatic change by proposing the system model for a 5G base station that takes advantage of the optimizations proposed in the conference paper and also its extension to LDPC codes that we propose in the current paper (no full-system simulations of a 5G base station were done in the conference version). In this work, we propose a set of novel online resource allocation algorithms for the base station and compare their performance (this was not done in the conference version).} broadband connectivity has become essential in the digital economy of developing countries. Rural businesses like farming, manufacturing, and green energy production can be monitored easily and at scale using digital and connectivity-based interventions~\cite{rural_broadband_ericcson}. With greater penetration of mobile phones in rural areas, the demand for high-speed broadband connectivity has increased dramatically, fueled by entertainment, gaming, education, and health applications~\cite{entertainment_microsoft, online_gaming_financial_express, remote_education_qualcomm, telemedicine_qualcomm, remote_surgery_business_insider}. The 5G new radio (5G-NR) communication system includes rural enhanced mobile broadband (rural eMBB) standard as one of its use cases~\cite{ruralembb}. The importance of good rural broadband is underscored by the fact that worldwide, 47\% of the people in developing countries live in rural areas~\cite{developing_country_perspective}. Due to high capital expenditure per inhabitant, revenue uncertainty, and high maintenance costs, telecom companies are reluctant to install base stations in rural and remote areas~\cite{business_model}. A possible solution is to use a single base station to serve multiple nearby villages spread over large areas~\cite{lmlc}, as shown in Figure~\ref{fig:intro_combined}(a). To support rural applications like remote surgery and remote education, which use virtual reality and rich multimedia~\cite{virtual_reality_surgery, video360degree}, the performance of the existing wireless systems needs to be improved~\cite{polarcode_JSCD}. Concomitantly, any performance-improving approach should not increase the cost of base stations.


We consider an edge-computing scenario (see Figure~\ref{fig:intro_combined}(b)) where the requests made by the user equipment (UE) are processed by an edge-compute node at the base station. As a case study, let us consider the HPE EL8000 converged edge system, which is a well-suited general-purpose compute infrastructure at the edge of a cellular network~\cite{EL8000}. The EL8000 system can support up to four server blades~\cite{EL8000_datasheet}. An example of a server used for edge computing is the HPE ProLiant DL110 Gen10 Plus telco server. This server is customized for edge applications and radio access network (RAN) workloads~\cite{vDU}; it can support up to four high-performance accelerator cards~\cite{hpe_telco_server, DUX100_accelerator_card}. Hardware accelerator cards are used to primarily offload the processing of LDPC and polar code forward error protection (FEC) functions. For example, Kaltenberger et al.~\cite{T1_offload} offload the baseband processing for the OpenAirInterface 5G-NR software~\cite{OpenAirInterface} to a Xilinx T1 telco accelerator card~\cite{T1_product_brief}, which includes forward error correction; they observe a significant performance improvement. Now, note that deployment of 5G services is a capital-intensive business~\cite{business_model} where up to 50\% of the total cost of a telecom company is spent on buying telecom equipment and setting up the infrastructure for the radio access network~\cite{EL8000, cost_base_station}. In our work, we show that by using the available hardware resources on a 5G base station optimally, we can save on the compute-node requirements of the base station. Using our proposed approach, the compute requirements of a 5G base station, in our case study, can be reduced from 12 hardware accelerator cards to 10 hardware accelerator cards. This is a 17\% reduction in the equipment cost of a 5G base station; it has huge implications for telecom operators. For a countrywide 5G deployment in a large developing country (for example, India), more than a million base stations need to be installed~\cite{number_base_station}. Hence, a 17\% reduction in the cost of a 5G base station adds up to several billion dollars of cost savings for a telecom operator.

\subsection{Contributions in the Paper}

Our aim is to reduce the cost of a 5G base station while ensuring that the performance of the 5G system is not degraded. 

\noindent\circled{1} We extend the performance-enhancing approach of using the unequal error protection property of polar codes, originally proposed by Shreshtha et al.~\cite{polarcode_IPSN}, to also accommodate LDPC codes and then perform a detailed characterization of the achieved gain in performance across multiple channel conditions.
	
\noindent\circled{2} To optimally utilize the hardware resources on the 5G base station, we use the idle time of the polar code encoders for error correction in the data plane of the 5G-NR system. This allows the base station to serve the UE requests faster and provides an opportunity to reduce the cost of the base station.
	
\noindent\circled{3} The data processed at the base station needs to be intelligently distributed between the available compute nodes. A good resource allocation algorithm is critical to the performance of a 5G base station. We propose a novel reinforcement learning (RL) based algorithm and two new heuristics. We implement a total of four online resource allocation algorithms and compare their performance using Monte Carlo simulations.
	
\noindent\circled{4} We demonstrate a performance improvement of up to 24\% as compared to a conventional 5G-NR system. Simultaneously, our proposed approach leads to a reduction in the cost of the base station by 17\%. This indicates that it is possible to reduce the cost of a 5G base station while also ensuring that the 5G communication system's performance is enhanced.

The rest of the paper is organized as follows. We provide the necessary background in \cref{sec:background}. The motivation for our proposed optimizations is discussed in \cref{motivation}. The proposed web page transmission framework, utilizing the unequal error protection property of LDPC codes, is presented in \cref{proposedApproach1}. \cref{proposedApproach2} explains the system model of a 5G base station that uses both polar and LDPC encoders for the FEC (forward error correction) of the 5G-NR data plane. \cref{proposedApproach3} discusses the four resource allocation algorithms that we compare for workload scheduling at a base station. The experimental setup is described in detail in \cref{evaluationSetup}. The results for the characterization of the LDPC and the polar code-based web page transmission frameworks are presented in \cref{char_results}. These results motivate our design choices. \cref{sec:opt_results} reports the evaluation results for the four resource allocation algorithms. The evaluation results for a 5G base station that incorporates our proposed optimizations are presented in \cref{systemResults}. We discuss related work in \cref{sec:relatedwork}, and finally conclude in \cref{conclusion}.


\section{Background}
\label{sec:background}

We give a background of approximate communication, polar codes, LDPC codes, and reinforcement learning, especially the multi-armed bandit algorithm. Our proposed web page transmission framework is an approximate communication system. The 5G new radio (5G-NR) communication system utilizes the LDPC and the polar codes for forward error correction. We present the problem of workload scheduling at a 5G base station as an online resource allocation algorithm, and one of the approaches to solve it is using a multi-armed bandit (MAB) formulation of the problem, which itself is a part of the reinforcement learning (RL) paradigm.

\subsection{Approximate Communication}

Approximate computing is a computation paradigm for applications that can tolerate limited amounts of error but still give acceptable results. These applications are related to human perception and need not give 100\% accurate results. Approximate computing has been used in various domains such as pattern recognition, machine learning, image processing, and scientific computing~\cite{approximate_communication}. Traditionally, data networks aim to achieve 100\% accuracy when transmitting data from the sender to the receiver. This is achieved using repeated retransmissions. As a consequence, the network consumes excess power, and the latency of the network also increases. However, this is unnecessary for approximate computing applications. For such applications, we can relax the accuracy constraint on the data during transmission. As a result, the energy consumption of the network reduces, and it also has a positive impact on the network latency.

\subsection{Polar Codes}

\begin{figure}[!t]
	\centering
	\includegraphics[width=6cm]{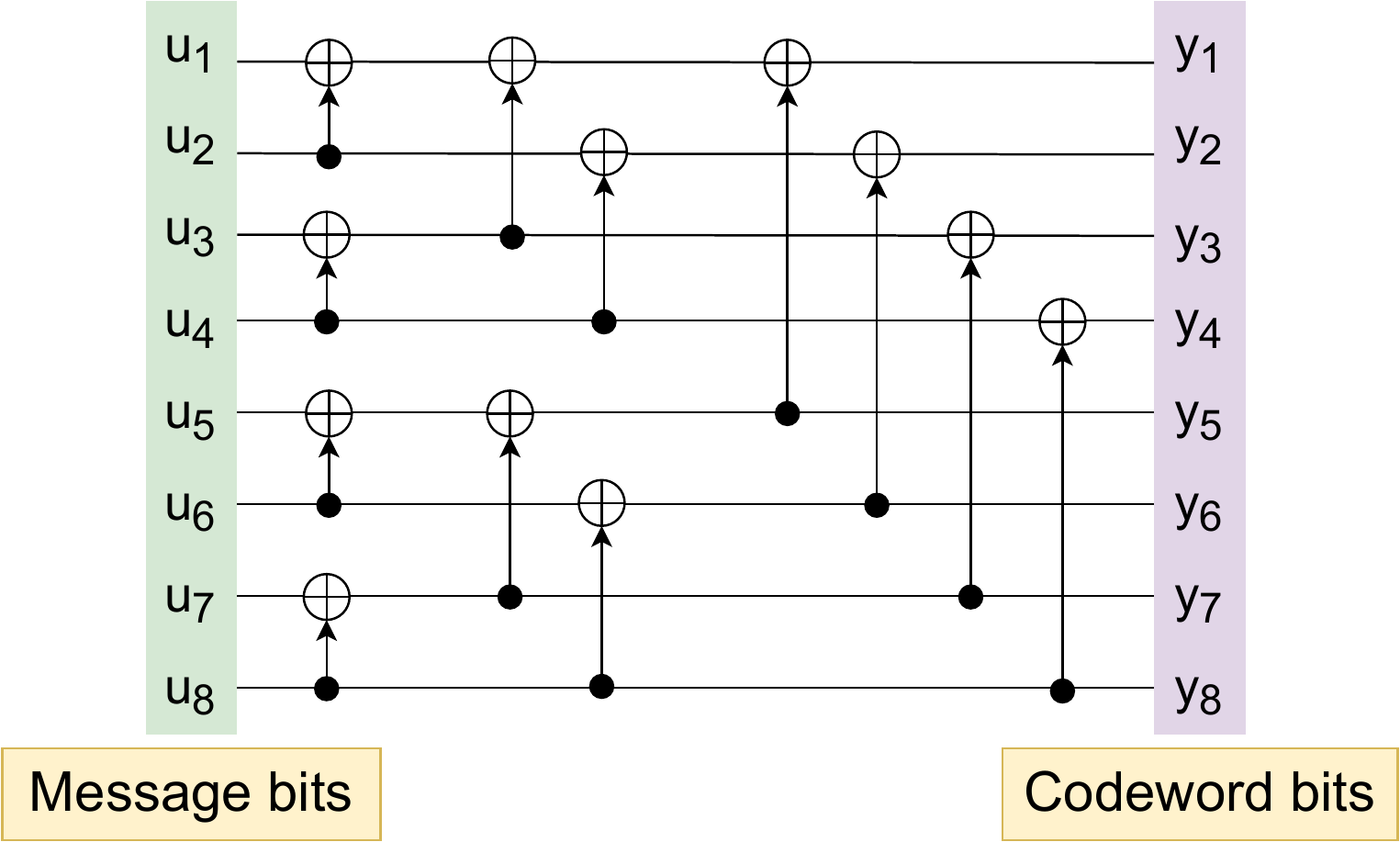}
	\caption{Polar code encoding for an 8-bit codeword}
	\label{fig:polarencoding}
\end{figure}

Polar codes are block error-correction codes used for error detection and correction in the control channel of the eMBB use case of the 5G-NR communication system. The control plane of the 5G communication system handles the signaling between the user equipment and the base stations. Polar codes are provably capacity-achieving at infinite length, and the computation complexity of both the encoder and the decoder is $\mathcal{O}(n\log{}n)$. These properties make polar codes a desirable candidate as the channel coding scheme for a communication system. The construction of polar codes is based on the channel polarisation phenomenon. The polarization transform is given by the kernel \textbf{F}. The transform for larger code lengths is obtained by calculating the Kronecker product of the kernel \textbf{F} with itself in a recursive fashion~\cite{arikan}.


\begin{equation}
	\textbf{F} = \left(\begin{IEEEeqnarraybox*}[][c]{,c/c,}
		1 & 0 \\
		1 & 1%
	\end{IEEEeqnarraybox*}\right)
\end{equation}

In the polar code encoding operation, K information bits and (N-K) frozen bits are interleaved. Here, N is the total number of bits in a codeword. These N bits are then combined using the XOR operator. The encoding process for a codeword of length 8 bits is shown in Figure~\ref{fig:polarencoding}. The polar code encoding algorithm splits a single physical channel into multiple virtual channels with different error probabilities. The information bits are transmitted on the highly reliable bit positions of the codeword. The highly unreliable bit positions are treated as frozen bits and are set to 0. Both the encoder and the decoder know the frozen bits~\cite{frozen_bits}. The main decoding algorithms used for polar codes are the Successive Cancellation (SC) decoding~\cite{arikan}, the Successive Cancellation List (SCL) decoding~\cite{list_decoding}, and the CRC-aided Successive Cancellation List decoding (CA-SCL)~\cite{crc_aided_list_decoding} algorithms. The 5G-NR system uses the CA-SCL algorithm for polar code decoding~\cite{polar_code_matlab}. In our implementations, we use the same algorithm for channel decoding.

\subsection{LDPC Codes}

Low-density parity-check (LDPC) codes are capacity-approaching block error-correcting codes. LDPC codes are widely used for forward error correction in WiFi and, more recently, in the data channel of the 5G-NR communication system. The data plane of the 5G communication system handles user traffic. The voice, multimedia, and other user data moving through the network infrastructure constitute the data plane. LDPC codes have a low decoding complexity, and LDPC decoders have been designed that work at Gbps data rates~\cite{ldpc_decoder_gbps}. LDPC block codes are constructed using a sparse parity check matrix. A sparse parity check matrix contains mostly 0s and a small number of 1s. \textbf{H} is an example of a parity check matrix.


\begin{equation}
	\textbf{H} = \left(\begin{IEEEeqnarraybox*}[][c]{,c/c/c/c/c/c,}
		1 & 1 & 0 & 0 & 1 & 0 \\
		1 & 0 & 1 & 1 & 0 & 0 \\
		0 & 0 & 1 & 0 & 1 & 1 \\
		0 & 1 & 0 & 1 & 0 & 1%
	\end{IEEEeqnarraybox*}\right)
\end{equation}

\begin{figure}[!t]
	\centering
	\includegraphics[width=5.6cm]{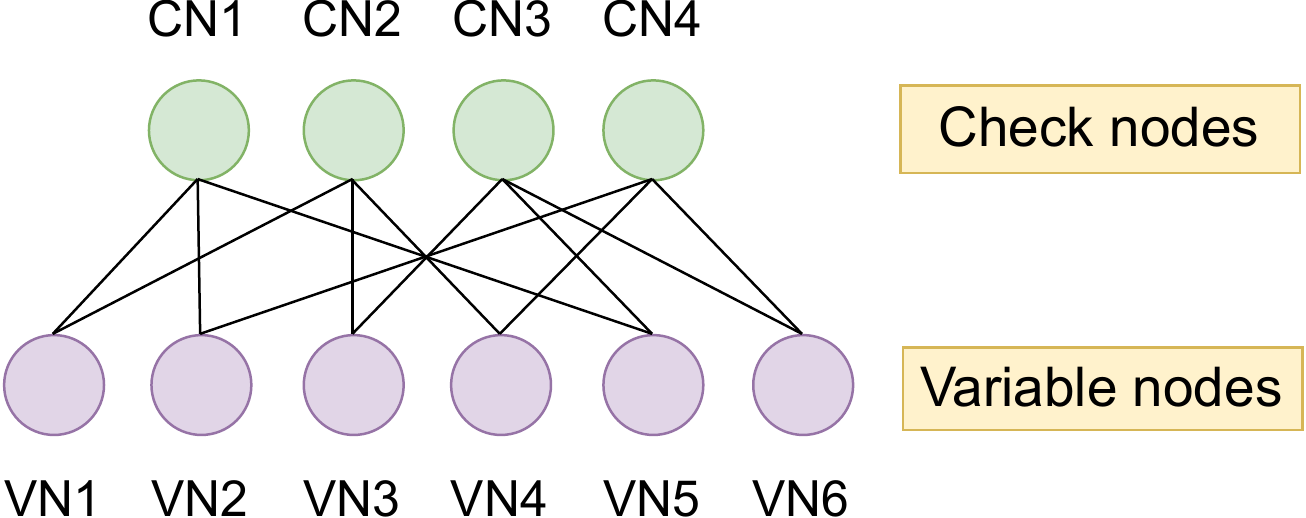}
	\caption{Tanner graph representation for LDPC codes}
	\label{fig:LDPCtanner}
\end{figure}

LDPC codes are easily represented as a Tanner graph. Each variable node corresponds to a codeword symbol, and each check node corresponds to a parity-bit symbol. If a variable node $i$ (codeword symbol) satisfies a constraint given by the check node $j$ (parity symbol), we add an edge between the two nodes in the tanner graph. This also means that the element $h_{ij}$ of the \textbf{H} matrix is equal to 1. The tanner graph representation for the LDPC code with the parity check matrix as H is shown in Figure~\ref{fig:LDPCtanner}. The structure of the tanner graph is used to design the computational architecture of LDPC decoders~\cite{tanner_decoder_arch}. A sparse encoding matrix leads to an iterative decoding scheme with the decoding complexity increasing linearly with the codeword length~\cite{ber_convolution_turbo_ldpc_polar}. Belief Propagation (BP) is a low-complexity, iterative decoding algorithm for LDPC codes~\cite{gallager_ldpc}. The LDPC decoding operation can be easily implemented as a highly parallel implementation~\cite{ldpc_parallel_decoder}. This makes LDPC codes suitable as error correction codes for high data rate systems like WiFi and the data plane of the 5G-NR communication system. Comparatively, it is difficult to parallelize the implementation of polar code decoders. The bit error rate (BER) performance of LDPC decoders is better than that of the SC decoding algorithm for polar codes. The SCL decoding algorithm for polar codes matches and even improves upon the BER performance of LDPC decoders~\cite{ldpc_polar_decoder_comparison}.

\subsection{Reinforcement Learning (Multi-Armed Bandit)}

\begin{figure}[!t]
	\centering
	\includegraphics[width=7.5cm]{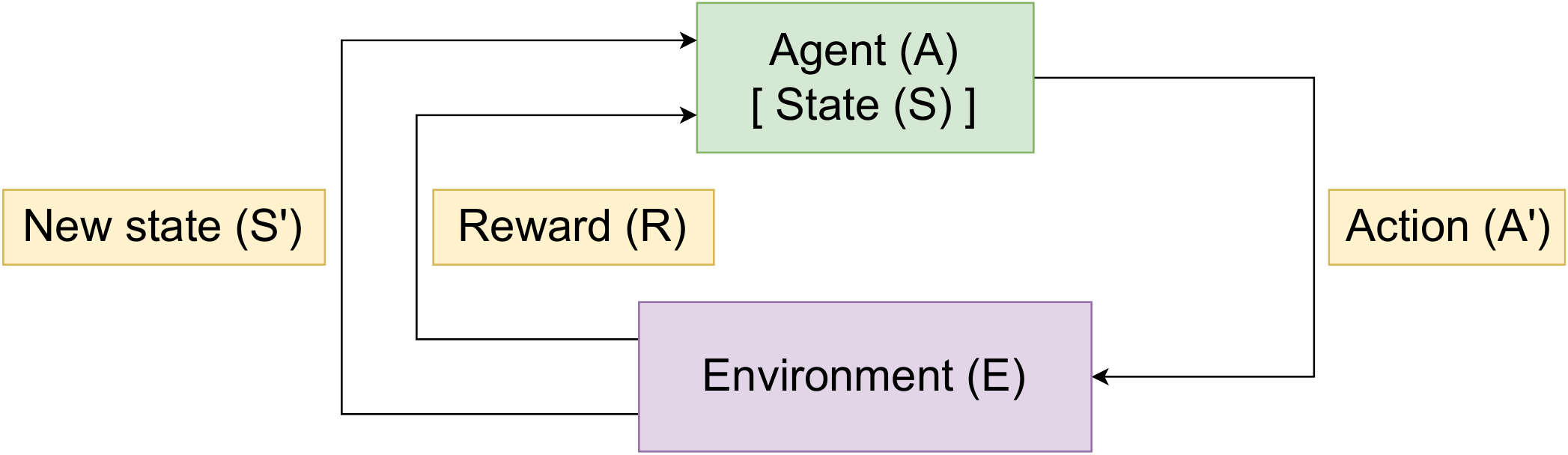}
	\caption{Block diagram of reinforcement learning}
	\label{fig:reinforce}
\end{figure}

Reinforcement Learning (RL) algorithms are a class of algorithms used to train an intelligent agent to maximize the \emph{reward} it earns while operating in an environment with an inexact mathematical model of the environment. The main components of an RL system are shown in Figure~\ref{fig:reinforce}. We have an \emph{intelligent agent} operating in an environment \emph{E}. The agent's position in the environment is described using a set of state variables \emph{S}. In any given state \emph{S}, the agent has a set of actions that it can take. This set of actions forms the action set denoted by \emph{A}. The permissible actions that the agent can take depend on the current state of the agent. The agent's purpose is to interact with the environment over several time steps. At each time step, the agent performs an action \emph{A'}. Depending on the current state \emph{S} and the action \emph{A'} performed by the agent, the agent receives a reward \emph{R} from the environment. After taking the action \emph{A'} the agent moves into a new state \emph{S'}. The agent's aim is to maximize the total reward it earns over a large period of time. Because the agent has no or limited information about the environment, it does not know beforehand the optimal action to be taken in any given state to earn the highest reward. So, the agent first takes random actions from the available set of actions in action set \emph{A}. From this, it learns an estimate of the state-action pair that leads to the maximum reward. Then, in later steps, the agent exploits the learned estimates to maximize the earned rewards.

The fundamental part of an RL algorithm is the exploration vs. exploitation trade-off. This has been extensively analyzed using the multi-armed bandit problem formulation. A multi-armed bandit is a slot machine with \emph{K} levers. Every lever has an associated reward distribution that decides the probability of getting a reward when that particular lever is pulled. The objective of an agent with a limited number of lever pulls is to maximize the earned reward over all the pulls. Here, the agent has to decide at every time step whether to try a new arm (explore) to gain more information or to keep playing the best-known arm till now (exploitation). This problem formulation has been used to model and solve resource allocation problems in wireless communication systems~\cite{multiarmed_bandit_resource_allocation}. The simple multi-armed bandit problem formulation does not use any information about the state of the environment. The contextual multi-armed bandit problem formulation extends the model by changing the state of the system after every level pull and making the reward earned from a lever pull dependent on the current state of the system~\cite{contextual_mab}.

\section{Motivation}
\label{motivation}

\subsection{Unequal Error Protection in LDPC and Polar Codes}

\begin{figure}[!t]
	\centering
	\subfloat[\centering LDPC code]{{\includegraphics[width=8cm]{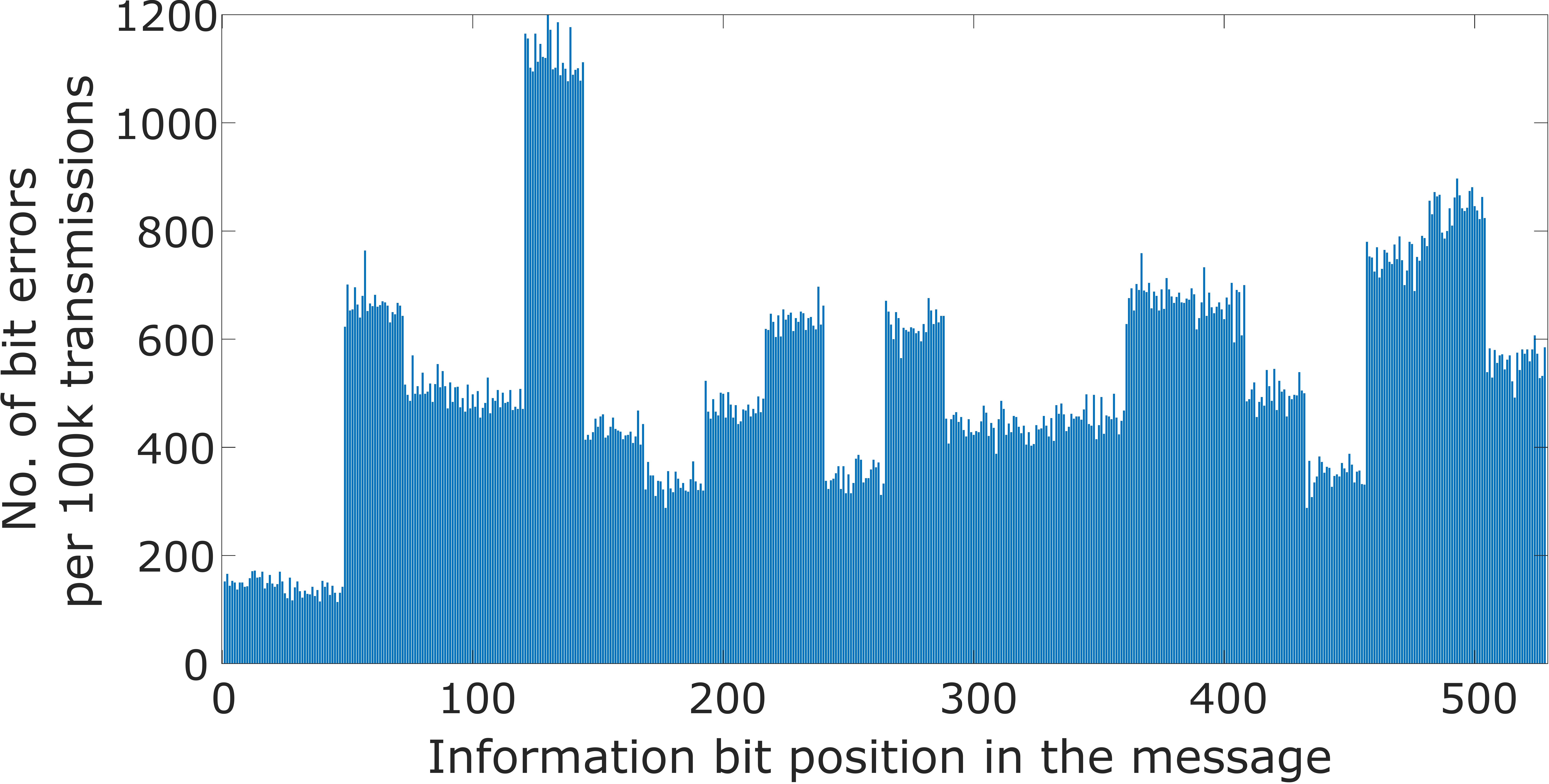} }}%
	\qquad
	\subfloat[\centering Polar code]{{\includegraphics[width=8cm]{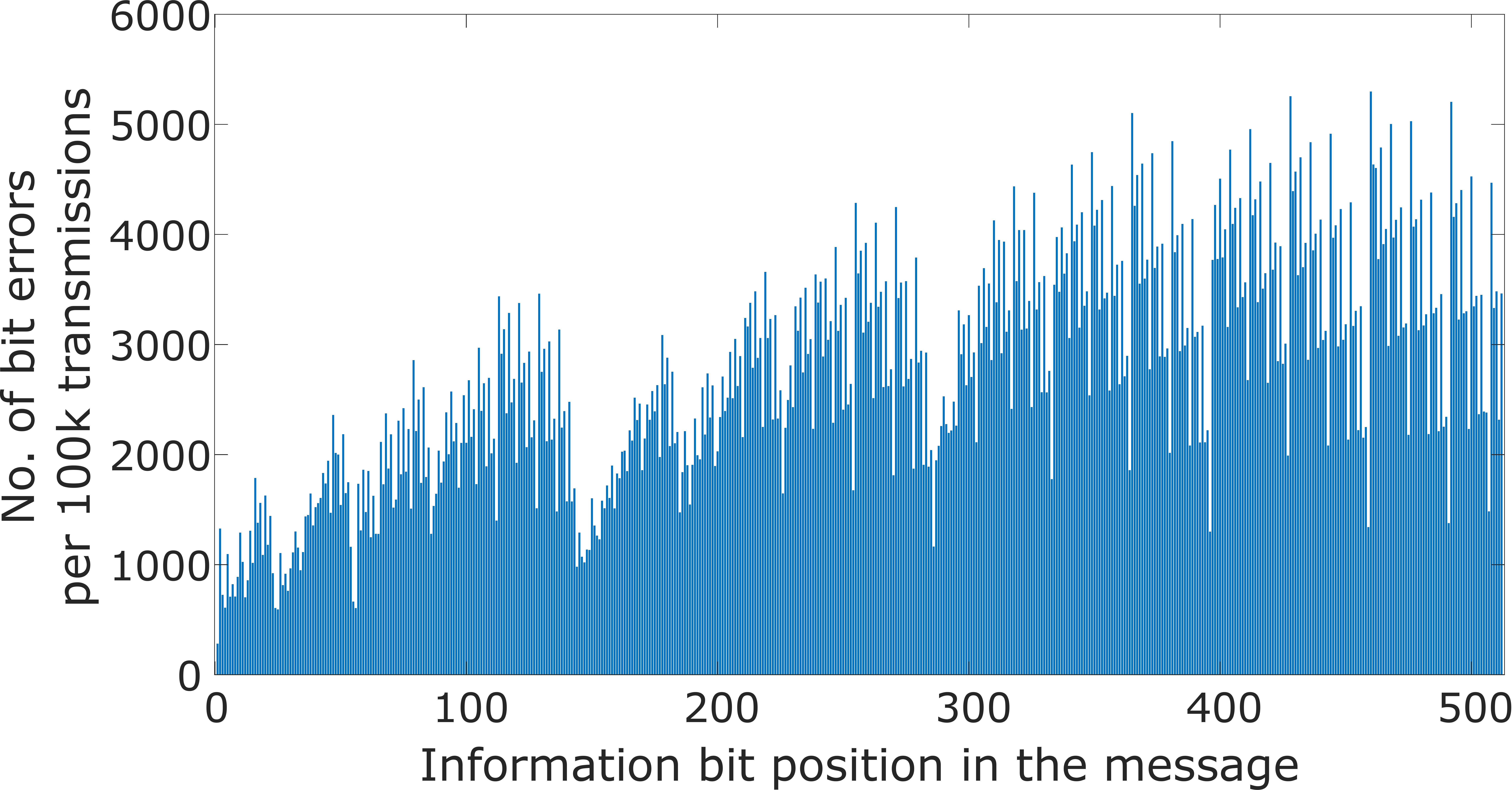} }}%
	\caption{Unequal bit error rate in error-correction codes}%
	\label{fig:uep_combined}%
\end{figure}

We conduct experiments using MATLAB simulations to investigate the unequal error protection (UEP) property in the LDPC and polar forward error correction (FEC) schemes in a 5G-NR system. Specifically, we consider randomly generated messages transmitted over an AWGN channel after forward error correction. The receiver decodes the received codewords. This process of message transmission is repeated $10^5$ times. The simulation parameters are summarized in Table~\ref{table:simulation_parameters}. Due to the noise added by the channel, some of the message bits may be received in error. For all the information bit positions in the message, we plot a graph of the total number of transmissions in which the bit at that position was received in error. The graph for both the LDPC and polar codes is presented in Figure~\ref{fig:uep_combined}.

From the graphs, we observe that LDPC codes show strong UEP behavior. The initial 50 bit positions are strongly protected compared to the other bits. Some of the bits are weakly protected and are received with very low reliability at the decoder. In the case of polar codes, the initial bit positions are more reliable, and the reliability of the bits decreases as we move to higher bit positions. For polar codes, the reliability of the bit positions follows a linear pattern with some deviation.

\begin{table}[!t]
	\centering
		\begin{tabular}{ |c|c| }
			
			\hline
			\rowcolor{Grayd} \textbf{Parameter} & \textbf{Value} \\
			\hline\hline
			
			Error correction code & (512,1024) LDPC and polar code \\
			\hline
			Code rate & 0.5 \\
			\hline
			Channel & AWGN ($E_b/N_o$ = 2 dB) \\
			\hline
			Number of transmissions & 100,000 \\
			\hline
			
		\end{tabular}
	\caption{Simulation parameters}
	\label{table:simulation_parameters}
\end{table}

\subsection{Computational Resource Utilization of Polar Code Encoder}

In downlink transmission, the data and control signals are transmitted from the base station to the UE. The physical downlink shared channel (PDSCH) is the physical channel that carries the downlink data. The control signals are transmitted through the physical downlink control channel (PDCCH). In the enhanced mobile broadband (eMBB) use case of the 5G-NR communication system, the user data or the data plane is protected using the LDPC error correction codes. Polar codes are used for error correction of the control signals or the control plane. Wang et al.~\cite{polar_util} measure the CPU utilization and the memory usage of the radio access network (RAN) and the mobile edge computing (MEC) workloads running as 5G edge-network virtual functions. The authors report that for downlink transmission, the CPU utilization of the data plane encoding function is around seven times the CPU utilization of the control channel processing function. The polar code encoding function is an essential and computation-intensive part of the control channel processing function. These results imply that the polar code encoder has a significantly lower ($\approx$ 86\%) utilization of the computation resources on the base station than the LDPC encoder. 

\begin{figure}[!t]
	\centering
	\includegraphics[width=8cm]{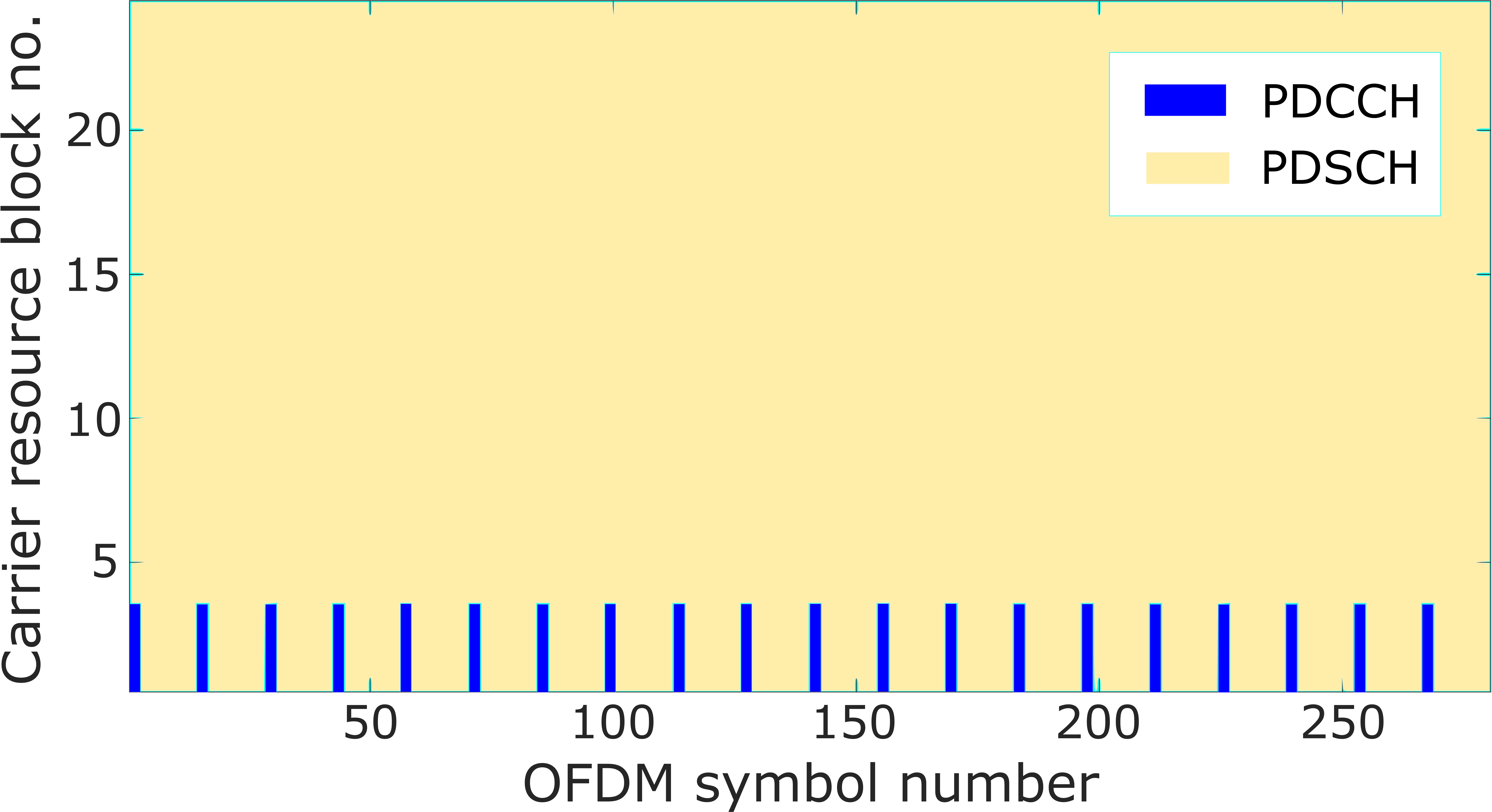}
	\caption{PDSCH and PDCCH resource block utilization - the PDCCH channel is scheduled intermittently}
	\label{fig:resource_util}
\end{figure}

We analyze this observation in more detail. The MATLAB 5G Toolbox~\cite{matlab_5G_toolbox} supports standard-compliant waveform generation. The 5G-NR test model (NR-TM) waveform is used for conformance testing of base stations. In Figure~\ref{fig:resource_util} we show the resource block utilization graph for the NR-TM waveform with the following specifications; model name: NR-FR1-TM1.1, bandwidth: 10 MHz (FR1), subcarrier spacing: 30 kHz (FR1), duplexing mode: FDD. In a resource block utilization graph, we have the OFDM symbol number on the x-axis and the carrier resource block number on the y-axis. Essentially, the resource block utilization graph is a frequency versus time graph that depicts what data is transmitted at any given time and the frequency bands used for that data transmission. From the graph in Figure~\ref{fig:resource_util}, we observe that the PDCCH channel is scheduled intermittently. The PDSCH channel is scheduled for a significantly longer duration of time and a larger frequency bandwidth. The polar code encoder function is called within the time intervals when the PDCCH channel is scheduled, while the LDPC code encoder is called within the time intervals when the PDSCH channel is scheduled.

The results from the paper by Wang et al.~\cite{polar_util} and the resource utilization graph in Figure~\ref{fig:resource_util} indicate that there exists a significant amount of time in the whole transmission process when the polar code encoder is sitting idle. These are the time intervals when the PDSCH channel is scheduled, but the PDCCH channel is not scheduled. This leads to lower computational resource utilization for the polar code encoder on the base station. Real-life deployment of 5G base stations uses hardware accelerator cards to offload the LDPC and polar code FEC functions~\cite{ldpc_accelerator, polar_accelerator}, and here the problem of lower resource utilization becomes more acute.

\section{Approximate Communication System for Multimedia Web Pages}
\label{proposedApproach1}

The authors in~\cite{polarcode_IPSN} show that by using the unequal error protection property of polar codes and the idea of approximate communication, we can improve the performance of multimedia web page transmission over TCP by 60-70\%. It is feasible and even advantageous to use polar codes for error correction in the data plane of 5G systems. Figures~\ref{fig:uep_combined}(a) and~\ref{fig:uep_combined}(b) show that the unequal error protection (UEP) property is intrinsic to the LDPC and the polar codes in a 5G communication system. In the first part of this work, we implement an LDPC code-based web page and video transmission system using the unequal error protection property of LDPC codes and the idea of approximate communication. Let us elaborate.

\subsection{Throughput Model}
\label{timing_model}

To measure the total time taken for data transmission, we use the mathematical model given by Padhye et al.~\cite{tcp_timing_model}. They have proposed a model of TCP throughput as a function of the packet loss probability, \emph{Th(p)}. The formula for \emph{Th(p))} is given as follows:

\begin{equation*}
	a = \frac{W_{max}}{RTT}, b = RTT*\sqrt{\frac{2bp}{3}} + RTO*3\sqrt{\frac{3bp}{8}}p(1+32p^2)
\end{equation*}

\vspace{-5px}

\begin{equation}
	Th(p) = min \Bigg( a, \frac{1}{b} \Bigg)
\end{equation}

Here, $W_{max}$ is the maximum value of the sender's congestion window size. The congestion window size is used to determine how many packets can be sent by the sender at a time. RTT is the round-trip time. The round-trip time (RTT) is the time taken for a packet sent by a sender to reach its destination, plus the time it takes for the sender to receive an acknowledgment of that packet. The retransmission timeout (RTO) is the time set for the acknowledgment of a packet to be received. After the RTO, the packet is assumed to be lost and retransmitted. The factor $b$ is added to account for the delayed acknowledgments. Delayed acknowledgments is a technique where acknowledgments for several packets are combined into a single response. This reduces the overhead of the data transmission protocol.

QUIC is another transport layer protocol developed and deployed globally by Google~\cite{QUIC}. QUIC allows for faster connection establishment than TCP. QUIC supports multiple streams in one connection. This allows it to avoid the head-of-line blocking problem caused because of TCP's sequential delivery. One of the major advantages of QUIC over TCP is that QUIC is located in the user space while TCP is implemented in operating system kernels. Hence, making major changes to TCP is very difficult, but QUIC does not have this limitation. QUIC is built on top of UDP. The flow control and congestion control mechanisms of QUIC are similar to that of TCP~\cite{QUIC_similar_TCP}. Hence, both have the same throughput model~\cite{QUIC_compare_TCP_2}. QUIC is also a reliable transport protocol (reliability is achieved using retransmissions). So, taking the reliable QUIC or TCP as the baseline, where all the data bits are protected using retransmissions, would lead to a similar analysis and results~\cite{QUIC_compare_TCP}. Our approach is independent of the transport layer protocol and can be used for both TCP and QUIC-based transport.

\subsection{Proposed Approach}

The data content of a web page can be divided mainly into two categories. One is the textual content of the web page. This includes the HTML tags, CSS files, and the web page's text. The textual content of the web page needs to be received without any errors. This is essential for the web page to load correctly. For example, if some essential HTML tag is downloaded incorrectly, there will be an error in rendering the website. The second type of content is multimedia content. This includes the images and videos on the web page. Multimedia data can be received with a limited amount of error. The quality score for images and videos, as experienced by humans, deteriorates very slowly until a certain point. We operate on the safer side of the knee of this curve, and so we can safely tolerate errors in the received images and videos.

\begin{figure}[!t]
	\centering
	\subfloat[\centering Web page transmission]{{\includegraphics[width=6.0cm]{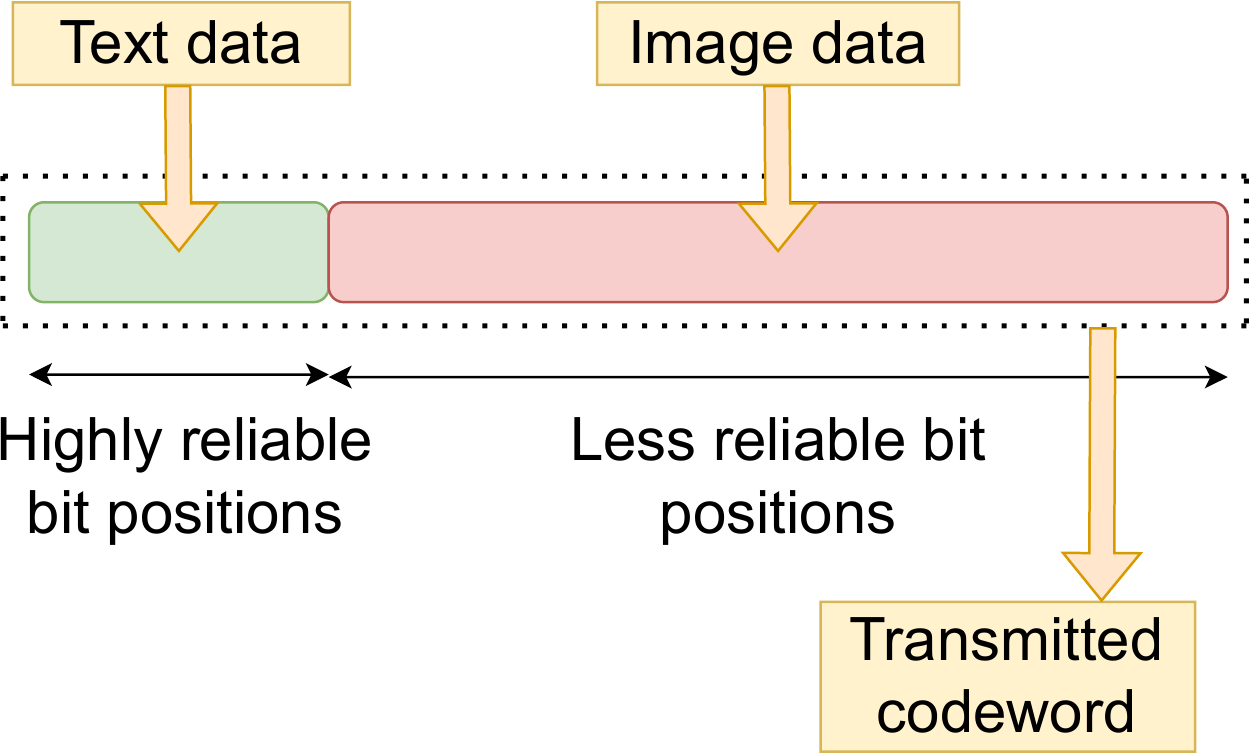} }}%
	\qquad \vspace{5px}
	\subfloat[\centering Video transmission]{{\includegraphics[width=6.0cm]{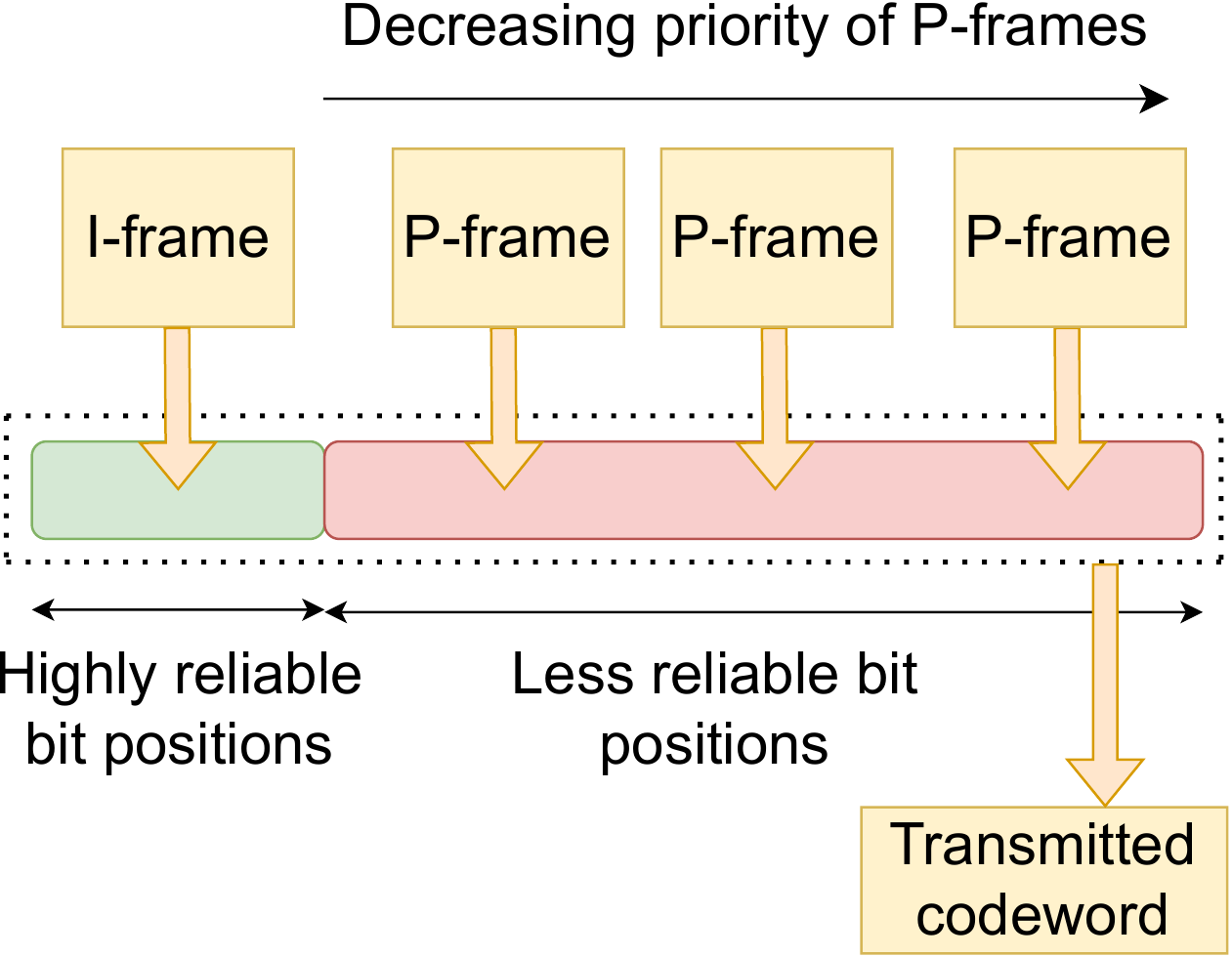} }}%
	\caption{Proposed approach}%
	\label{fig:proposed_approach}%
\end{figure}

\subsubsection{Web Page Transmission}

In the baseline method, the text data and the images are transmitted using separate codewords. We retransmit if any codeword is received in error to ensure 100\% accuracy of both types of data. We suggest a different approach. We consider that each codeword contains both text and image data. We treat the text data as high-priority data and the image data as low-priority data. The text data is mapped to the more protected bit positions of the codeword, and the image data is mapped to the less protected bit positions. The mapping is shown in Figure~\ref{fig:proposed_approach}(a). We ensure that the text data is received with 100\% accuracy. This is done by retransmitting the message if there is any error in the text data part of the codeword. For the image data, we consider two different scenarios: 

\begin{enumerate}
	\item The message is not retransmitted even if there is an error in the image data part of the codeword.
	\item The message is retransmitted if there is an error in the first \emph{k} MSB bits of any image data pixel.
\end{enumerate}

\subsubsection{Video Transmission}

In a modern video encoding algorithm like H.264/AVC, the complete video is arranged in a group of pictures (GOP) structure. The GOP is a collection of successive image frames in a video. The first frame of the GOP is called the I-frame, and the rest of the frames are called the P-frames. An I-frame (Intra-frame) is a complete JPG image, whereas a P-frame (Predicted-frame) contains only the changes compared to the previous frame. For good decoding accuracy of the video at the receiver, the I-frame is of the highest priority. The P-frame following the I-frame is the next most important frame. The priority of the P-frames keeps decreasing as we move further away from the I-frame. There is a clearly defined priority order of the different frames in the GOP. We take advantage of this priority order. In the conventional video transmission approach, all the frames in a GOP are equally protected using retransmissions. We suggest a different approach. We mix the data from the I-frame and the P-frames into a single codeword. The I-frame data is of the highest priority and is mapped to the most protected bit positions of the codeword. The data from the P-frames is mapped to the less protected bit positions in the order of priority. Figure~\ref{fig:proposed_approach}(b) shows the I-frame and the P-frame mapping onto the transmitted codeword. In our implementation, we consider 14 P-frames. During transmission, the I-frame is always protected using retransmissions. The number of protected P-frames is varied from 0 to 14, and we measure the number of retransmissions, and the received video quality score in each case.

In our work, we have proposed application-specific protocols that break the conventional internet protocol stack (TCP/IP protocol stack) as the behavior of the lower layer is decided by a higher layer (application layer). This is a common practice in wireless networks. Heinzelman et al.~\cite{application_spec_protocol} have previously proposed application-specific protocols for sensor networks and wireless delivery of videos. 5G networks support network slicing using software-defined networking (SDN) and network function virtualization (NFV)~\cite{network_slicing}. Network slicing allows the network operator to support multiple use cases on the same physical network and also run different protocols for each use case. Thus, modifying or even replacing the conventional protocol stack to improve network performance has gained momentum in 5G networks~\cite{replacement_TCP_IP_5G, ETSI_protocol}. We show that our approach leads to improved performance for multimedia web page transmission. We perform the following experiments:  

\begin{enumerate}
	\item Measure the gain in performance and the received image or video quality score when using polar codes as the error correcting codes. The experiments are performed for multiple channel conditions.
	
	\item Measure the gain in performance and the received image or video quality score when using LDPC codes as the error correcting codes. The experiments are performed for multiple channel conditions.
	
	\item Comparison of the results obtained for polar codes and LDPC codes.
\end{enumerate}

\section{System Model}
\label{proposedApproach2}

\begin{figure*}[!t]
	\centering
	\includegraphics[width=12cm]{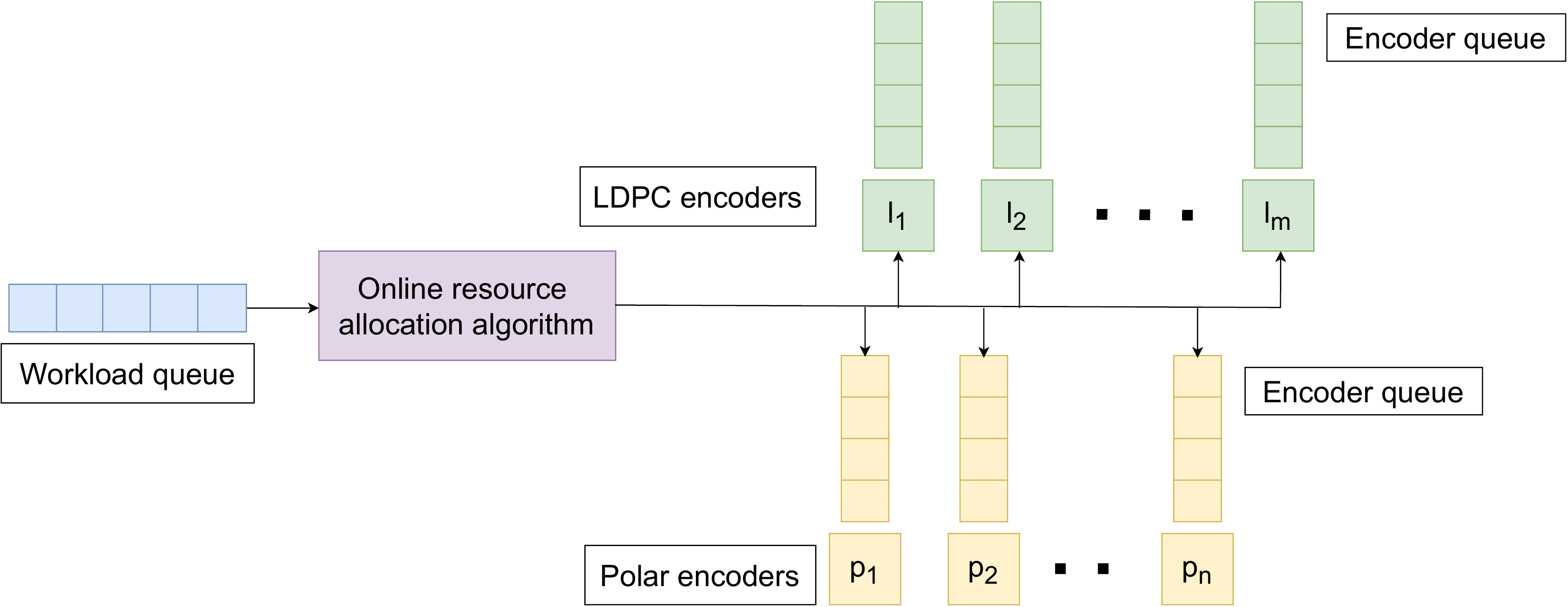}
	\caption{An overview of the system}
	\label{fig:system_model}
\end{figure*}

\begin{table*}[!t]

	\centering
	\begin{tabular}{ |c|c| } 
		\hline
		\rowcolor{Grayd} \textbf{Parameter} & \textbf{Description} \\
		\hline \hline
		\cellcolor{Grayl}job $j$ & A workload in the \emph{workload queue}\\ 
		\hline
		\cellcolor{Grayl}$s_j$ & Size of the workload $j$ (in MBs) \\
		\hline
		\cellcolor{Grayl}$r_j$ & Ratio of the text content size to the image content size of the workload $j$ \\
		\hline
		\cellcolor{Grayl}$q_j$ & Quality level of the workload $j$ \\
		\hline
		\cellcolor{Grayl}$m$   & Number of LDPC encoders \\
		\hline
		\cellcolor{Grayl}$n$   & Number of polar code encoders \\
		\hline
		\cellcolor{Grayl}$tl_j$ & Processing time when job $j$ is scheduled on one of the LDPC encoders \\
		\hline
		\cellcolor{Grayl}$tp_j$ & Processing time when job $j$ is scheduled on one of the polar code encoders \\
		\hline
		\cellcolor{Grayl}$E_b/N_o$ & Measure of channel quality (in dB) \\
		\hline
		
	\end{tabular}
	
	\caption{Description of the system parameters}
	
	\label{table:paramter_system_model}
	
\end{table*}

Consider an edge-compute node that processes the data requests received by a 5G base station. The edge node offloads the L1 processing operations to hardware accelerator cards connected using the PCIe protocol. During the downlink transmission of data from the base station to the UE, two computationally heavy L1 functions are the LDPC and polar encoding functions, respectively. Hence, the LDPC encoders and the polar code encoders are implemented as hardware-accelerated functions in the base station.

All the workload requests processed by the base station are stored in a queue. We call this the \emph{workload queue}. Every job $j$ in the \emph{workload queue} has some properties. The first one is the size of the workload, denoted by $s_j$ and measured in megabytes (MBs). A web page workload also has a certain text content size to image content size ratio value, denoted by $r_j$. It also has a quality level $q_j$. A quality level of 0 means that no bit in the image content of the web page is protected using retransmissions. Subsequently, a quality level of $k$ implies that the first $k$ MSB bits of the image data pixels are protected using retransmissions. The higher the quality level of a workload, the better the quality of the received image or video content. On the downside, a higher quality level leads to a lower gain in performance over the baseline approach. So, there is a trade-off between the received quality and the achieved gain in performance, and this trade-off can be controlled for every workload using its quality level. The quality level is decided based on the application and the required quality of service (QoS). All the parameters of the system model are listed in Table~\ref{table:paramter_system_model}.

For the compute nodes that process the workload requests, we have $m$ identical LDPC encoders, denoted as the set of processing nodes $\{l_1, l_2, .. l_m \}$. We also have $n$ identical polar code encoders, denoted as the set of processing nodes $\{p_1, p_2, .. p_n \}$. Every processing node has its own \emph{encoder queue} that stores the workloads scheduled to be processed on that node. Figure~\ref{fig:system_model} shows a diagram of the system model. A job $j$, if scheduled on any of the $m$ LDPC processing nodes requires $tl_j$ units of processing time, while when scheduled on any of the $n$ polar code nodes, it requires $tp_j$ units of processing time. The processing time of a workload on an LDPC or a polar code encoder depends on the size of the workload, the gain in performance achieved by the LDPC or the polar code encoder, the channel quality (decided by the $E_b/N_o$ of the channel), and the quality level of the workload. The rate at which requests are received by the base station is determined by the \emph{workload injection probability}. The \emph{workload injection probability} is the probability that a new workload request will be generated by any of the UEs in any given time interval.

In a conventional 5G-NR communication system, forward error correction for the user data is done using LDPC codes. As discussed in \cref{motivation}, polar code encoders are idle for a significant amount of time during the downlink transmission of data from the base station to the UE. In our work, we propose to use the polar code encoders also for encoding the user plane data. By using the idle time of the polar code encoders for error correction in the data plane, we optimally utilize the hardware resources on a 5G base station. It provides an opportunity to reduce the cost of the base station by reducing the number of encoders (accelerator cards) and makes it possible for the base station to serve the UE requests faster. Every request in the \emph{workload queue} must be assigned to one of the LDPC or the polar code encoders. We need an efficient scheduling algorithm to allocate the workload optimally. We use an online resource allocation algorithm to assign the current job $j$ to one of the processing nodes $\{l_1, l_2, .. l_m \}$ or $\{p_1, p_2, .. p_n \}$. The algorithm makes the decision based on the properties of the job $j$ and the current size of the \emph{encoder queues}. It tries to minimize the time taken to process all the workloads serviced by the base station.

\section{Online Resource Allocation Algorithms}
\label{proposedApproach3}

We implement four online resource allocation algorithms. One of these is a novel reinforcement learning (RL) based algorithm, while two others are new heuristics. The algorithms assign a workload from the \emph{workload queue} (see table~\ref{table:paramter_system_model}) to one of the LDPC or polar code encoders. For optimal use of the computation resources at the base station, we need an efficient resource allocation algorithm. So, we perform Monte Carlo simulations and compare the performance of these four algorithms. The algorithms are compared based on four important scheduling metrics: average throughput, average wait time, average flow time, and makespan. The average throughput is the throughput at which the base station processes the workloads, averaged over all the workloads. The wait time of a workload is the time it has to wait in the \emph{encoder queue} of the compute node it is assigned to. A workload should preferably be assigned to an \emph{encoder queue} that minimizes its wait time, i.e., to a compute node with fewer workloads already assigned to it. The flow time of a workload is the sum of the wait time of the workload and the time required to process the workload by the compute node to which it is assigned. The makespan of a set of workloads is the total time taken by the base station to service all the individual tasks in the \emph{workload queue}. All four metrics are important for comparing the performance of the resource allocation algorithms~\cite{metrics_algorithm}. We now describe the four algorithms.

\subsection{Weighted Flow Time Minimization Algorithm (WFTM)}

Anand et al.~\cite{weighted_flowtime} explore the online job scheduling problem on multiple machines. The authors propose a theoretically provable greedy algorithm that minimizes the weighted sum of flow time of the jobs. This is the best algorithm for weighted flow time minimization that the traditional optimization algorithms community has to offer. We implement this algorithm as the first candidate algorithm for workload scheduling at the base station.

We have a set of $m + n$ machines and a set of jobs that arrive over time. A job $j$ requires $p_{ij}$ units of processing time and has a weight $w_{ij}$ if it is scheduled on machine $i$. In our implementation, $p_{ij}$ is equal to $tl_j$ or $tp_j$ depending on whether the workload is assigned to an LDPC encoder or a polar code encoder, respectively. The weight of a workload is the gain in performance observed by the workload. This again depends on which type of encoder the workload is assigned to. The weighted flow time of a job $j$ being processed on machine $i$ is $w_{ij}$ times the flow time of the job. When a new job arrives in the \emph{workload queue}, the algorithm immediately dispatches it to one of the machines. The dispatch algorithm is a greedy algorithm. When a new job $j$ arrives, the algorithm computes the following: for every machine $i$, what will be the increase in the weighted flow time if job $j$ is dispatched to machine $i$. Then, the algorithm assigns job $j$ to the machine for which the increase in the weighted flow time is the minimum. A limitation of this algorithm is that it assumes that the processing times and the weights of the jobs are exactly known.

\subsection{Stochastic Multi-Armed Bandit Algorithm (SMAB)}

The second algorithm that we implement is the stochastic multi-armed bandit algorithm. We propose this novel algorithm by formulating the resource allocation problem at a 5G base station as a multi-armed bandit problem, as has previously been done by Feki et al.~\cite{multi_armed_bandit} for LTE networks.

The number of arms of the multi-armed bandit equals the total number of machines. So, $k = m + n$. We use the upper confidence bound (UCB)~\cite{multi_armed_bandit_ucb} strategy to solve the multi-armed bandit problem. At every time step $t$, we perform the action of choosing one of the $m + n$ machines to run the current workload. After every action, we receive a reward that is unknown beforehand. If a job $j$ is allocated to run on machine $i$, the reward ($\mathcal{R}$) is equal to: 

\begin{equation}\label{equation:reward_mab}
	\mathcal{R} = \frac{\text{$Gain\_Perf_{ij}$}}{\text{$Enc\_Occ_{i}$ + 1}}
\end{equation}

Here, $Gain\_Perf_{ij}$ is the gain in performance for job $j$ when running on machine $i$, and $Enc\_Occ_{i}$ is the current \emph{encoder queue} occupancy of machine $i$. The $Gain\_Perf_{ij}$ value is obtained by performing a characterization of the LDPC and polar code encoders. The characterization results will be presented in \cref{char_results}. The \emph{encoder queue} occupancy of the machines determines the current state of the system and changes every time a job is allocated to one of the machines. As the reward function is dependent on the state of the system, this algorithm is a contextual bandit variant of the multi-armed bandit algorithm. The overall objective of the resource allocation system is to maximize the performance of the system. So, when the algorithm chooses to run the job $j$ on a machine that provides a higher performance improvement, we get a larger reward. Deciding on the reward based on only the performance improvement of the machine runs the risk of the algorithm assigning all the jobs to the machine that provides the maximum performance improvement. This will decrease the overall performance of the system. So, we add the current \emph{encoder queue} occupancy of the machine $i$ to which job $j$ is assigned in the denominator of the reward function. This ensures that the workloads are distributed equally between the encoders such that all the encoder nodes on the hardware accelerator cards have high utilization. We add 1 to the denominator to handle situations when the \emph{encoder queue} length is zero. The gain in performance for job $j$ when running on machine $i$ is not fixed and follows a Gaussian distribution around a mean value. We sample the gain in performance from a Gaussian distribution with the mean equal to the mean observed value of the gain in performance. This leads to the stochastic behavior of the algorithm. The UCB algorithm tries to maximize the total reward earned after all the tasks are scheduled.

\subsection{Random Allotment Algorithm (Random)}

We implement a random heuristic algorithm for resource allocation. This algorithm randomly assigns the current job $j$ to any of the $m + n$ machines. This algorithm is pretty straightforward, and we use this as a baseline for comparison.

\subsection{Min-Queue Allotment Algorithm (Min-Queue)}

The fourth candidate algorithm is the min-queue heuristic algorithm. This algorithm tracks the \emph{encoder queue} lengths. For a job $j$, if $tl_j < tp_j$ (See Table~\ref{table:paramter_system_model}), the job is assigned to one of the LDPC encoders. Among the $m$ LDPC encoders, the job is assigned to the one that has the minimum \emph{encoder queue} occupancy. On the contrary, if $tl_j \ge tp_j$, the job is assigned to one of the polar code encoders. Among the $n$ polar code encoders, the job is assigned to the encoder with the minimum \emph{encoder queue} occupancy.

\section{Evaluation Setup}
\label{evaluationSetup}

\begin{figure}[!t]
	\centering
	
	\includegraphics[width=7.5cm]{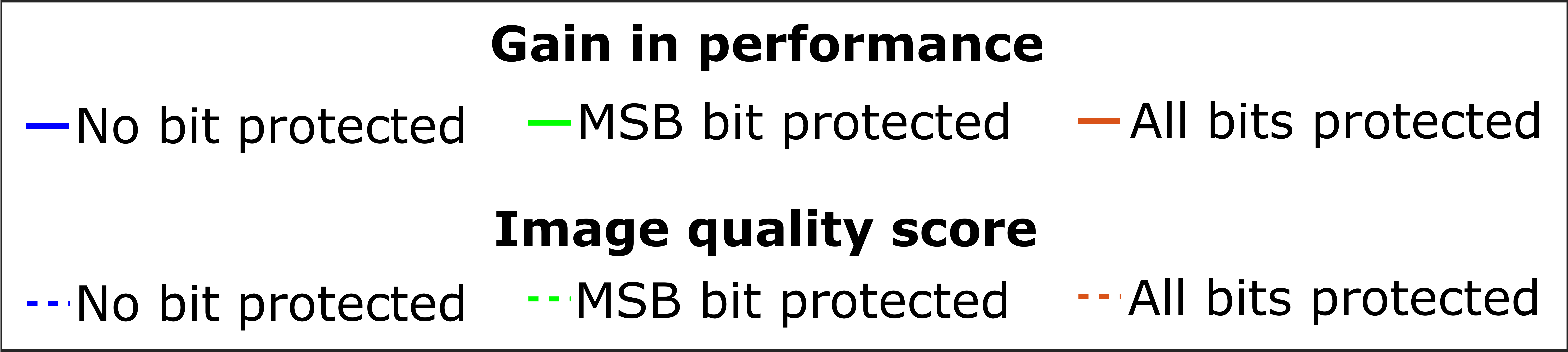} \\
	
	\subfloat[\centering $E_b/N_o$ = 1.5 dB]{{\includegraphics[width=7.5cm]{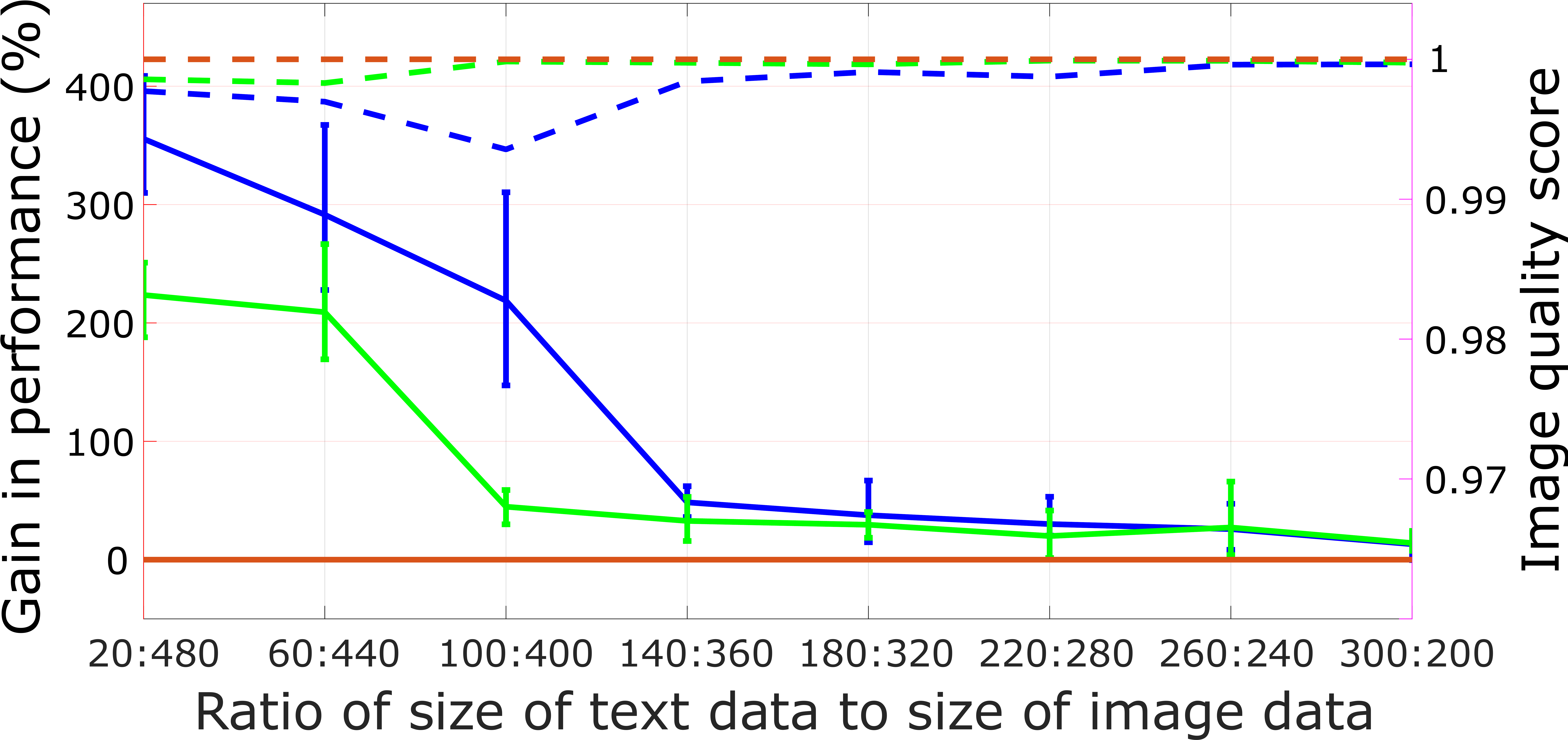} }}%
	\qquad \vspace{2px}
	\subfloat[\centering $E_b/N_o$ = 2 dB]{{\includegraphics[width=7.5cm]{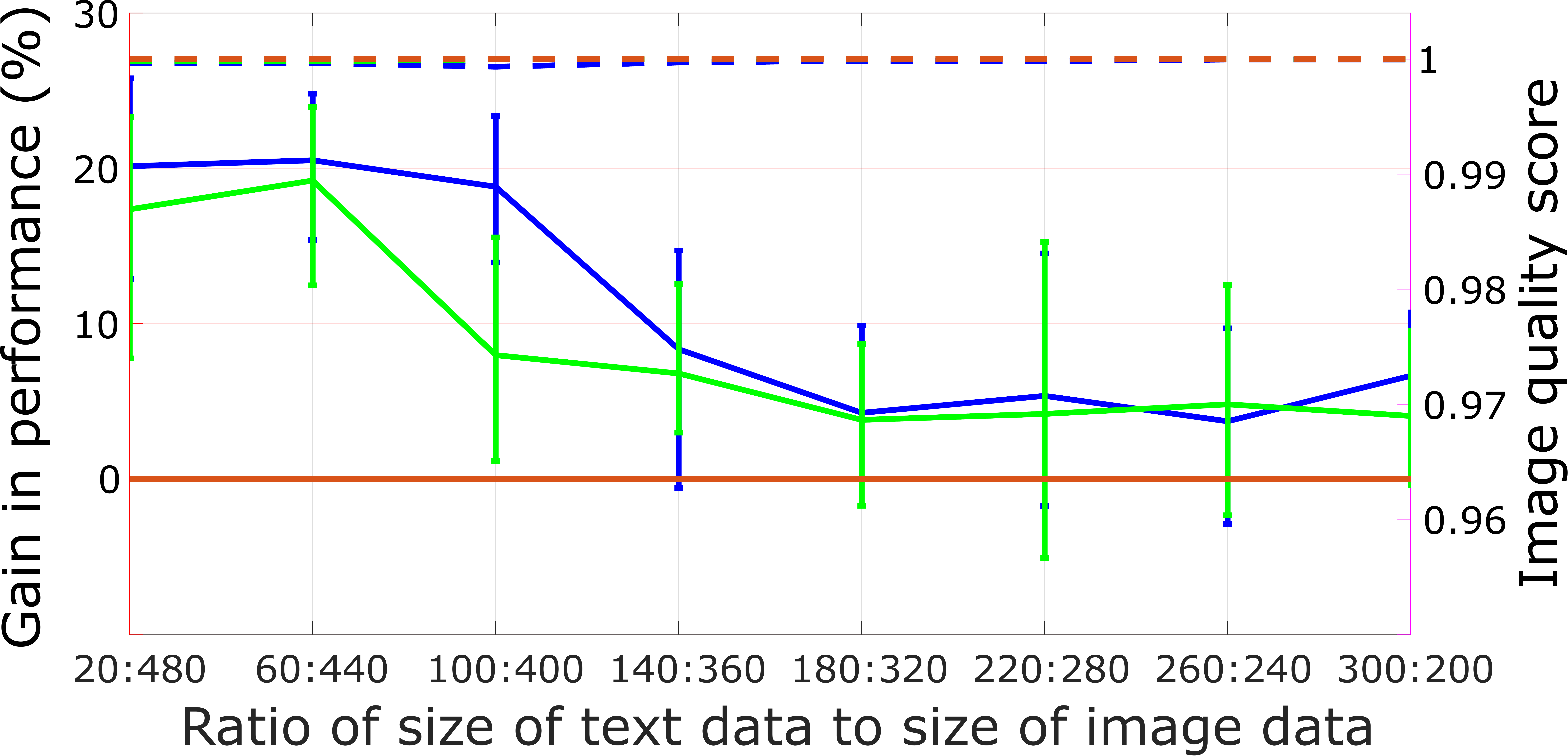} }}%
	\caption{Characterization results for web page transmission using LDPC codes}%
	\label{fig:char_webpage_ldpc}%
\end{figure}

\subsection{Setup}

We simulate the proposed approximate communication system for multimedia web pages in Matlab (version R2021a). The link delay for a 5G network is found using the Netsim v12.0 simulator~\cite{netsim}. We use the Matlab 5G Toolbox~\cite{matlab_5G_toolbox} to generate a standard-compliant 5G-NR downlink waveform. The Monte Carlo simulations used to compare the four resource allocation algorithms are implemented using Python 3. All the experiments are done on an $11^{th}$ Gen Intel i7-1165G7 machine running the Windows 11 operating system.

\subsection{Metrics and Datasets}

The received image quality is measured using the MS-SSIM score~\cite{ms_ssim}. The video quality score is the average of the MS-SSIM scores for all the image frames in the GOP (group of pictures)~\cite{gop}. For video data, we use the Big Buck Bunny video~\cite{big_buck_bunny}. Using Matlab simulations, we find the number of transmissions and retransmissions required to send the downlink data from the base station to a UE. This data is then fed into the timing model (\cref{timing_model}) to determine the total transmission time. This timing model has previously been used to estimate data transmission throughput~\cite{tcp_timing_literature, polarcode_IPSN}. The system's performance is defined as the reciprocal of the total transmission time.

\begin{figure}[!t]
	\centering
	
	\includegraphics[width=7.5cm]{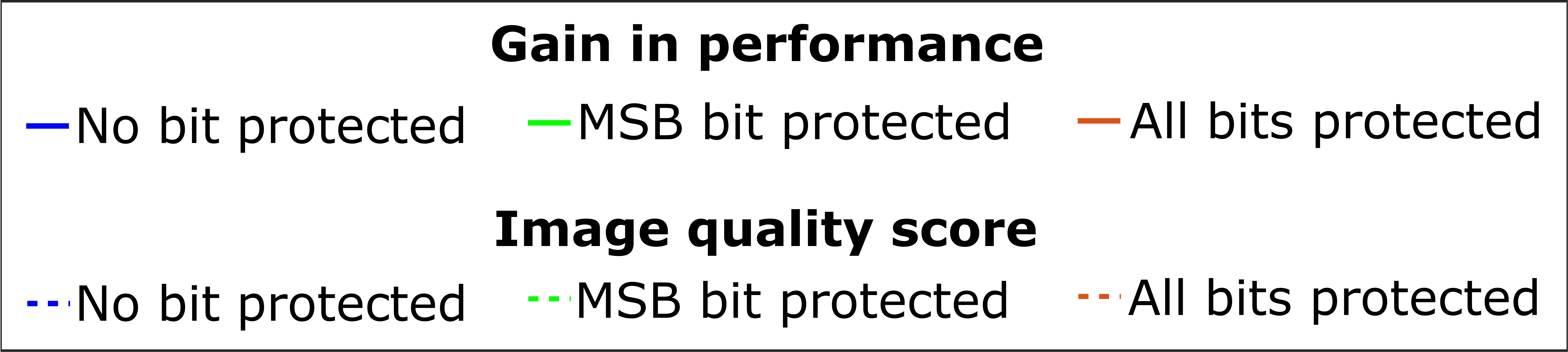} \\
	
	\subfloat[\centering $E_b/N_o$ = 1.5 dB]{{\includegraphics[width=7.5cm]{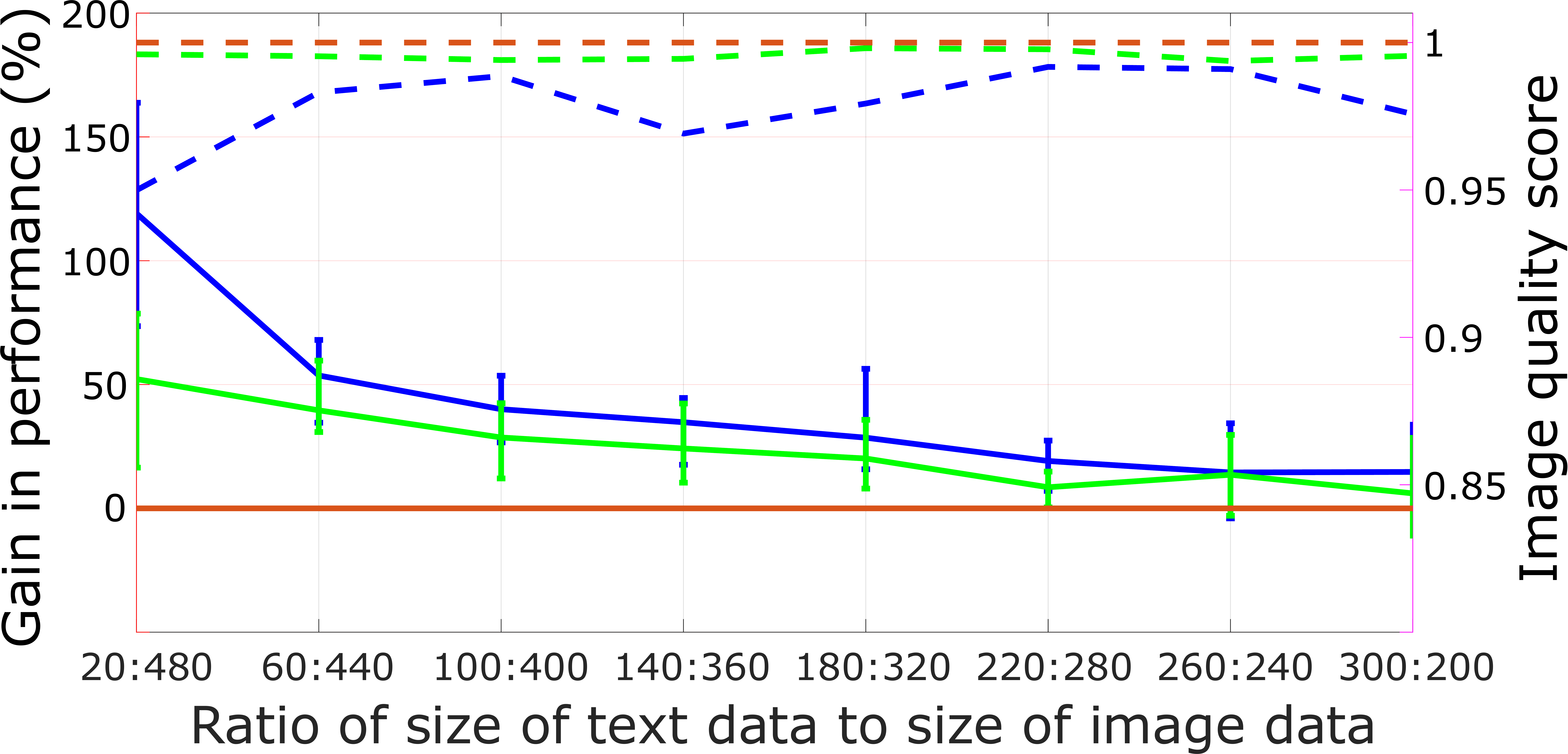} }}%
	\qquad \vspace{2px}
	\subfloat[\centering $E_b/N_o$ = 2 dB]{{\includegraphics[width=7.5cm]{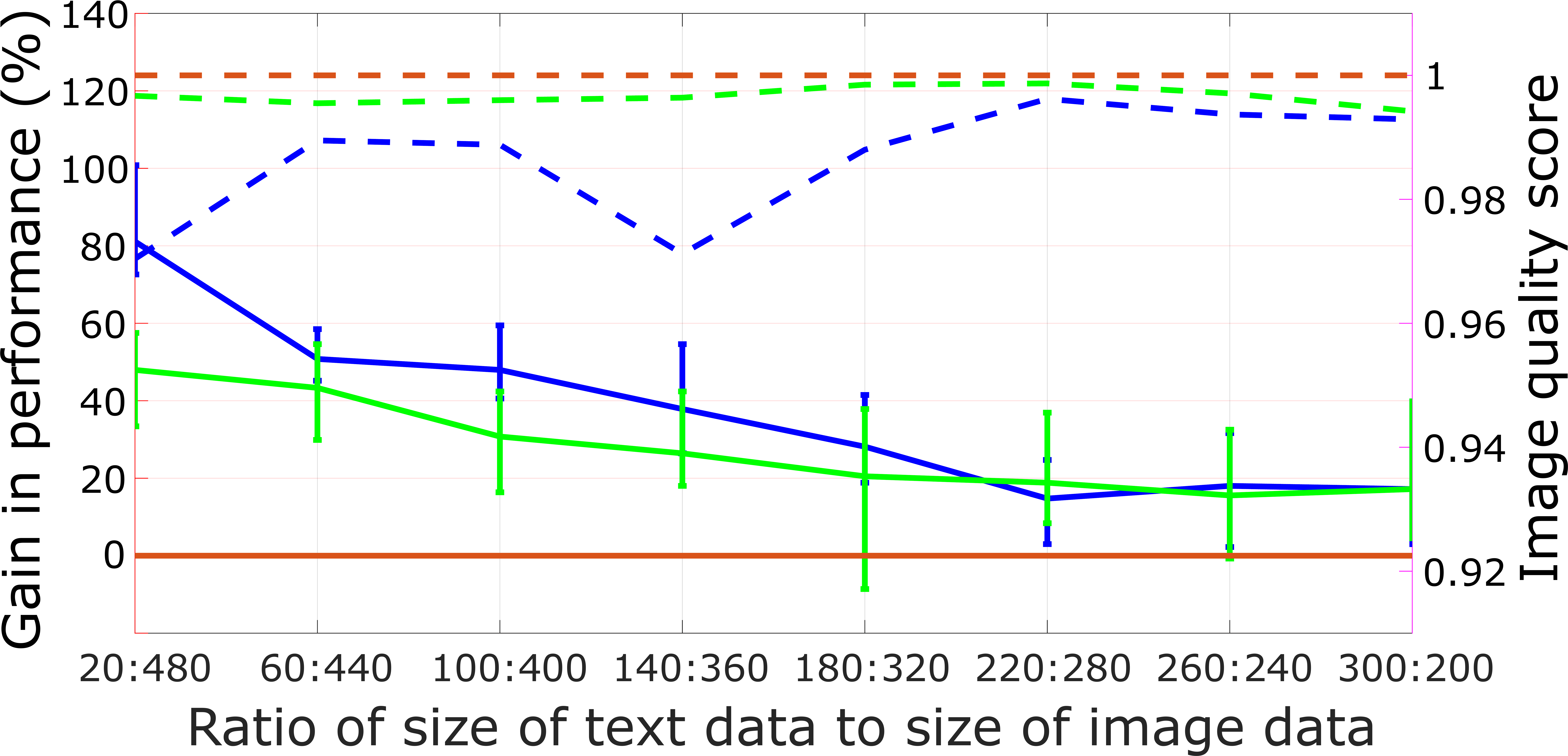} }}%
	\caption{Characterization results for web page transmission using polar codes}%
	\label{fig:char_webpage_polar}%
\end{figure}

\subsection{Hyperparameters}

\begin{itemize}
	
	\item Using the NetSim simulator~\cite{netsim} for the 5G-NR network and setting the single-hop distance to 190 m; the link delay is found to be 1.38 milliseconds. Using this value of link delay, we calculate the retransmission timeout (RTO) and the round-trip time (RTT) values for the timing model discussed in \cref{timing_model}. 
	
	\item For web page transmission, the ratio of the text data size to image data size is a hyperparameter. The performance of the system and the quality of the received image vary with this \emph{ratio}. We perform a design space exploration by varying the \emph{ratio} value from 20:480 to 300:200 in our experiments. For video transmission, we assume that the GOP has 14 P-frames and measure the performance and video quality score by varying the number of protected P-frames from 0 to 14.
	
	\item We solve the stochastic multi-armed bandit formulation of the resource allocation algorithm using the upper confidence bound (UCB) strategy. The learning rate for the UCB strategy is set at 0.1, and the $c$ value for the strategy is set to 15. Using hyperparameter tuning, we find that the algorithm has the minimum \emph{regret} when the $c$ value is set to 15.
	
	\item When performing the full-system simulations for a 5G base station, we consider multiple channel conditions. The channel condition is characterized by the $E_b/N_o$ value of the channel. We consider the $E_b/N_o$ values in the range of 1 dB to 2 dB. This is representative of the channel conditions experienced in remote and rural areas because of the greater distance of the users from the base station as compared to urban areas~\cite{channel_EbNo_range}.  
	
\end{itemize}

\section{Characterization and Comparison of the LDPC and the Polar Code-Based Multimedia Web Page Transmission Systems}
\label{char_results}

\begin{figure*}[!t]
	\centering
	
	\includegraphics[width=13.2cm]{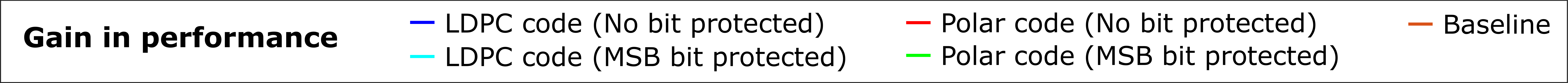} \\ \vspace{5px}
	
	\includegraphics[width=13.1cm]{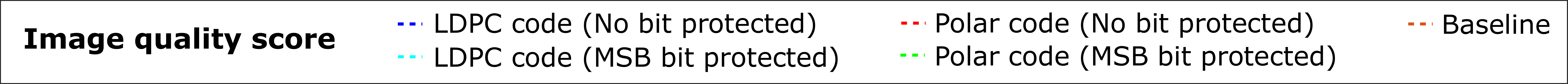} \\
	
	\subfloat[$Eb/No$ = 1 dB]{{\includegraphics[width=5.9cm]{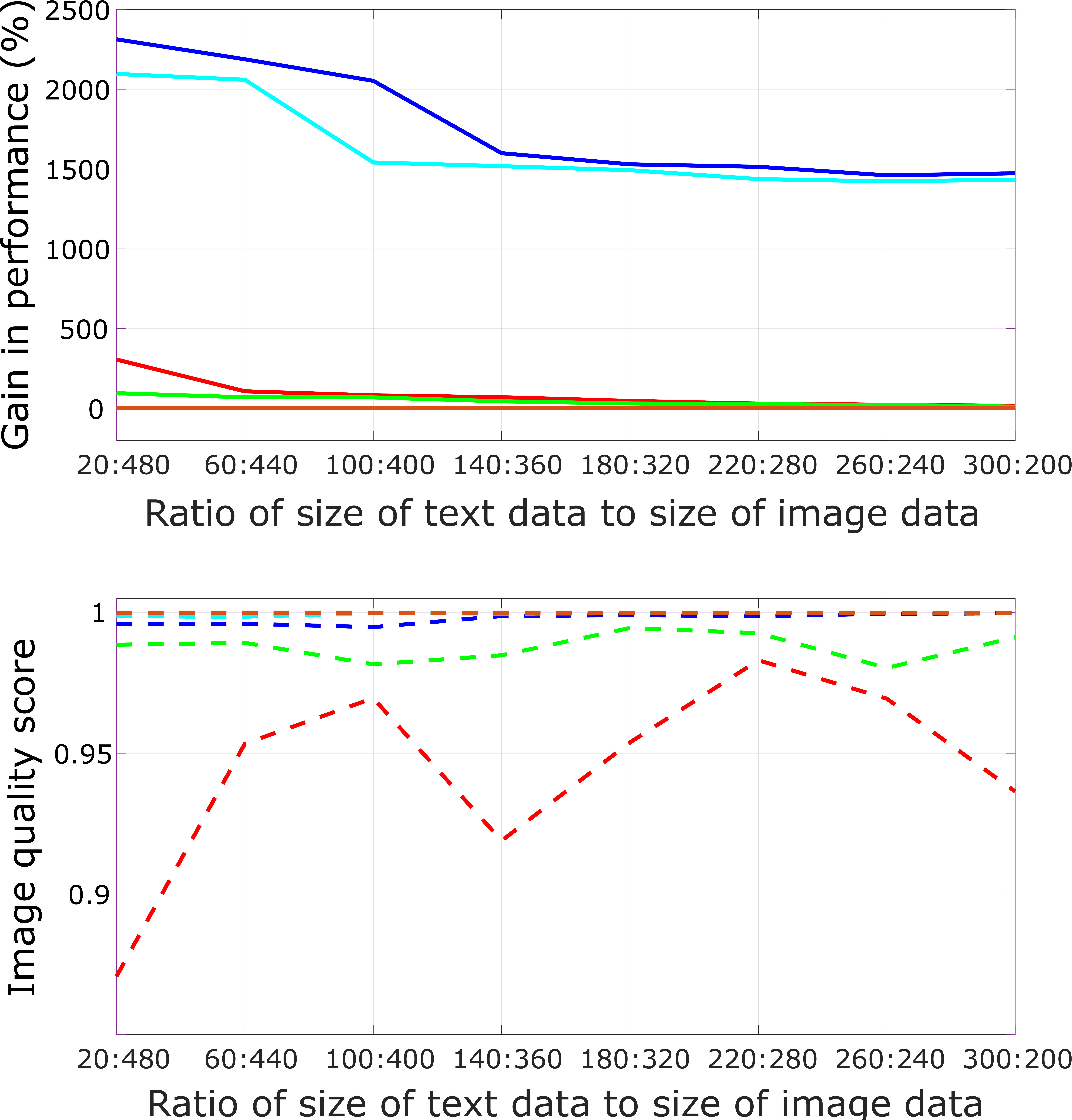} }}%
	\qquad \hspace{-20px}
	\subfloat[$Eb/No$ = 1.5 dB]{{\includegraphics[width=5.9cm]{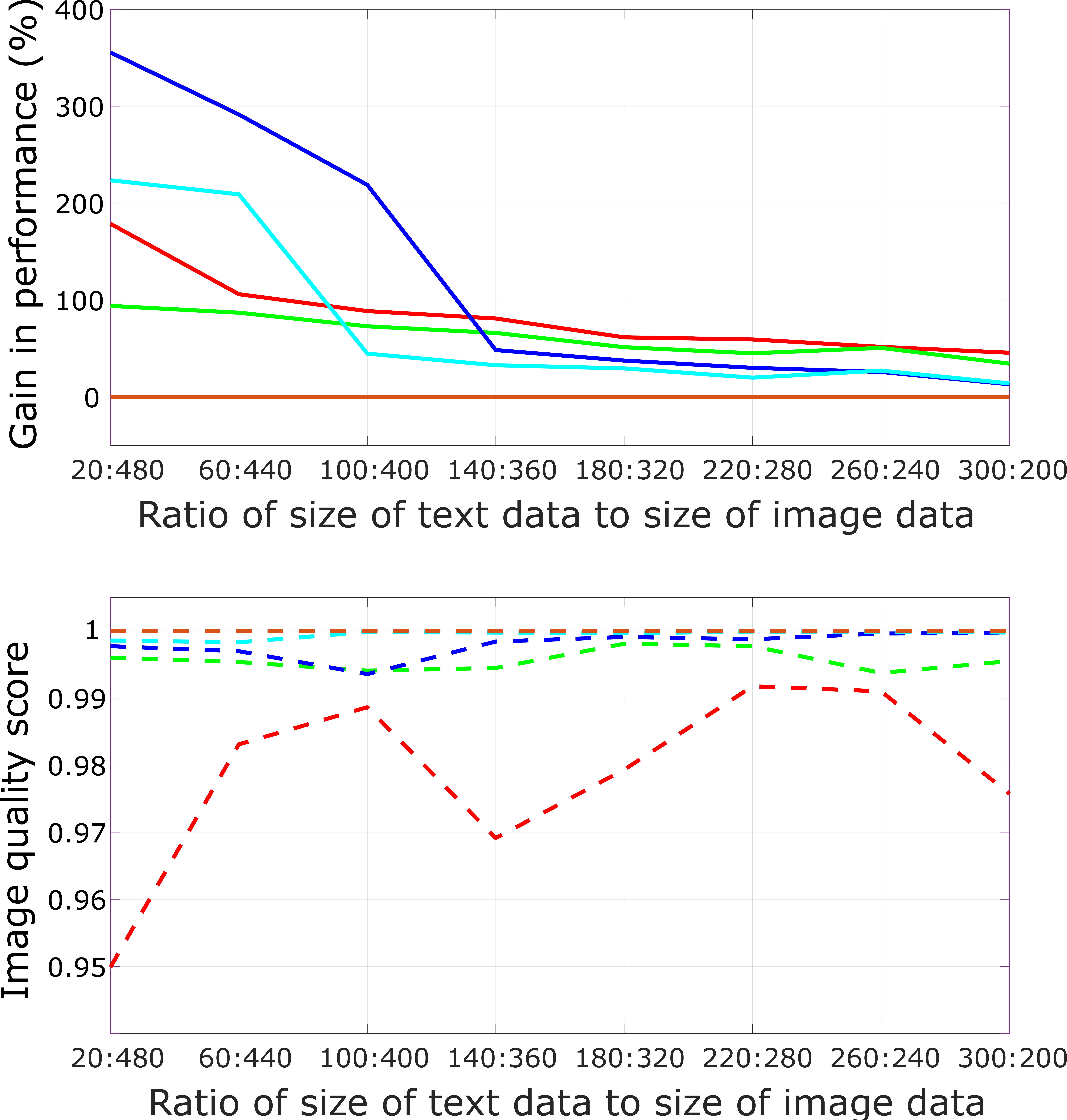} }}%
	\qquad \hspace{-20px}
	\subfloat[$Eb/No$ = 2 dB]{{\includegraphics[width=5.9cm]{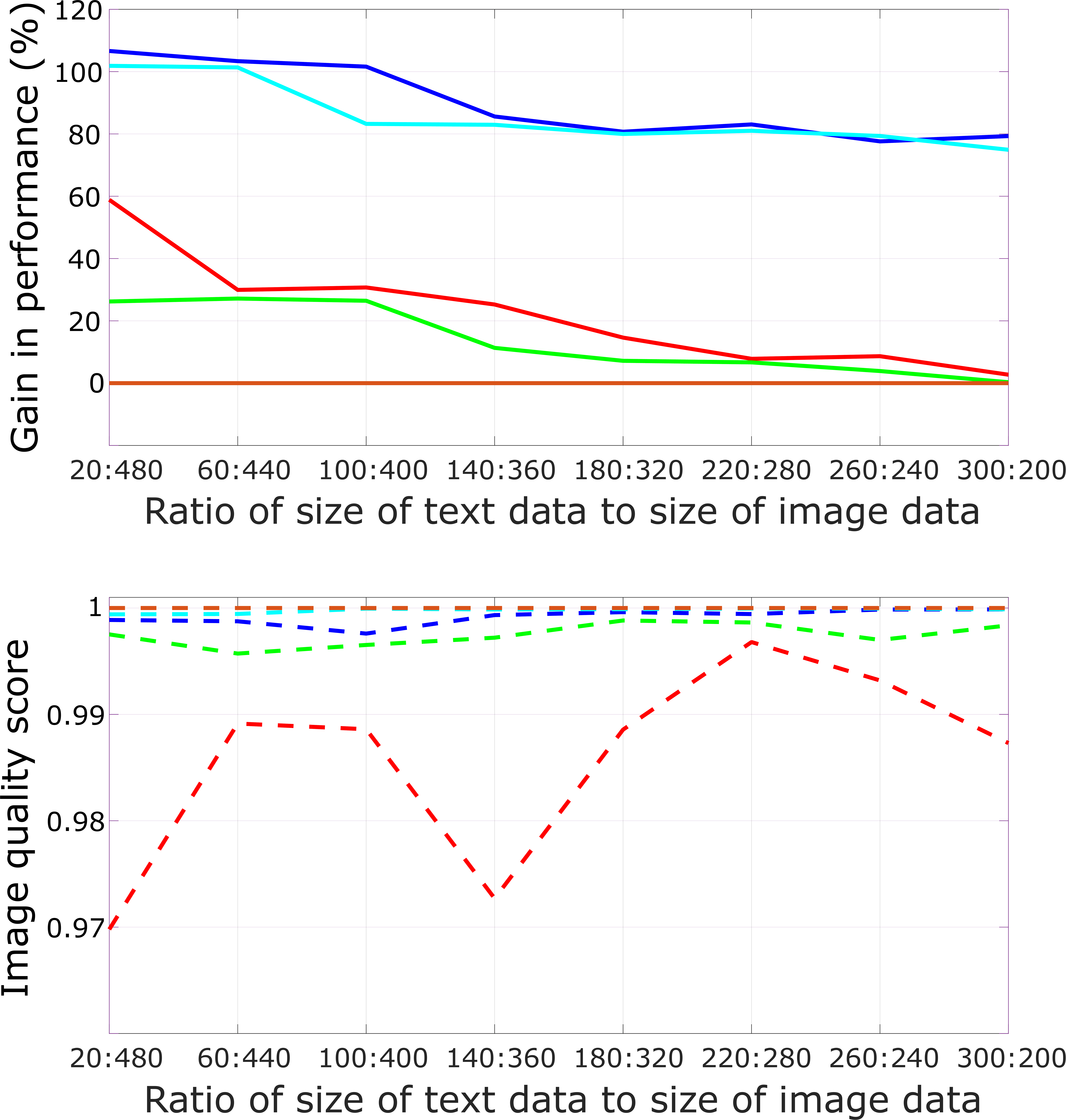} }}%
	 
	\caption{Characterization results for web page transmission (comparison of LDPC and polar codes) - polar code performs better than LDPC code at $E_b/N_o=$ 1.5 dB}%
	
	\label{fig:char_webpage_comparison}%
\end{figure*}


We perform MATLAB simulations to compare the gain in performance and the received image or video quality for data transmission using LDPC and polar codes. We find the time required for web page transmission using the Matlab simulations and the model discussed in \cref{timing_model}. The performance is defined as the reciprocal of the time taken to transmit a multimedia web page (with images or video) from the sender to the receiver (including retransmissions). The experiments are performed for multiple channel conditions. The channel condition is specified using the $E_b/N_o$ value of the channel. In the characterization results, we present the graphs for $E_b/N_o$ values of 1 dB, 1.5 dB, and 2 dB. These $E_b/N_o$ values are representative of the channel conditions in rural areas~\cite{channel_EbNo_range}.


\subsection{Web Page Transmission}


We plot the graph for the gain in performance using our proposed approach and the quality score of the received image when a web page is transmitted over a wireless channel. In the graphs, the gain in performance is plotted using solid lines, while the received image quality score is plotted using dashed lines. We plot the results for the baseline situation (where all the bits are protected using retransmissions), the situation when no image bit is protected, and the scenario when the first MSB bit of the image pixels is protected using retransmissions. The graphs with LDPC codes as the error correction code for the data plane are shown in Figure~\ref{fig:char_webpage_ldpc}. The graphs with polar codes as the error correction code for the data plane are shown in Figure \ref{fig:char_webpage_polar}.



To get a clear view of the comparison between transmission using LDPC code and transmission using polar codes, we plot additional graphs that show the results for both the LDPC and polar codes on the same graph. The comparison graphs for web page transmission are shown in Figure~\ref{fig:char_webpage_comparison}. The gain in performance is measured as compared to the baseline when all the bits in the transmitted codeword are protected using retransmissions. The error correction code with the lower performance is taken as the baseline to plot the graphs. We observe that at an $E_b/N_o =$ 1 dB, there is a large gap in the performance of the two codes. LDPC codes provide a significantly higher gain in performance. The image quality score is close to 1 for all the \emph{ratio} values for both LDPC and polar codes. The performance gap between the two codes reduces when $E_b/N_o =$ 1.5 dB. The performance for LDPC codes drops drastically for ratios greater than 100:400. This is because of the nature of the unequal error protection for the codeword bits. The most unreliable bits in an LDPC codeword are between the information bit positions 120 and 150, as shown previously in Figure~\ref{fig:uep_combined}(a). So, when we start protecting bits in this range using retransmissions, the performance takes a hit. On increasing the $E_b/N_o$ to 2 dB, we observe that the performance gap is less than for $E_b/N_o =$ 1 dB. Overall, we observe that in most situations, LDPC codes provide a higher gain in performance. However, when the $E_b/N_o =$ 1.5 dB and the \emph{ratio} value of the web page is more than 100:400, polar codes provide a higher gain in performance than LDPC codes. This shows that polar codes are a better choice for error correction codes in certain situations. This supports our argument to use polar code encoders along with the LDPC code encoders for error correction in the data plane of the 5G-NR communication system.


\subsection{Video Transmission}

\begin{figure}[!t]
	\centering
	
	\includegraphics[width=6.5cm]{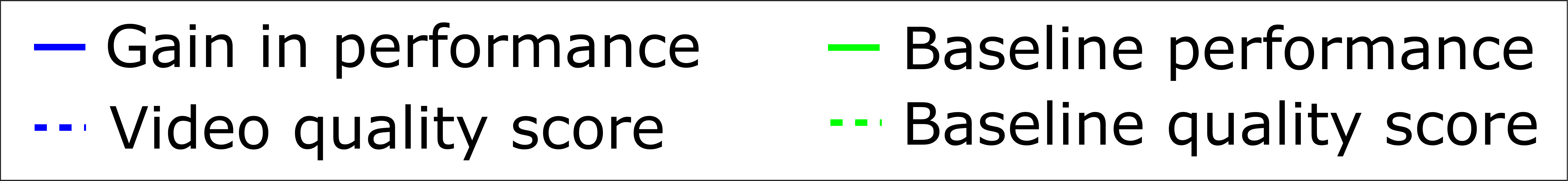} \\
	
	\subfloat[\centering $E_b/N_o$ = 1.5 dB]{{\includegraphics[width=7.5cm]{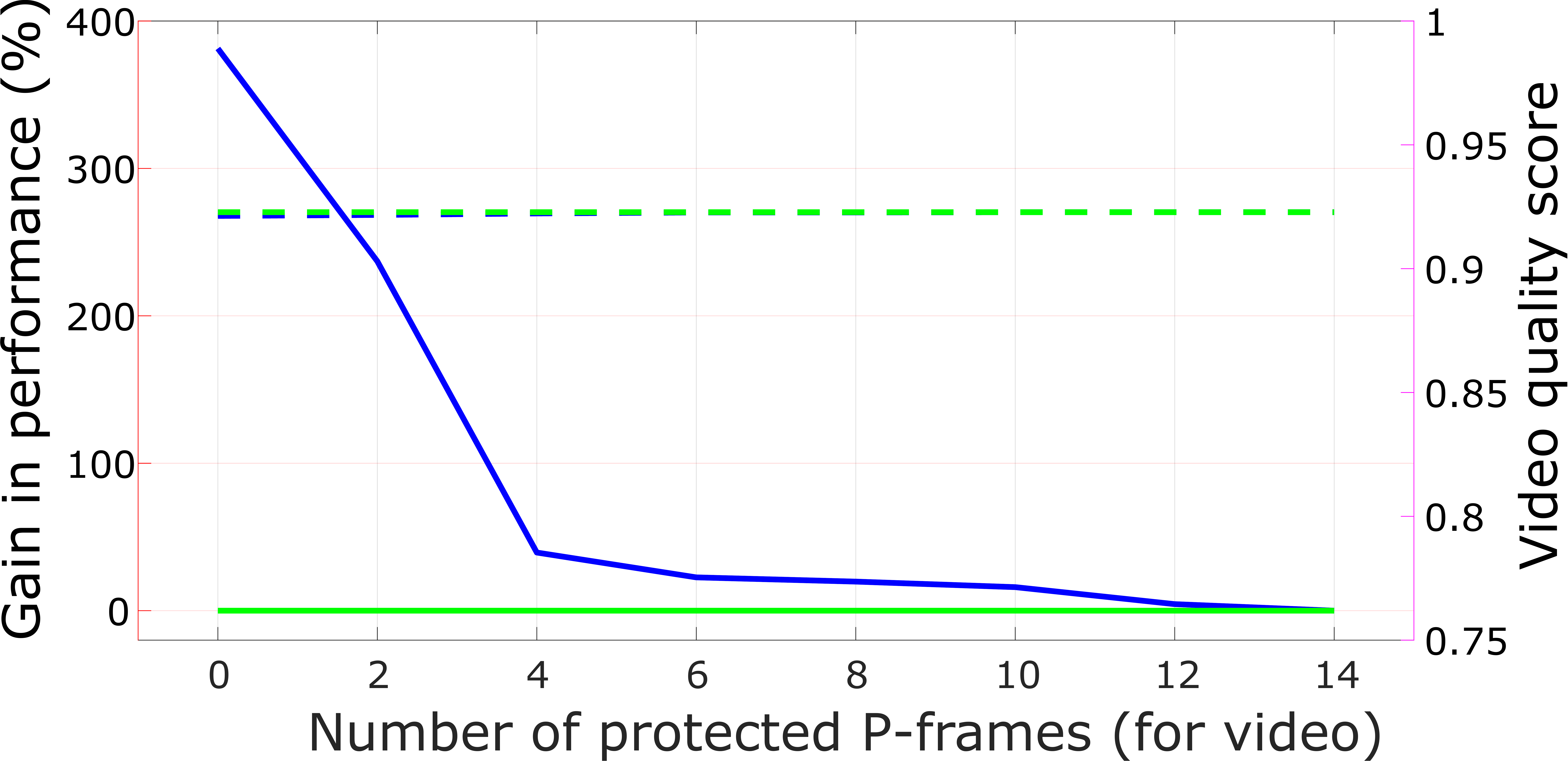} }}%
	\qquad \vspace{2px}
	\subfloat[\centering $E_b/N_o$ = 2 dB]{{\includegraphics[width=7.5cm]{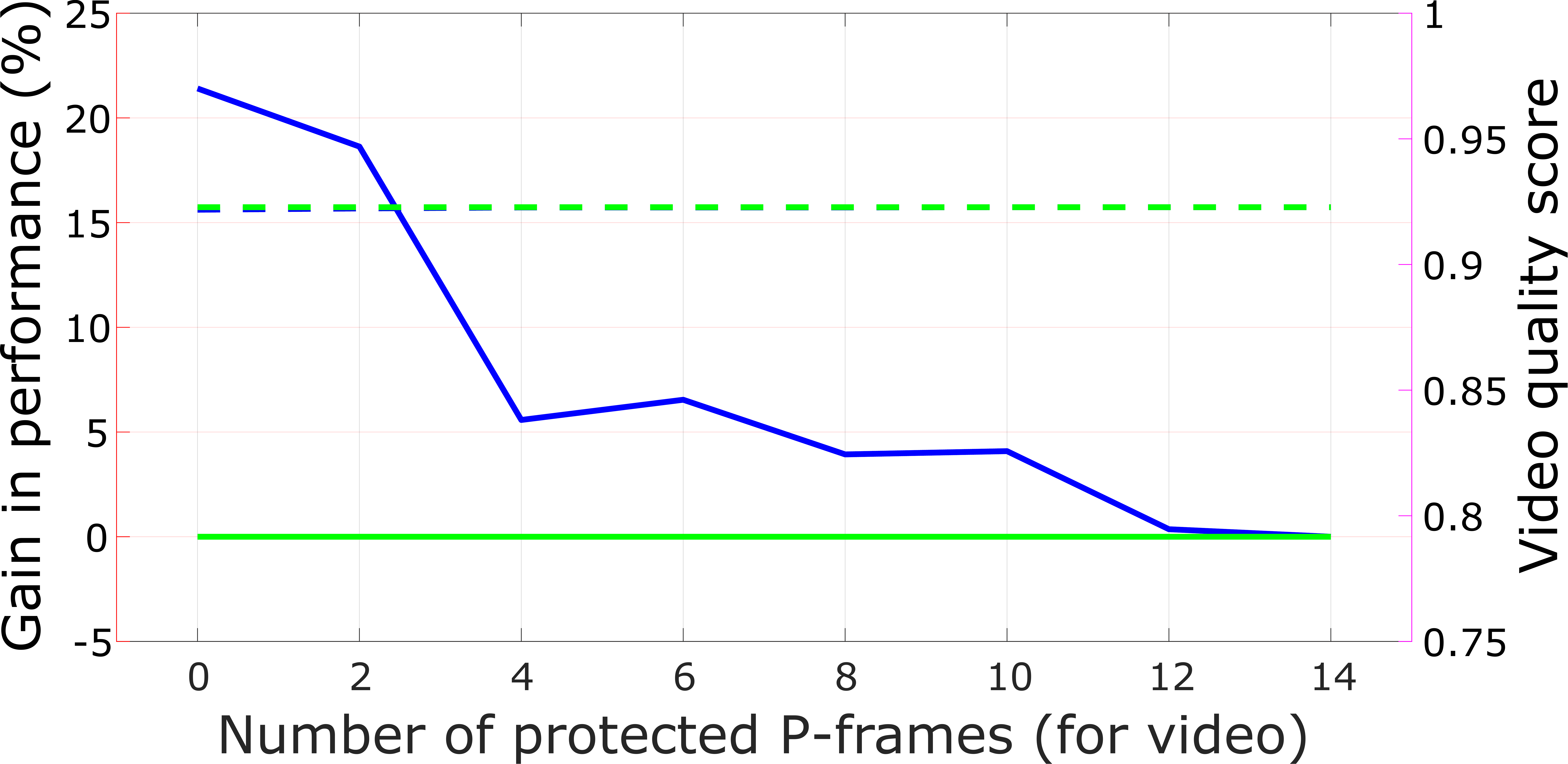} }}%
	\caption{Characterization results for video transmission using LDPC codes}%
	\label{fig:char_video_ldpc}%
\end{figure}

\begin{figure}[!t]
	\centering
	
	\includegraphics[width=6.5cm]{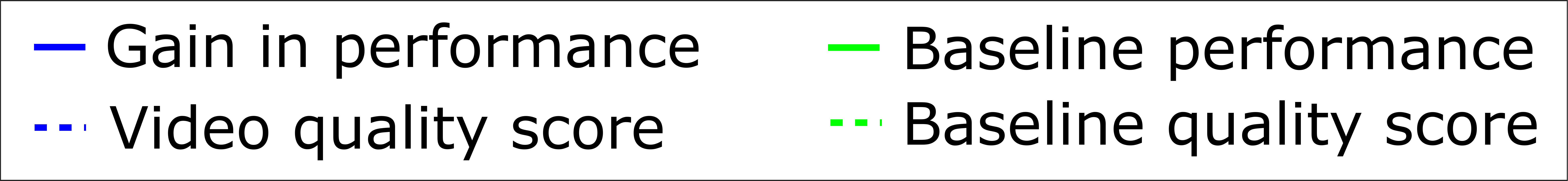} \\
	
	\subfloat[\centering $E_b/N_o$ = 1.5 dB]{{\includegraphics[width=7.5cm]{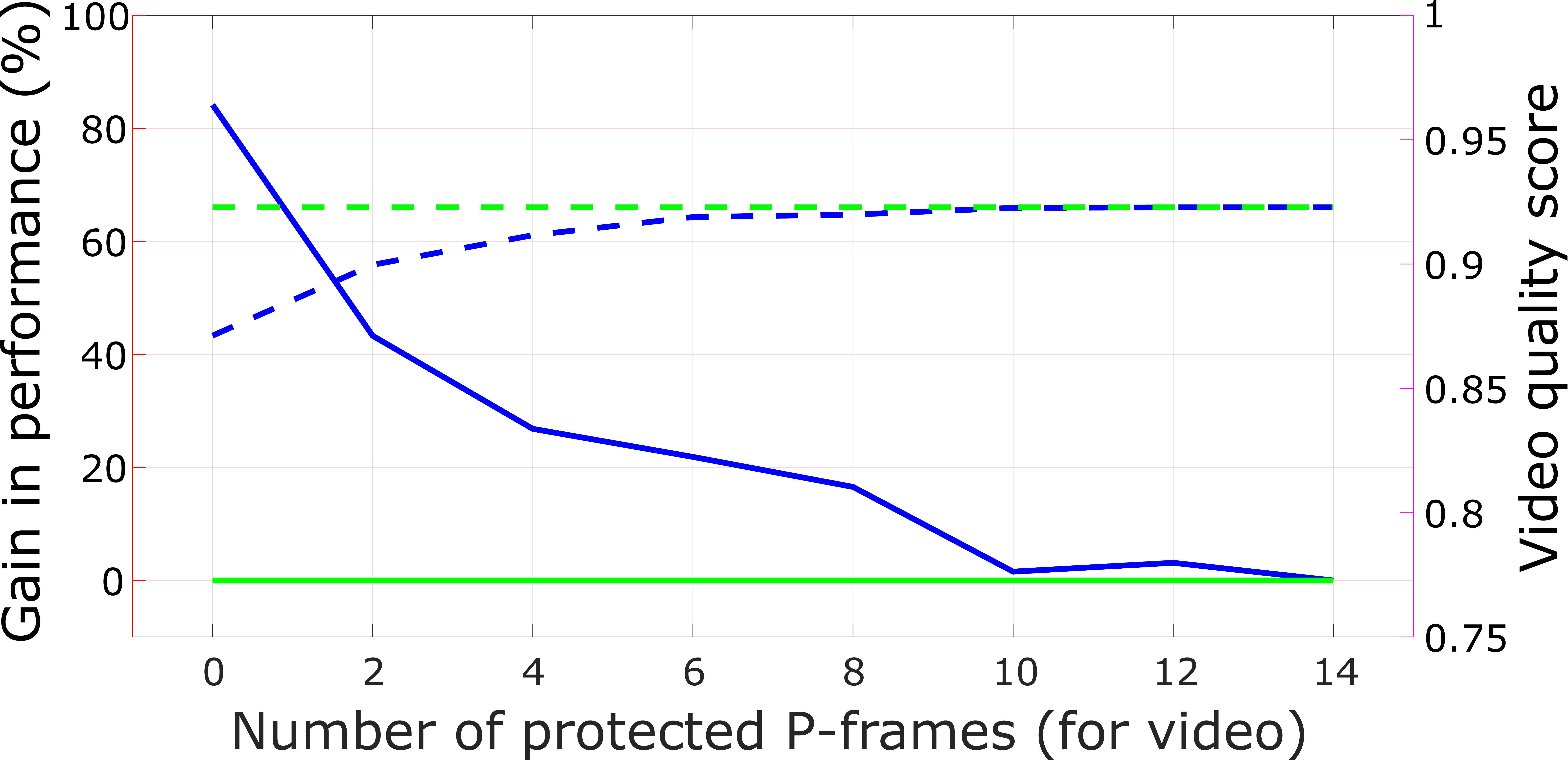} }}%
	\qquad \vspace{2px}
	\subfloat[\centering $E_b/N_o$ = 2 dB]{{\includegraphics[width=7.5cm]{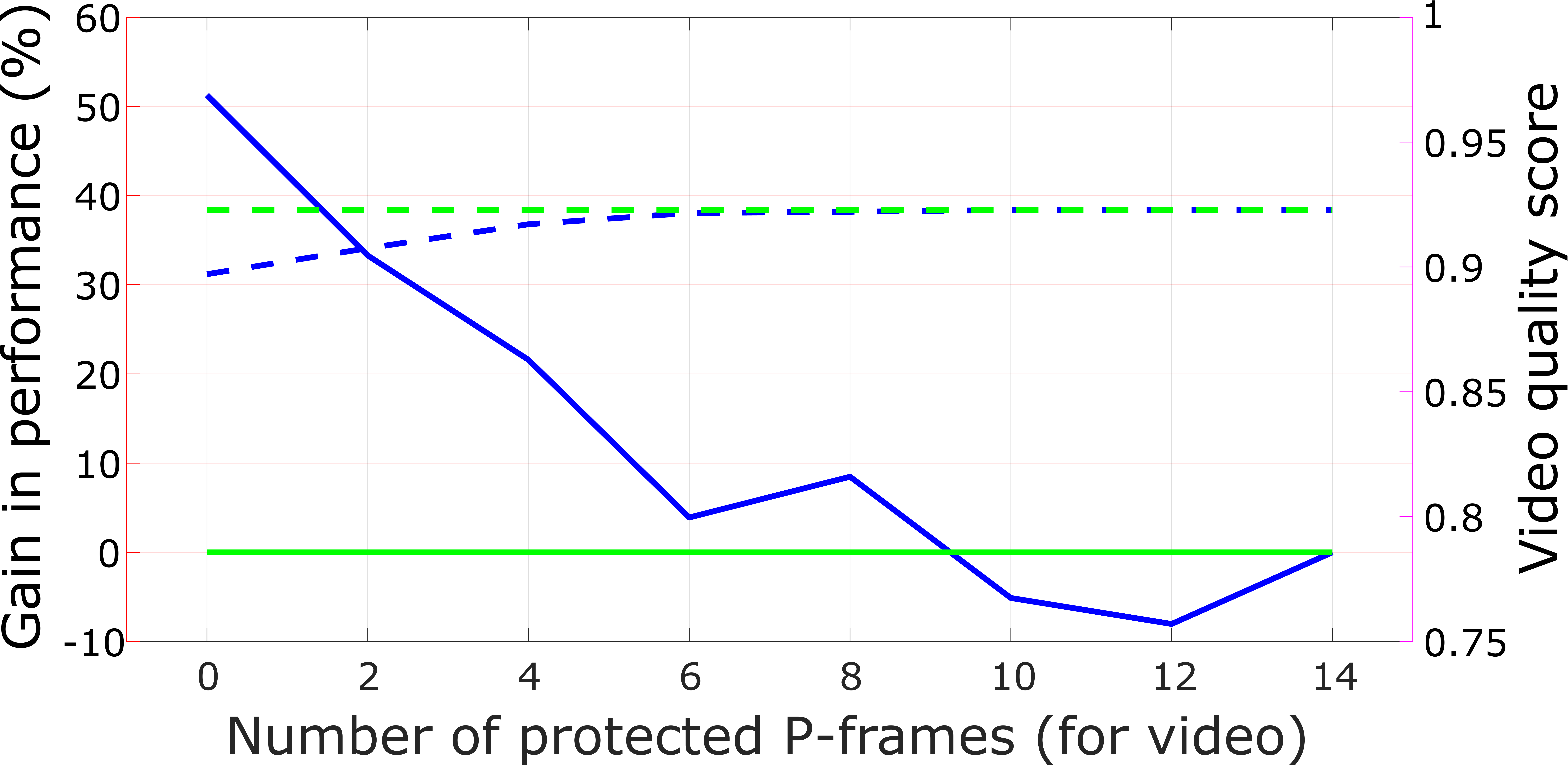} }}%
	\caption{Characterization results for video transmission using polar codes}%
	\label{fig:char_video_polar}%
\end{figure}

We perform experiments to measure the gain in performance and the received video quality score for video transmission using our proposed approach. The experiments are done by varying the number of protected P-frames in the video. For the situation when an embedded video is transmitted using LDPC code as the error correction code for the data channel, the graphs are shown in Figure~\ref{fig:char_video_ldpc}. Figure~\ref{fig:char_video_polar} presents the graphs for the situation when polar codes are used as the error correction code for the data plane.

\begin{figure*}[t]
	\centering
	
	\includegraphics[width=8.5cm]{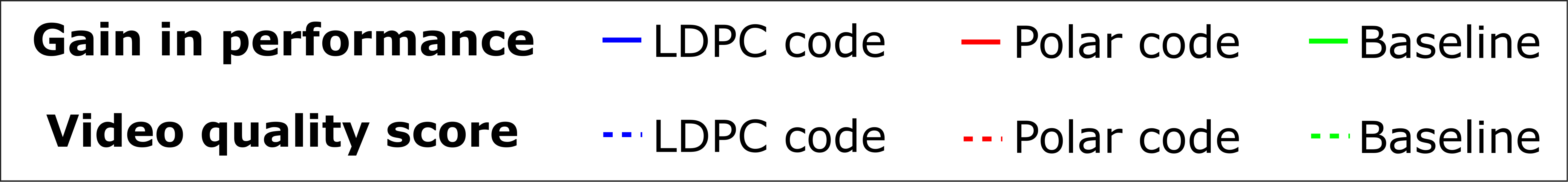} \\
	
	\subfloat[\centering $E_b/N_o$ = 1 dB]{{\includegraphics[width=5.8cm]{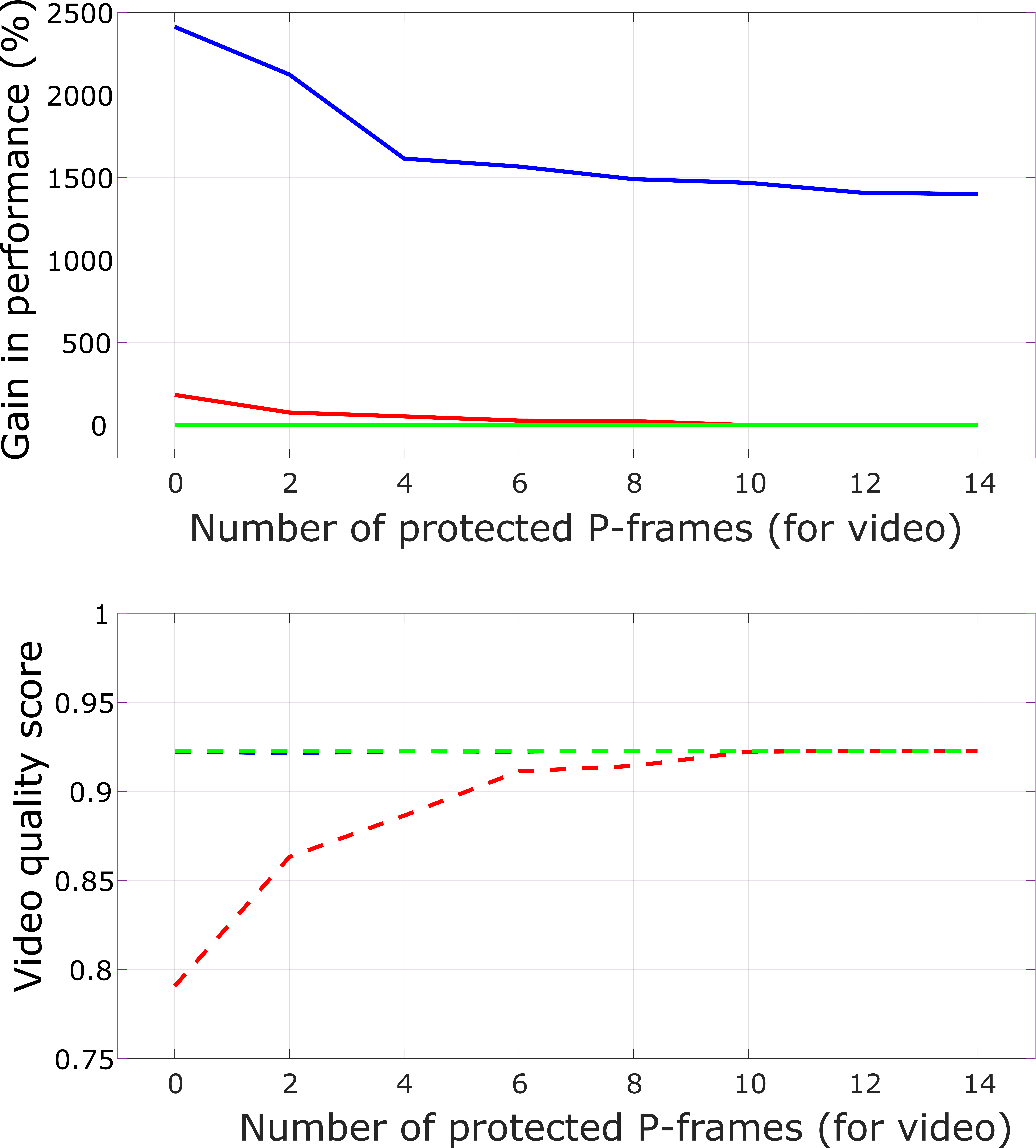} }}%
	\qquad \hspace{-15px}
	\subfloat[\centering $E_b/N_o$ = 1.5 dB]{{\includegraphics[width=5.8cm]{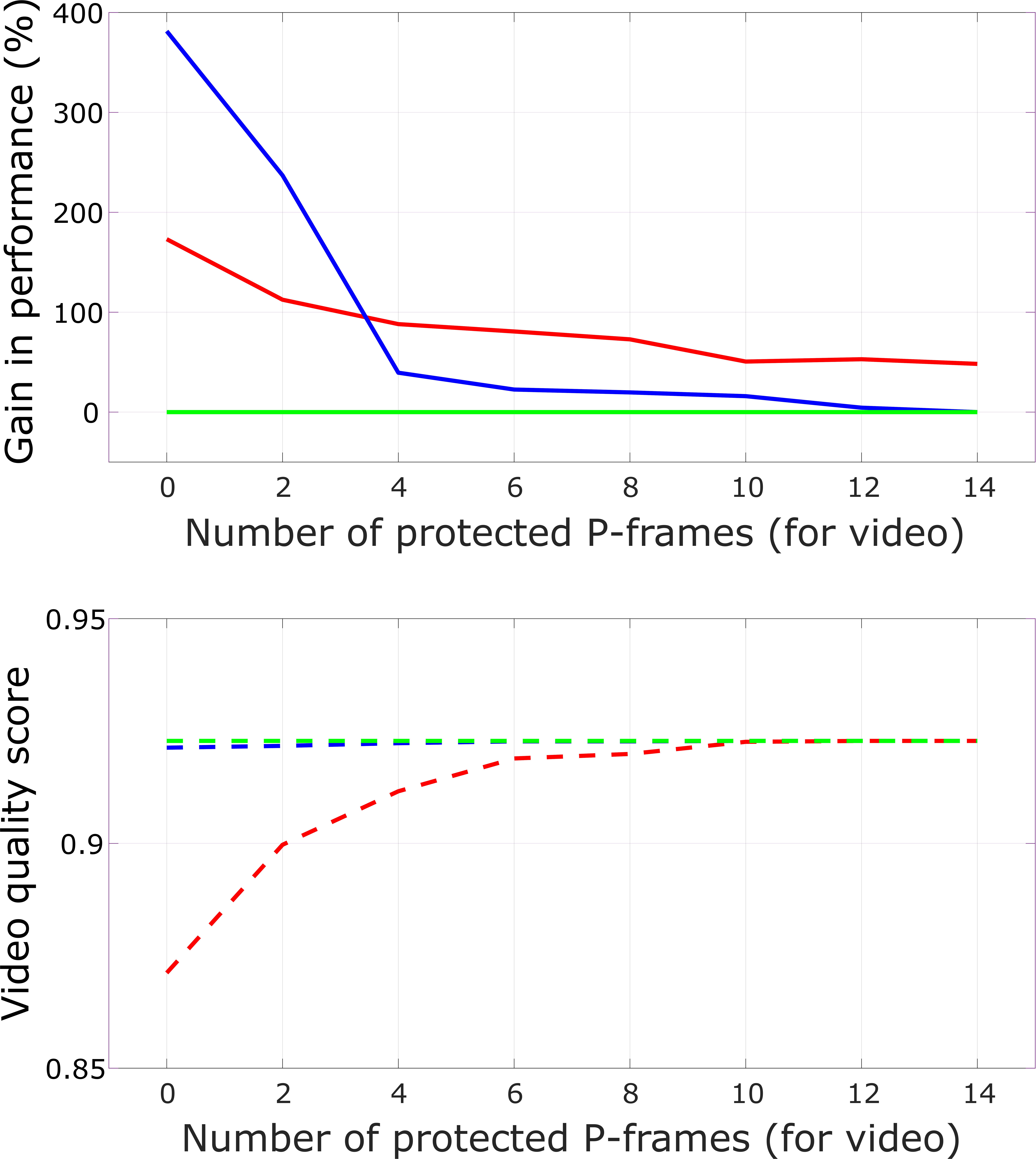} }}%
	\qquad \hspace{-15px}
	\subfloat[\centering $E_b/N_o$ = 2 dB]{{\includegraphics[width=5.8cm]{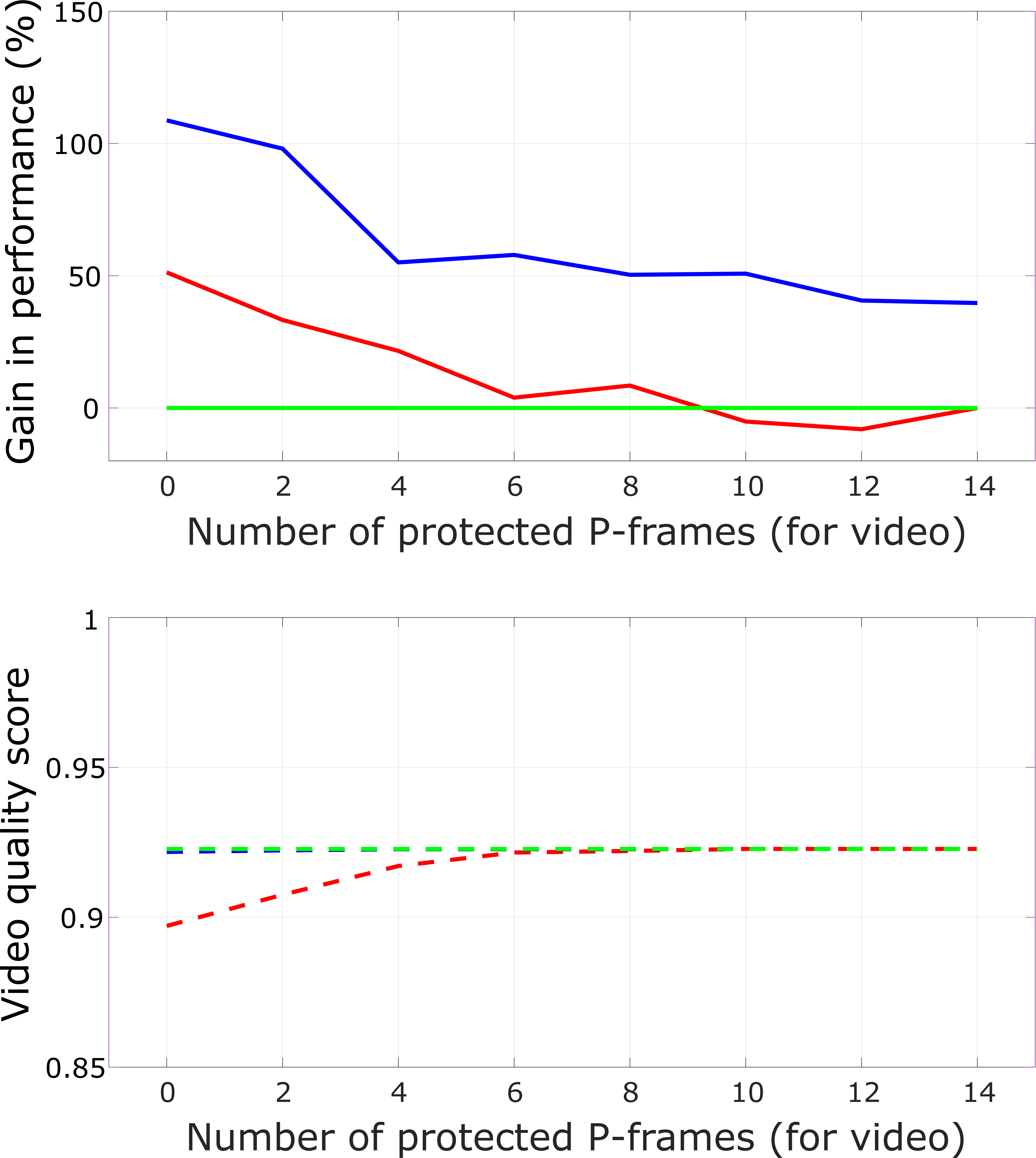} }}%
	
	\caption{Characterization results for video transmission (comparison of LDPC and polar codes) - polar code performs better than LDPC code at $E_b/N_o=$ 1.5 dB}%
	
	\label{fig:char_video_comparison}%
\end{figure*}

For a clear comparison, we plot the comparison graphs for the video transmission experiments. These are shown in Figure~\ref{fig:char_video_comparison}. For video transmission as well, we observe a very similar pattern as for the web page transmission case. To define the baseline in the graphs, we use the error correction code with a lower performance. The gap in the performance gains is large at $E_b/N_o =$ 1 dB. The gap in performance reduces at $E_b/N_o =$ 2 dB. Nevertheless, for both these cases, the LDPC code has a higher performance gain than the polar code. The situation is different at $E_b/N_o =$ 1.5 dB. When less than 4 P-frames are protected, LDPC codes provide a higher performance gain than polar codes. However, when we start protecting more than 4 P-frames, the performance gain is higher for polar codes than for LDPC codes. This happens because of the nature of the unequal error protection provided by the LDPC code. These graphs again show that in certain situations, using polar codes for protecting information in the data plane helps us achieve a higher performance gain than the conventional case of using LDPC codes for the data plane.

\section{Comparison of the Resource Allocation Algorithms using Monte Carlo Simulations}
\label{sec:opt_results}



\begin{figure}[!t]
	\centering
	\subfloat[\centering Total web page size]{{\includegraphics[width=4.2cm]{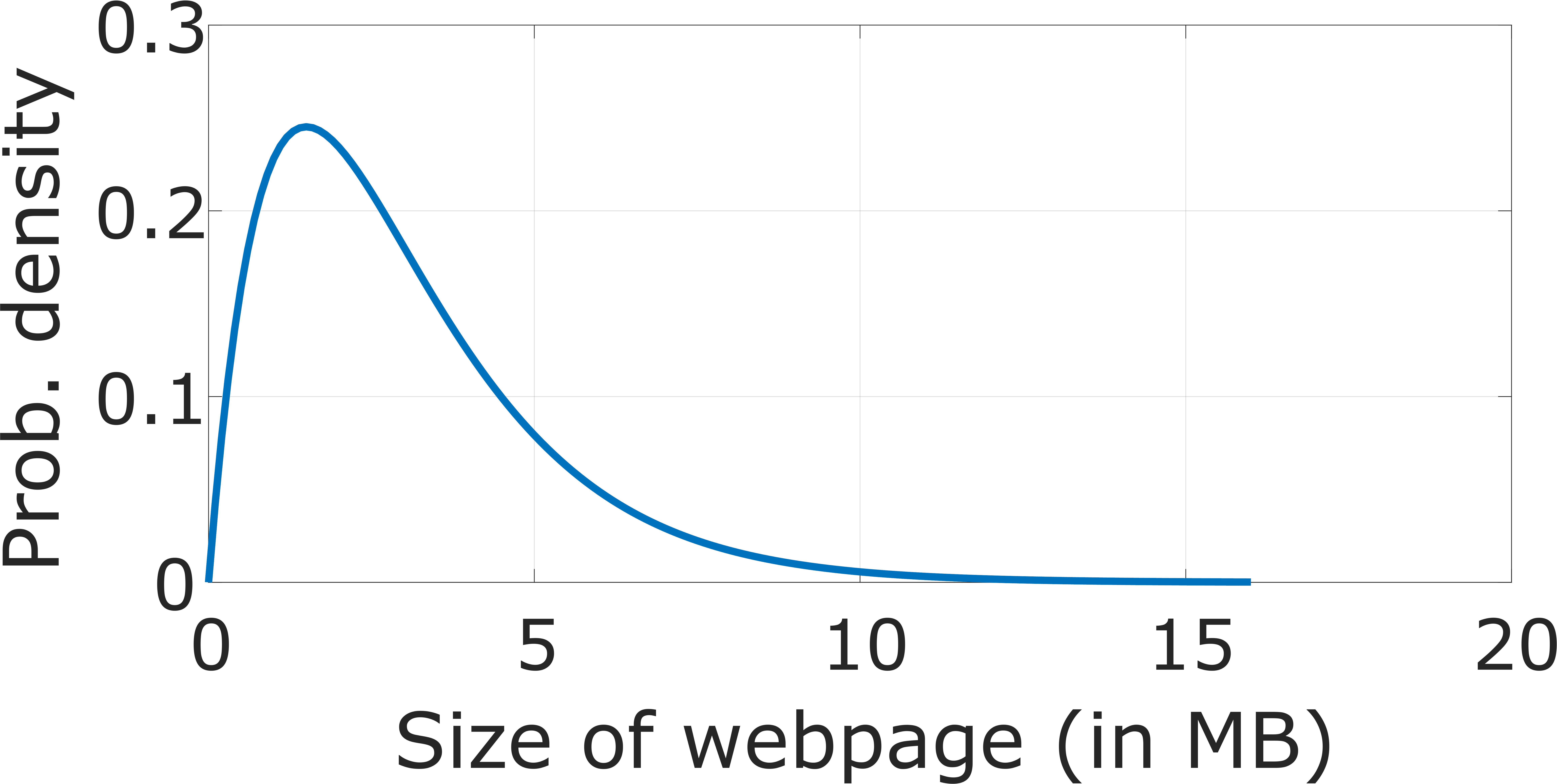} }}%
	\qquad \hspace{-15px}
	\subfloat[\centering Web page text size to image size ratio]{{\includegraphics[width=4.2cm]{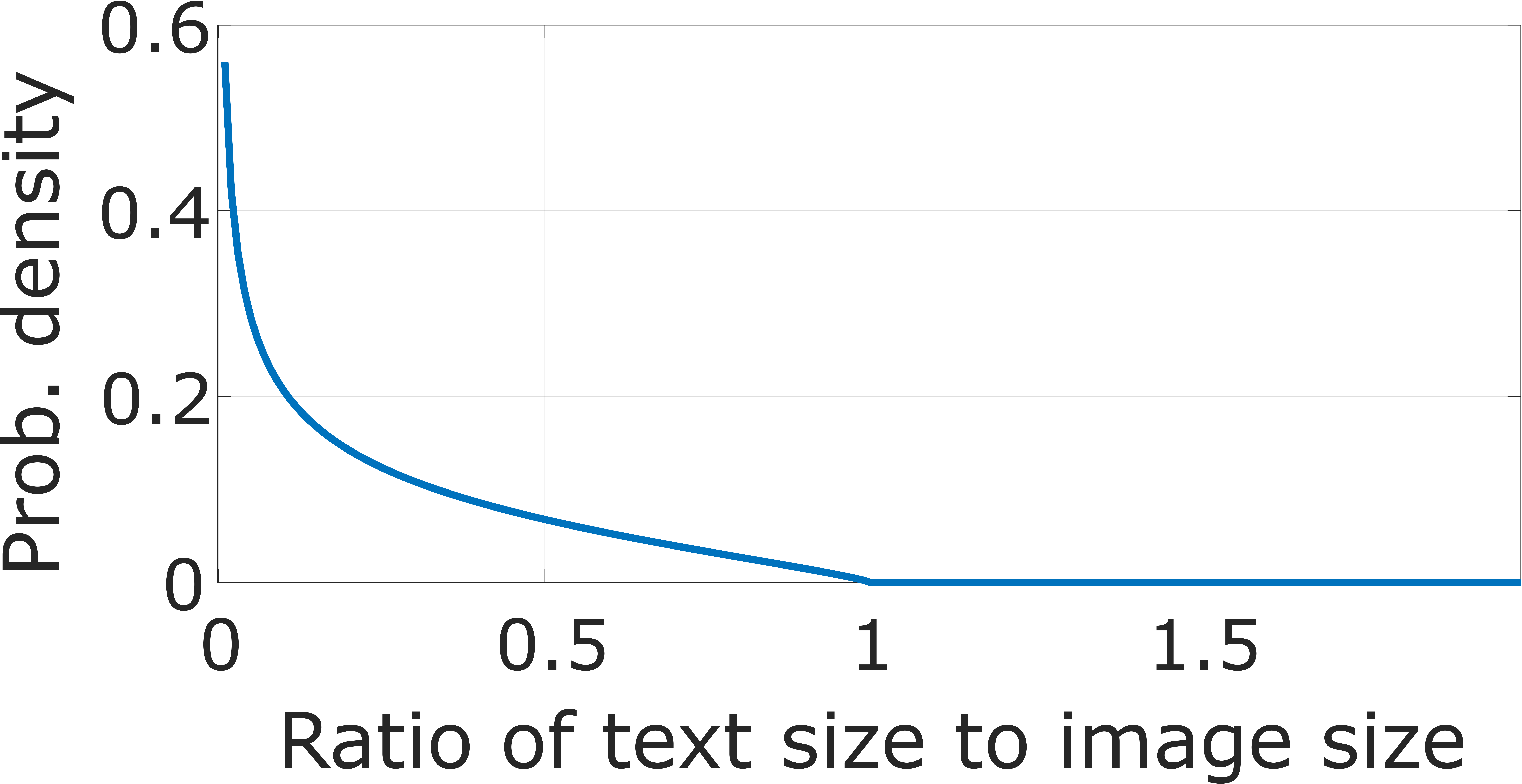} }}%
	\qquad	
	
	\caption{Probability distributions}%
	
	\label{fig:opt_prob_dist}%
\end{figure}

\subsection{Probability Distributions}

\subsubsection{Workload Size}

The size of web pages on the internet follows the gamma probability distribution with shape parameter ($k$) = 2, and scale parameter ($\theta$) = 1.5~\cite{webpage_size_distribution}. The probability distribution is shown in Figure~\ref{fig:opt_prob_dist}(a). In the graph, we have the web page size (in MBs) on the x-axis and the probability density of the web pages with that size plotted on the y-axis. The majority of web pages have a size in the range of 1 to 4 MB. Images and videos have become very common on modern websites. Because of this, the probability of web pages with a very small size is low. The probability distribution has a long tail with very few web pages larger than 10 MB.

\subsubsection{Variation of the Text/Image Size Ratio}

As seen in the graphs in \cref{char_results}, the gain in performance depends on the composition of the web page. In particular, the gain in performance achieved using our proposed approach depends on the text data size to image data size ratio of the web page. For the Monte Carlo simulations, we need the probability distribution of the \emph{ratio} value. The text data size of web pages follows a gamma distribution with $k$ = 0.6 and $\theta$ = 1.5, while the image size follows a gamma distribution with $k$ = 1.8 and $\theta$ = 1.5~\cite{webpage_ratio_distribution}. This implies that the \emph{ratio} value is the ratio of two gamma distributions. The ratio of two gamma distributions with shape parameter = $k_1, k_2$ and equal scale parameter = $\theta$, follows a beta distribution with parameters alpha ($\alpha$) = $k_1$ and beta ($\beta$) = $k_2$. Hence, the \emph{ratio} value follows a beta distribution with $\alpha$ = 0.6 and $\beta$ = 1.8. This probability distribution is shown in Figure~\ref{fig:opt_prob_dist}(b). Most web pages on the internet have a small \emph{ratio} value.

\begin{figure}[!t]
	\centering
	
	\includegraphics[width=6cm]{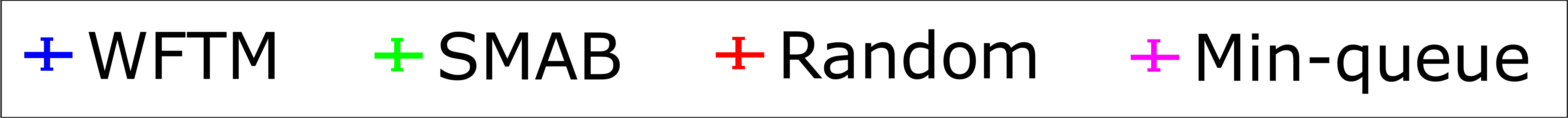} \\
	
	\subfloat[\centering Average throughput]{{\includegraphics[width=4.2cm]{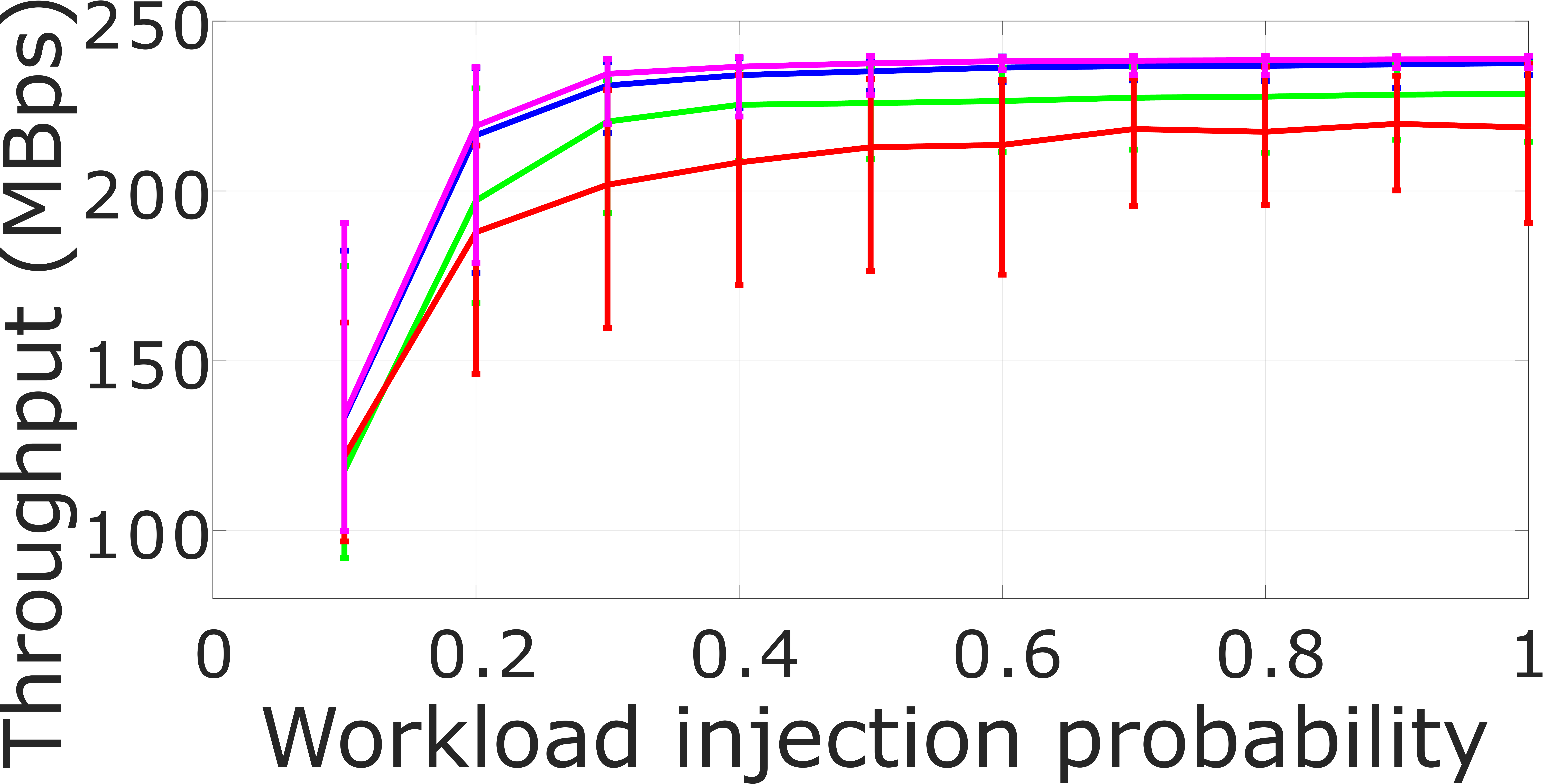} }}%
	\qquad \hspace{-15px}
	\subfloat[\centering Average wait time]{{\includegraphics[width=4.2cm]{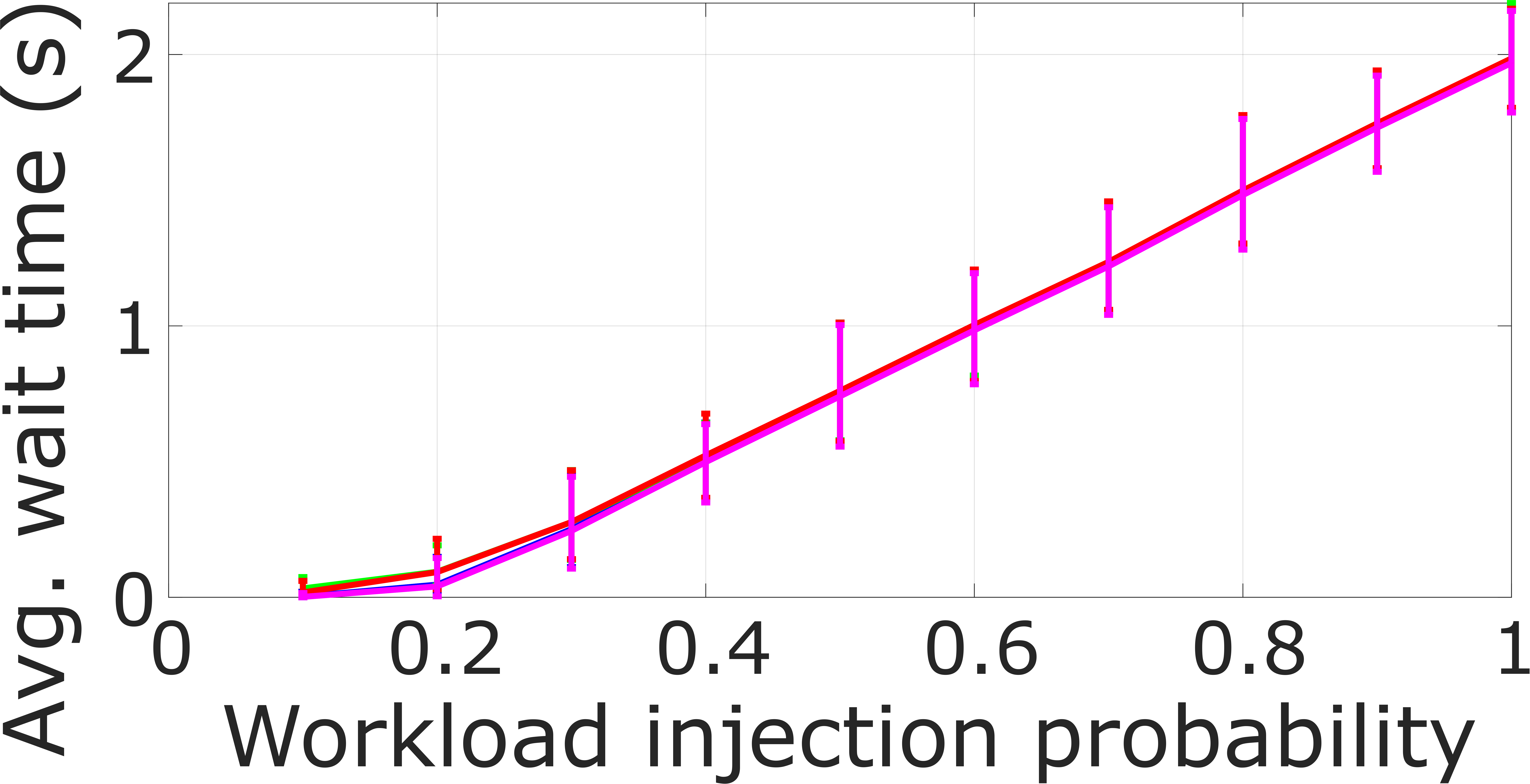} }}%
	\qquad \\
	\subfloat[\centering Average flow time]{{\includegraphics[width=4.2cm]{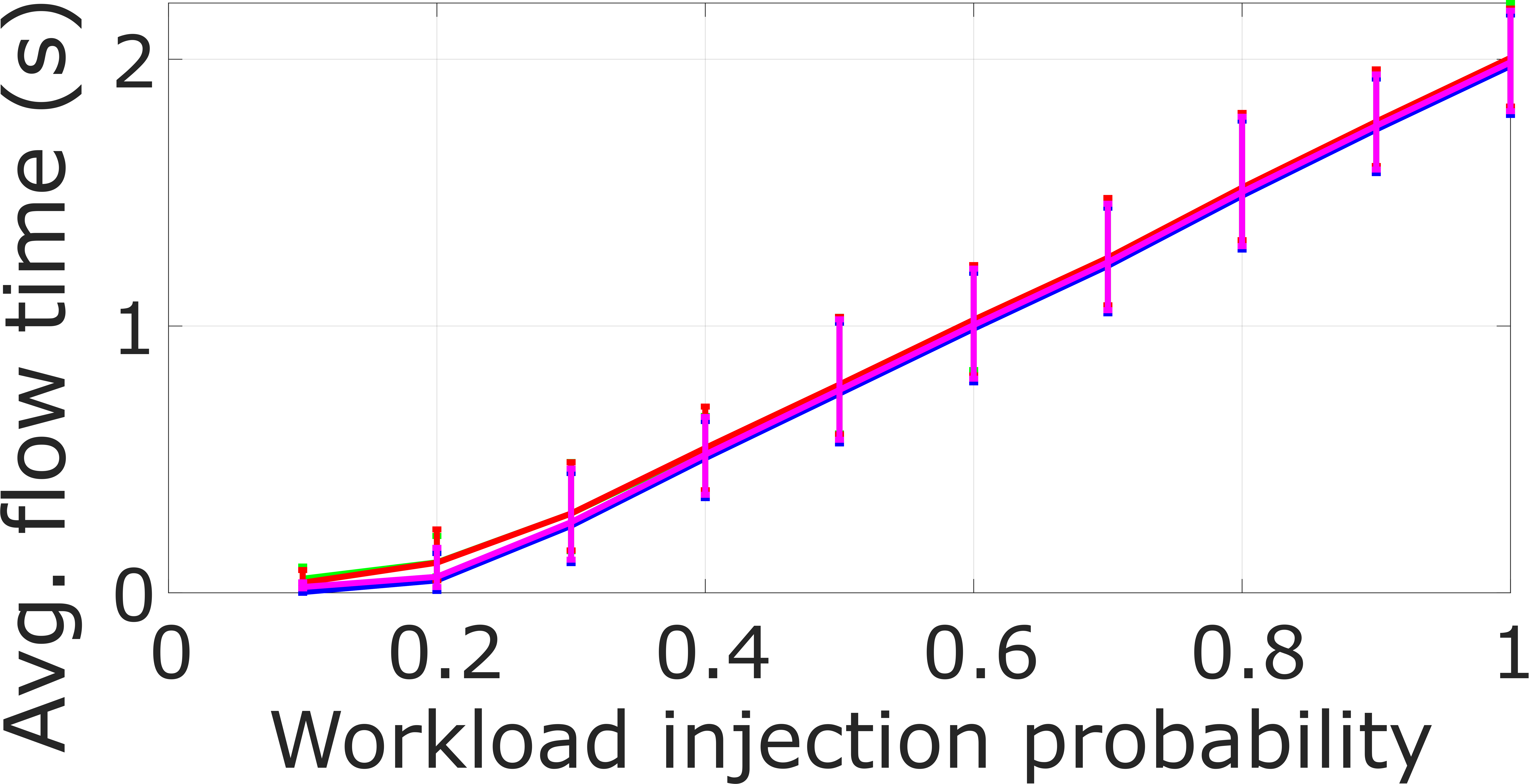} }}%
	\qquad \hspace{-15px}
	\subfloat[\centering Makespan]{{\includegraphics[width=4.2cm]{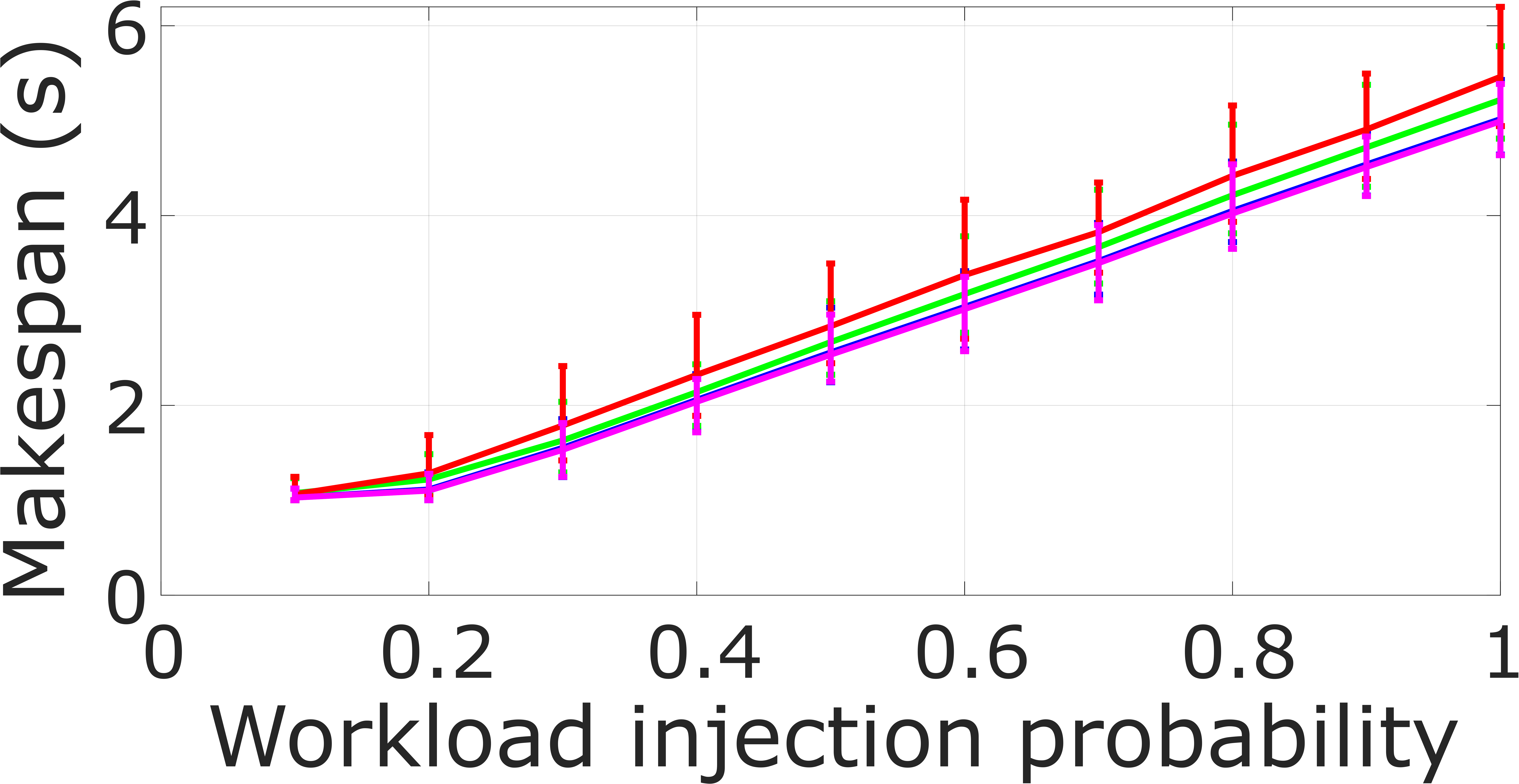} }}%
	\qquad
	
	\caption{Comparison of the four resource allocation algorithms (using the conventional approach) - min-queue heuristic is the best performing algorithm}%
	
	\label{fig:opt_4ldpc}%
\end{figure}


\subsection{Monte Carlo Simulations}

\begin{figure}[!t]
	\centering
	
	\includegraphics[width=6cm]{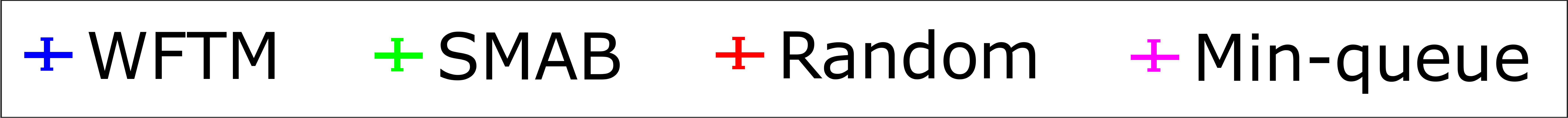} \\
	
	\subfloat[\centering Average throughput]{{\includegraphics[width=4.2cm]{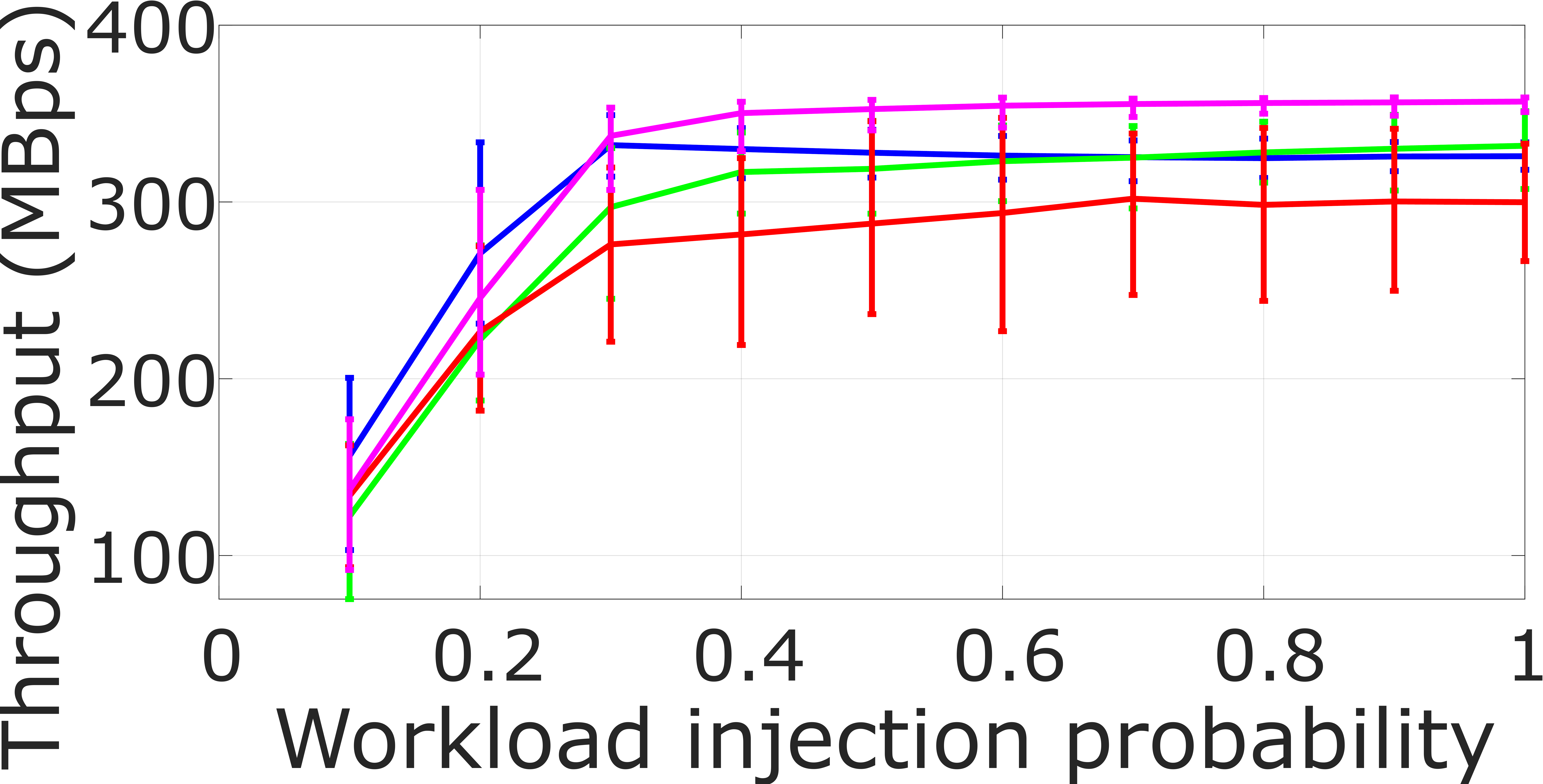} }}%
	\qquad \hspace{-15px}
	\subfloat[\centering Average wait time]{{\includegraphics[width=4.2cm]{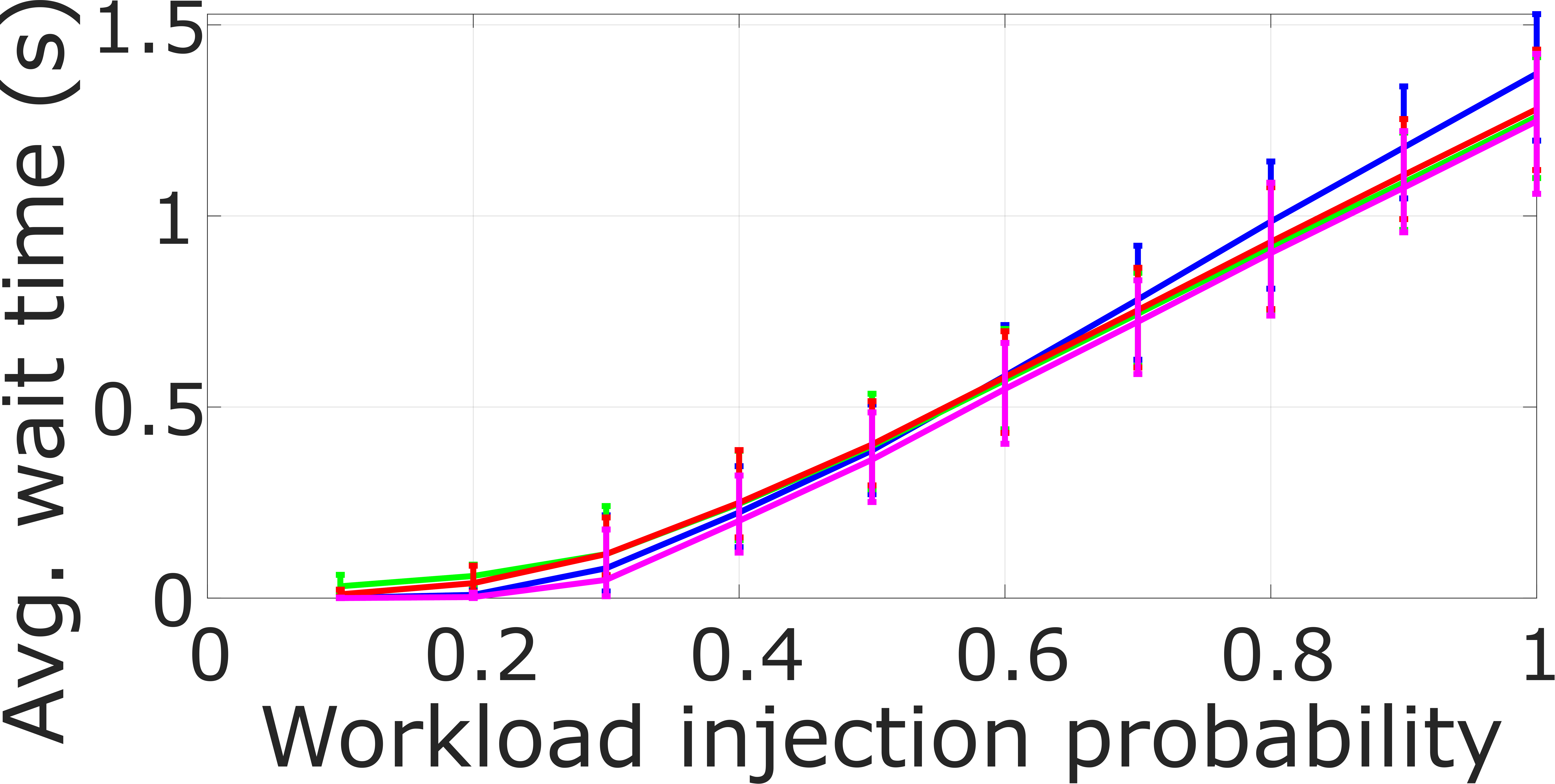} }}%
	\qquad \\
	\subfloat[\centering Average flow time]{{\includegraphics[width=4.2cm]{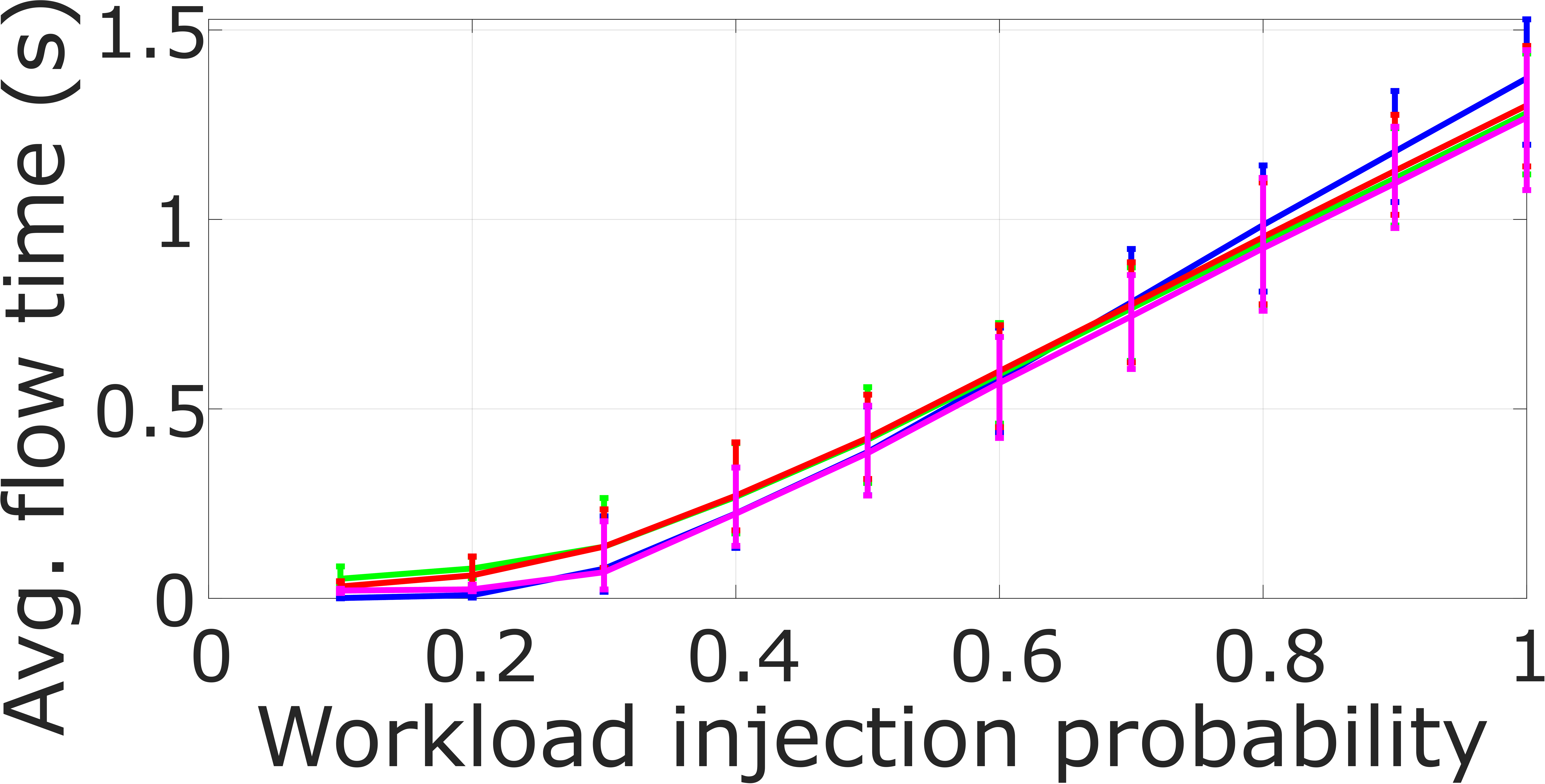} }}%
	\qquad \hspace{-15px}
	\subfloat[\centering Makespan]{{\includegraphics[width=4.2cm]{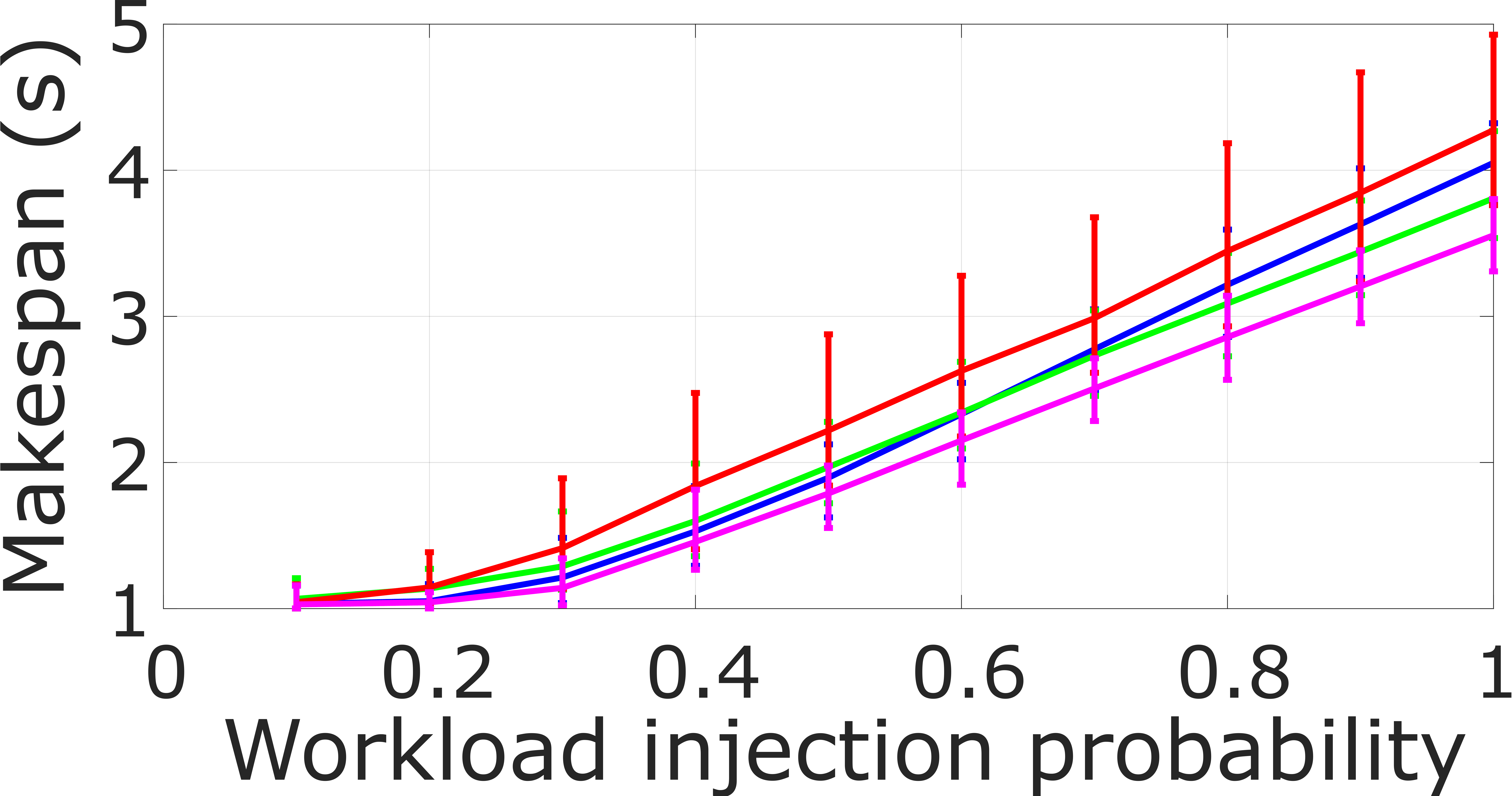} }}%
	\qquad
	
	\caption{Comparison of the four resource allocation algorithms (using our proposed approach) - min-queue heuristic is the best performing algorithm}%
	
	\label{fig:opt_4ldpc_2polar}%
\end{figure}

Using Monte Carlo simulations, we compare four resource allocation algorithms. The four algorithms are weighted flow time minimization (WFTM), stochastic multi-armed bandit (SMAB), random heuristic (random), and the min-queue heuristic (min-queue). We compare these algorithms using four metrics: average throughput, average wait time, average flow time, and makespan. We plot these four metrics for each resource allocation algorithm on varying the \emph{workload injection probability}. The \emph{workload injection probability} is the probability that a new workload request will be generated at the base station by any user equipment in any given simulation time interval. We simulate an edge-compute node that uses the Xilinx T1 telco accelerator card for baseband processing. With a total L1 encode throughput of 17.8 Gbps~\cite{T1_product_brief}, we consider that each T1 telco accelerator card implements four LDPC codecs and two polar code codecs. We consider a single hardware accelerator card for these simulations. The $E_b/N_o$ of the channel for these simulations is 1.5 dB. The reported throughput is the average throughput at which the base station processes the workloads, averaged over all the workloads. The wait time of a workload is the time it has to wait in the \emph{encoder queue} of the compute node it is assigned to. The flow time of a workload is the sum of the wait time of the workload and the time required to process the workload by the compute node to which it is assigned. The makespan of a set of workloads is the total time taken by the base station to service all the individual tasks in the \emph{workload queue}.

The first scenario is the conventional approach. Here, only the LDPC codes are used for data plane coding. Whenever the base station receives a new request, the resource allocation algorithm allocates the workload to one of the four LDPC encoders on the hardware accelerator card for processing. The graphs comparing the four algorithms in this scenario are shown in Figure~\ref{fig:opt_4ldpc}. The second scenario is our proposed approach, where we use the idle time of the polar code encoders for data plane encoding. Here, both the LDPC and polar codes are used for data plane coding. Whenever the base station receives a new request, the resource allocation algorithm allocates the workload to either one of the four LDPC encoders or one of the two polar code encoders for processing. The graphs comparing the four algorithms are shown in Figure~\ref{fig:opt_4ldpc_2polar}.

\begin{figure}[!t]
	\centering
	
	\includegraphics[width=6cm]{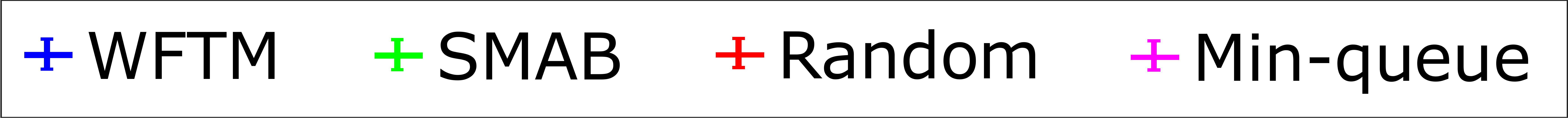} \\
	
	\subfloat[\centering Average throughput]{{\includegraphics[width=4.2cm]{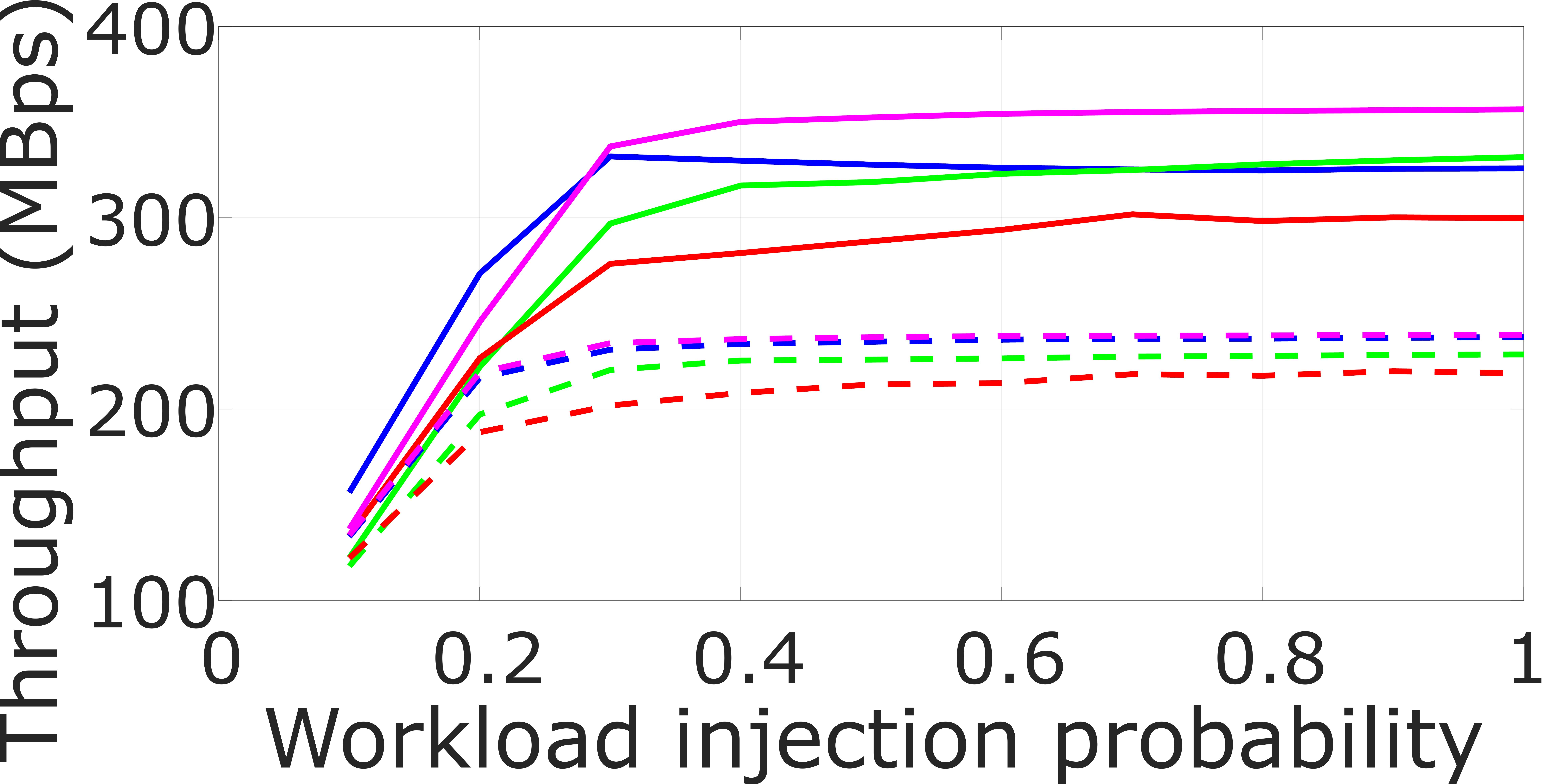} }}%
	\qquad \hspace{-15px}
	\subfloat[\centering Average wait time]{{\includegraphics[width=4.2cm]{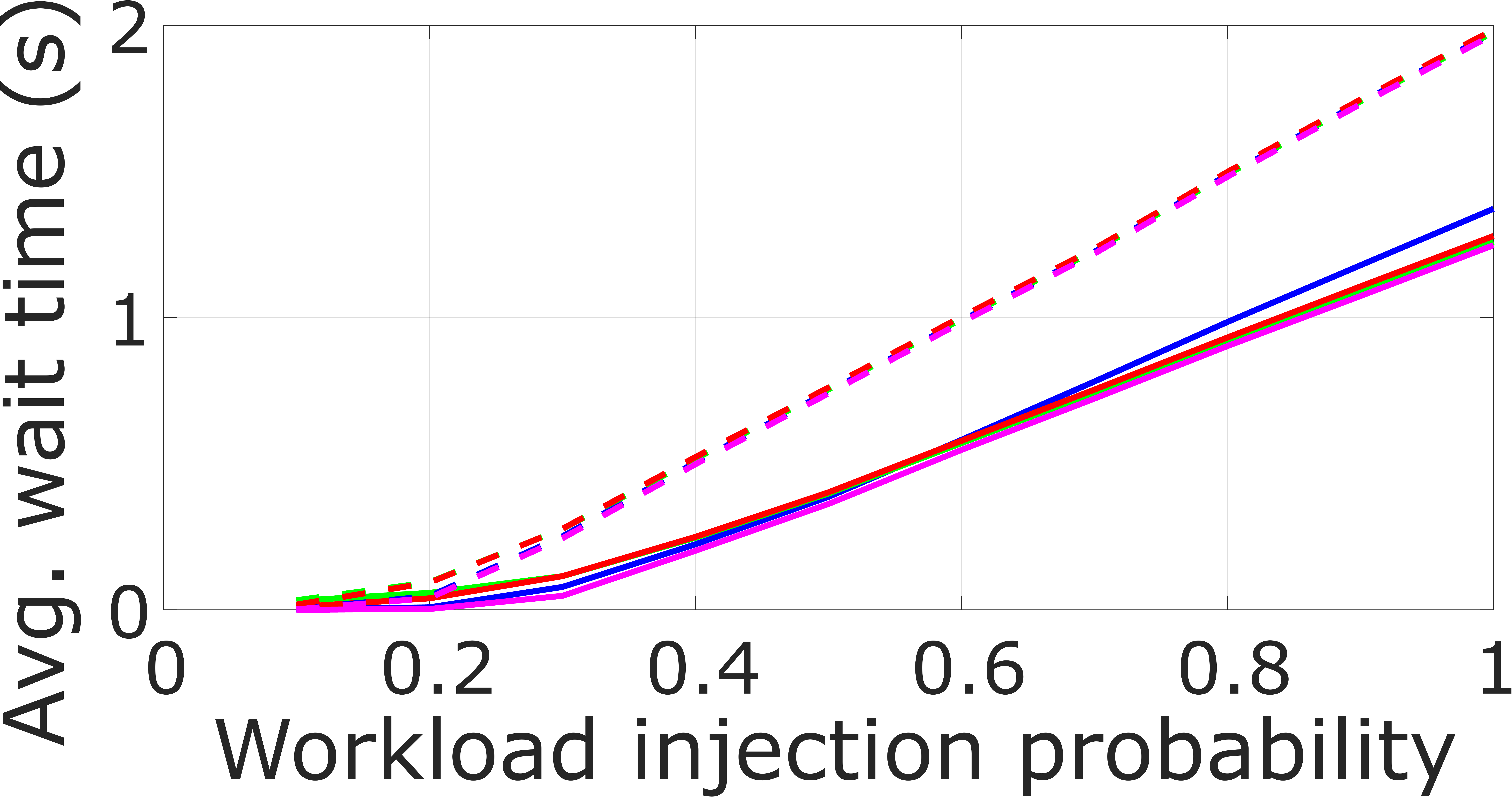} }}%
	\qquad
	\subfloat[\centering Average flow time]{{\includegraphics[width=4.2cm]{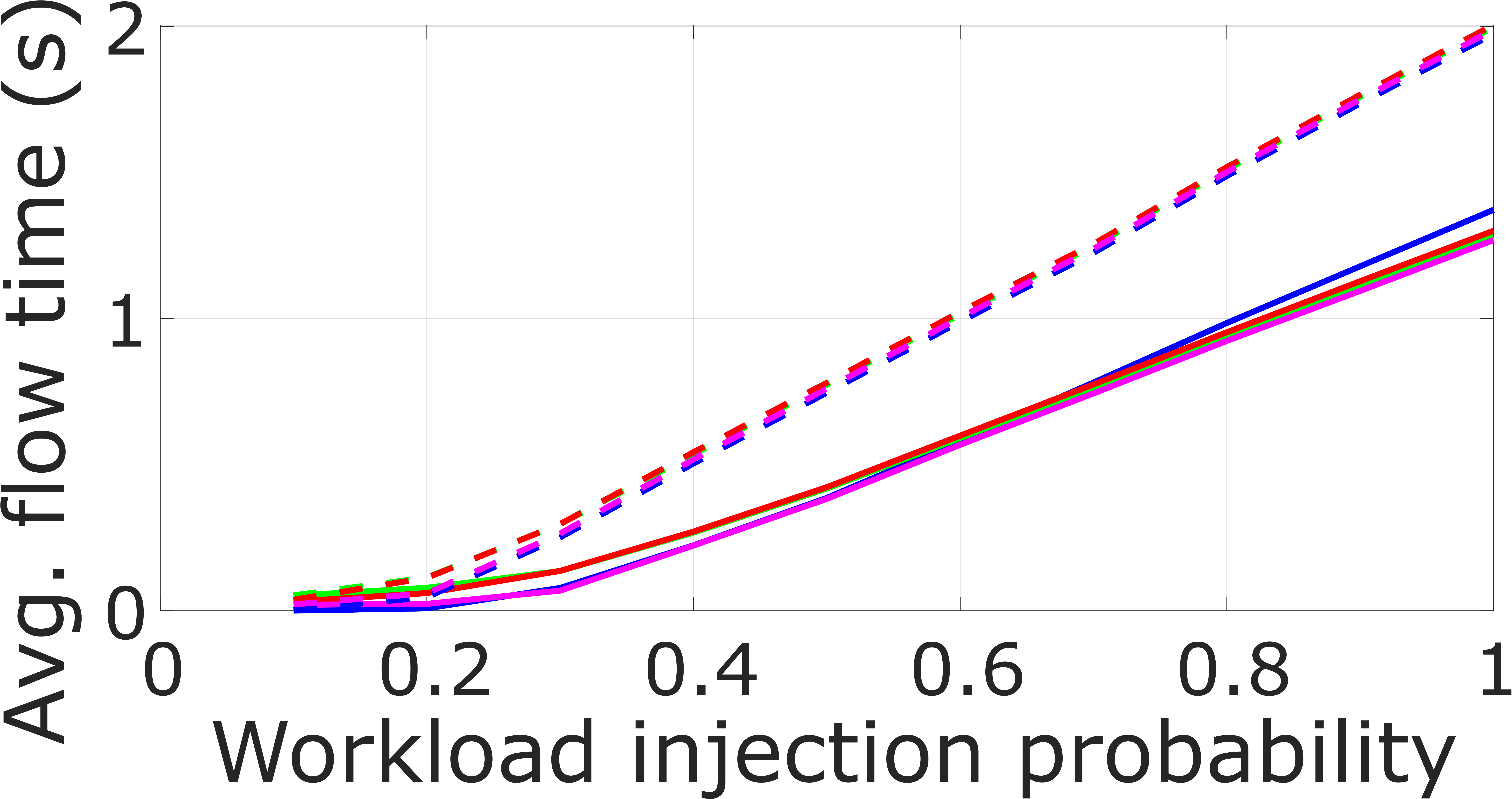} }}%
	\qquad \hspace{-15px}
	\subfloat[\centering Makespan]{{\includegraphics[width=4.2cm]{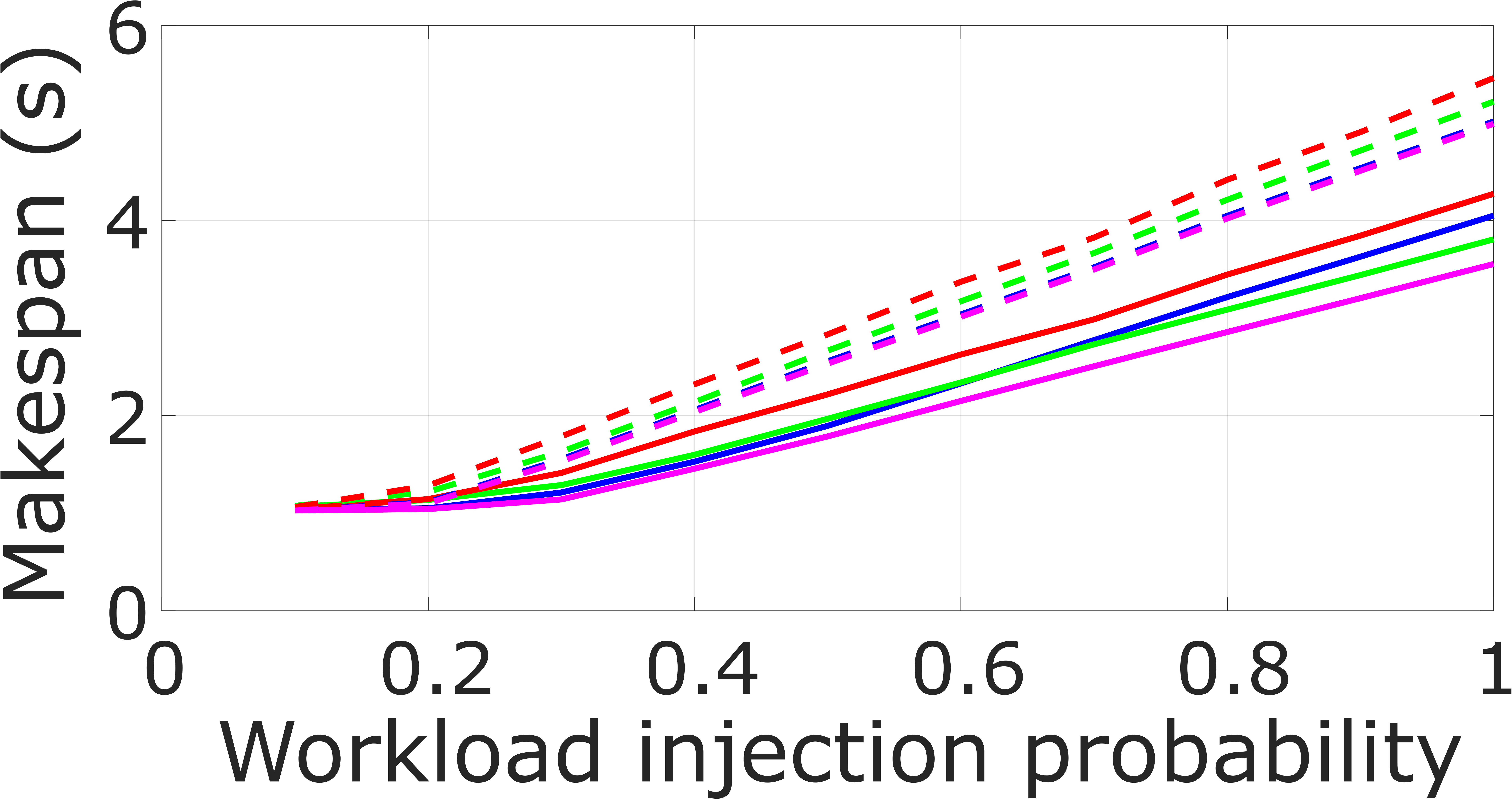} }}%
	\qquad
	
	\caption{Performance of the four resource allocation algorithms (comparison of the conventional and our proposed approach) - min-queue heuristic is the best performing algorithm}%
	
	\label{fig:opt_4l_4l2p_comparison}%
\end{figure}

To compare the conventional approach (using only LDPC code encoders) and our proposed approach (using both LDPC and polar code encoders), we plot the four metrics for both approaches on the same graph. The graphs are presented in Figure~\ref{fig:opt_4l_4l2p_comparison}. The graphs for the conventional approach are drawn using dashed lines, and the graphs for our proposed approach are drawn using solid lines. At low \emph{workload injection probability}, the processors are not fully utilized, leading to low throughput values. The throughput saturates after the injection probability becomes more than 0.5. An algorithm with a lower average wait time is better. We observe that the average wait time for all four algorithms is very similar. The min-queue heuristic has the least average wait time. For the average flow time graphs also, the algorithm with a lower average flow time is better. The weighted flow time minimization algorithm has the least average flow time. This is understandable as the algorithm specifically tries to minimize the flow time when allocating a resource to a workload. Makespan is the total time taken to execute all the workload requests. For both scenarios, we observe a significant gap in the makespan of the four algorithms, and the gap increases with increasing \emph{workload injection probability}. The min-queue heuristic algorithm has the least makespan, while the random heuristic algorithm has the highest makespan.

From the four comparison graphs in Figure~\ref{fig:opt_4l_4l2p_comparison} we have two important observations. Firstly, we observe that the min-queue heuristic algorithm is the best-performing algorithm in terms of throughput, average wait time, and makespan. The second important observation is that the average wait time and flow time for the proposed approach graphs (solid lines) are lower than for the conventional approach graphs (dashed lines). This is because of the more hardware resources available in our proposed approach, as we also use the polar code encoders also for error correction of the user data. Our proposed approach improves the system's performance in all four metrics.

\section{Full-system Simulation of a 5G Base Station}
\label{systemResults}

The graphs in Figure~\ref{fig:opt_4l_4l2p_comparison} show that the min-queue heuristic algorithm has the best performance for the resource allocation task. We now perform a full-system Monte Carlo simulation of a 5G base station using the min-queue heuristic algorithm for scheduling workloads. For these simulations, we define the system's performance as the reciprocal of the total time taken to process and transmit all the workloads from the base station to the UEs. We simulate an edge-compute node that uses the Xilinx T1 telco accelerator card for baseband processing. With a total L1 encode throughput of 17.8 Gbps~\cite{T1_product_brief}, we consider that each T1 telco accelerator card implements four LDPC codecs and two polar code codecs. Using three HPE ProLiant DL110 Gen10 telco server racks in one HPE EL8000 converged edge system, we have a total of 12 hardware accelerator cards for baseband processing. The \emph{workload injection probability} is varied from 0.1 to 1.0, and the $E_b/N_o$ of the channel is considered in the range of 1 dB to 2 dB. This is representative of the channel conditions in rural areas~\cite{channel_EbNo_range}. In the conventional case, only the four LDPC encoders are used to perform error correction in the data plane. In our proposed approach, we use the polar code encoders also for data plane error correction. Our aim is to reduce the hardware cost of the base station. So, we compare the system's performance for these three scenarios: (1) In an accelerator card, we use the four LDPC encoders for data plane channel coding (4L scenario). (2) In the accelerator card, three LDPC encoders and two polar code encoders are used for data plane encoding (3L2P scenario). (3) In the accelerator card, two LDPC encoders and two polar code encoders are used for data plane channel coding (2L2P scenario).

\begin{figure}[!t]
	\centering
	\includegraphics[width=8.5cm]{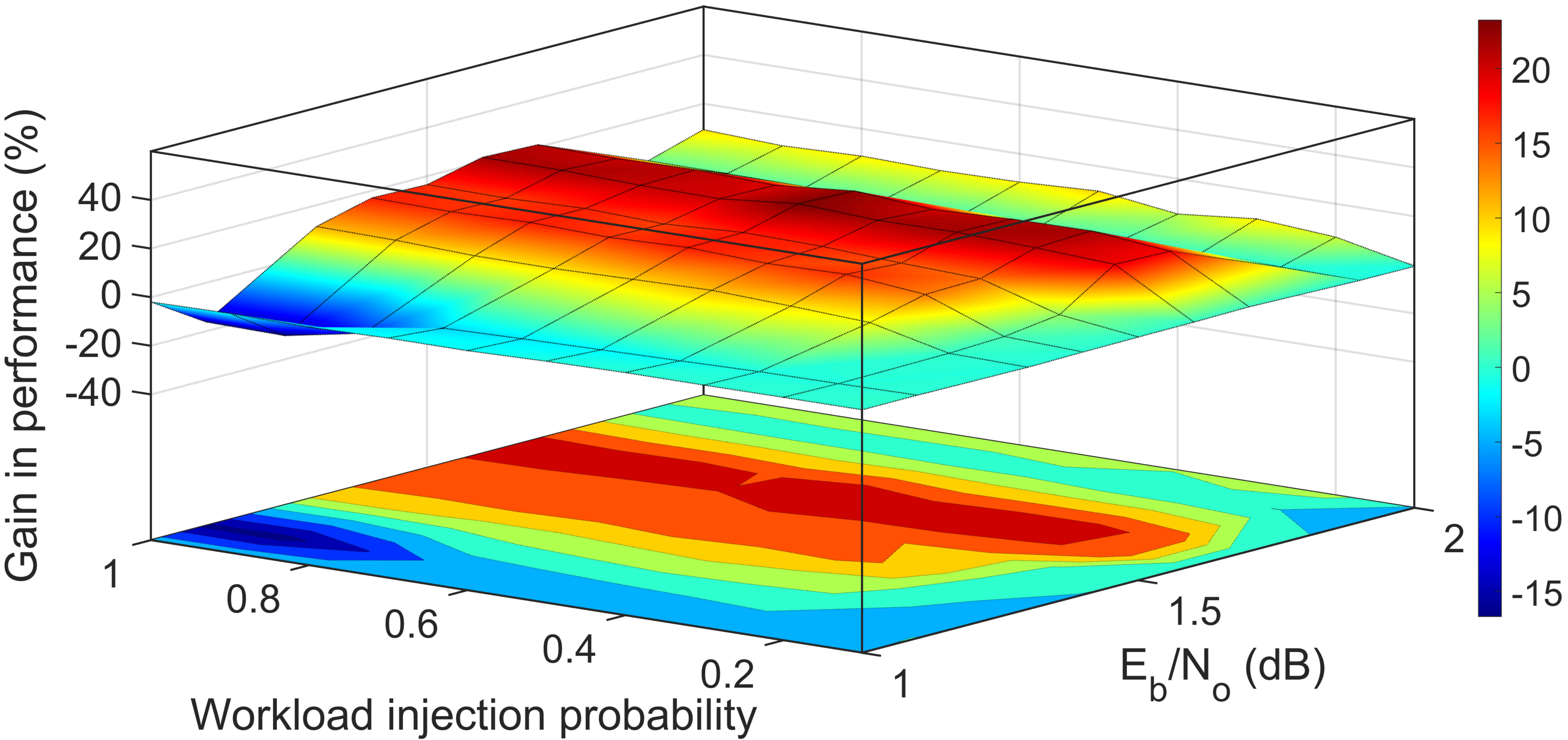}
	
	\caption{Performance of the system when using three LDPC and two polar code encoders compared to when using all the four LDPC encoders - up to 24\% performance improvement}
	
	\label{fig:fullsystem_3L2P}
\end{figure}

\begin{figure}[!t]
	\centering
	\includegraphics[width=8.5cm]{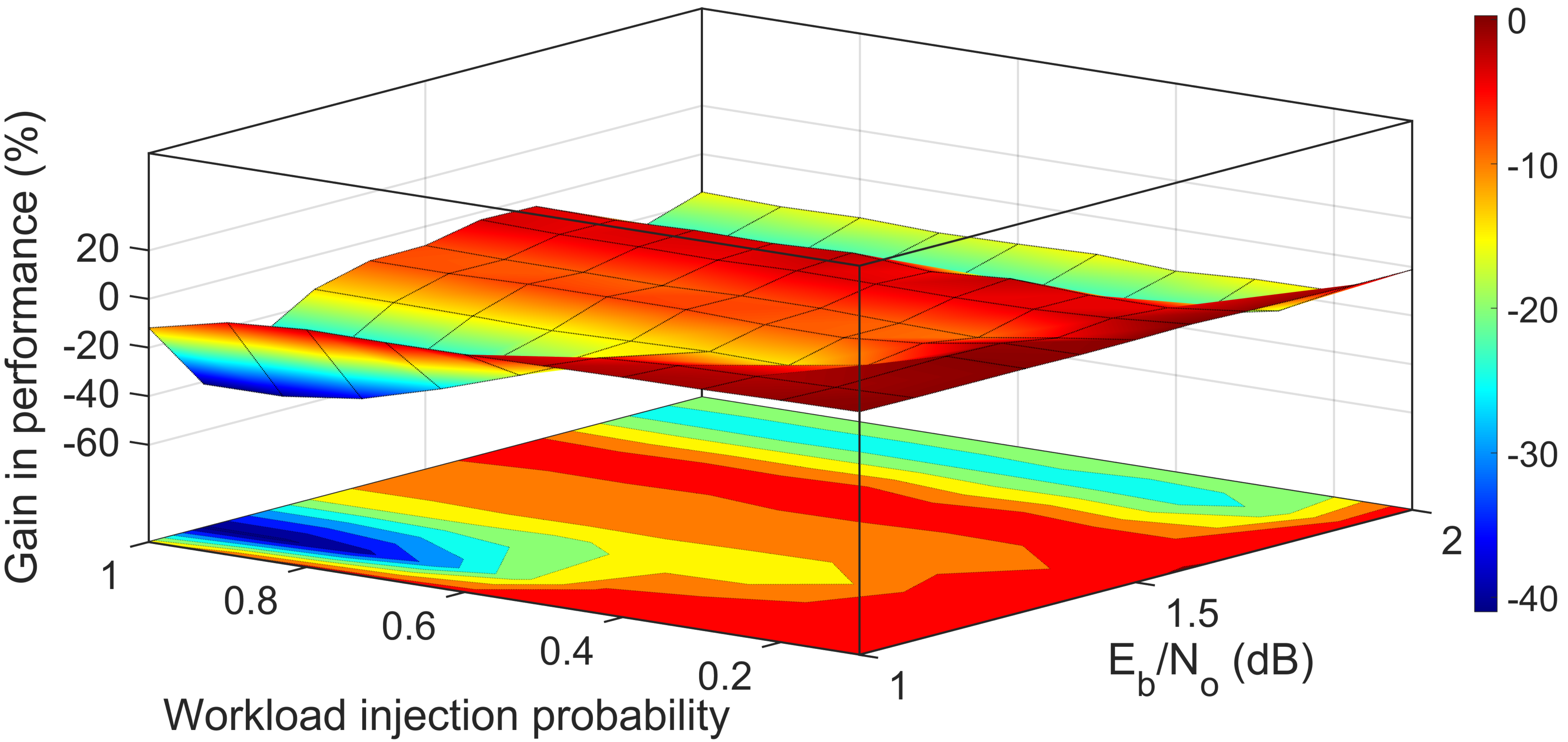}
	
	\caption{Performance of the system when using two LDPC and two polar code encoders compared to when using all the four LDPC encoders - up to 40\% decrease in performance}
	
	\label{fig:fullsystem_2L2P}
\end{figure}

In Figure~\ref{fig:fullsystem_3L2P}, we show the gain in performance of the system for the 3L2P scenario over the 4L scenario. The gain in performance of the 3L2P scenario over the 4L scenario is in the range of –5\% (degradation of 5\%) to +24\% (improvement of 24\%) for a major part of the $E_b/N_o$ range (1.2 dB to 2 dB). So, on average, the performance of the system in the 3L2P scenario is better than the conventional system, i.e., the 4L scenario. This means that we can use three LDPC and two polar code encoders in place of using four LDPC encoders, with significant improvement in the overall system performance. In Figure~\ref{fig:fullsystem_2L2P}, we show the gain in performance of the system for the 2L2P scenario over the 4L scenario. The performance of the 2L2P scenario shows considerable degradation over the 4L scenario, going up to a 40\% decrease in performance for some $E_b/N_o$ values. We, therefore, do not consider the 2L2P scenario because of this performance degradation. It is also essential to maintain the performance of the base station.

\subsection{Reducing the Cost of a 5G Base Station}

Using three HPE ProLiant DL110 Gen10 telco server racks~\cite{vDU} in one HPE EL8000 converged edge system~\cite{EL8000}, we have a total of 12 hardware accelerator cards for baseband processing at the edge-compute node. Here, each card implements four LDPC encoders and two polar code encoders, but the polar code encoders are not used for error correction of the data plane. From the performance comparison in Figure~\ref{fig:fullsystem_3L2P}, we find that using our proposed approach at the base station, we can use three LDPC and two polar code encoders in place of four LDPC encoders for the forward error correction of the user plane data. This means we can save on one LDPC codec in every hardware accelerator card. Subsequently, it allows us to reduce the required number of accelerator cards. For a base station that utilizes its hardware resources optimally (using our proposed approach), we need only ten hardware accelerator cards for the computational requirements of the base station (in place of 12 accelerator cards in the baseline situation). So, the number of hardware accelerator cards required at the base station is reduced by 17\%. This translates into a 17\% decrease in the cost of a 5G base station. This reduction in the cost of a 5G base station leads to significant cost savings for telecom operators.

To deploy our proposed approach to a real 5G base station, we need to factor in some overheads. The first overhead at a base station is the need to run an online resource allocation algorithm when transmitting the downlink data. The algorithm can be implemented as a virtualized network function (VNF) that can run on the CPU compute nodes that are already available on the edge server. This does not increase the hardware requirement of a base station~\cite{nfv_1, nfv_2}. A 17\% reduction in the cost of a 5G base station has much greater cost savings for a telecom operator compared to running an extra virtualized network function at the edge server. The telecom operators will be willing to accept this overhead. Another overhead that our proposed approach introduces is that we need to indicate to the UE whether LDPC or polar codes have been used for the forward error correction (FEC) of the workload that the UE is currently receiving. This can easily be done by adding a single bit before every workload. A bit value of 0 indicates that LDPC codes have been used for the FEC of the workload, and a bit value of 1 indicates that polar codes have been used. The overhead of adding a single bit is insignificant compared to the size of the workloads (in MBs). This overhead can be implemented by making a software change on the mobile phones (UEs). The 0/1 bit can be added as a header to the user data packet. Based on the bit value, the UE can then use the LDPC or the polar code decoding algorithm to decode the packet.

\section{Related Work}
\label{sec:relatedwork}

\subsection{Polar Codes for Forward Error Correction in the Data Plane}

Multiple papers have explored the possibility of using polar codes as the error correction codes in the data plane of futuristic 5G use cases. One such use case is the narrow-band internet of things (NB-IoT) systems~\cite{polar_nbiot}. Currently, NB-IoT uses turbo codes for data plane coding, but they have high computational complexity. The authors in~\cite{polar_nbiot} propose to use belief propagation-based polar codes to protect user data. Using their approach, they demonstrate reduced computational complexity and better error-rate performance. Our work uses polar codes in the data plane for multimedia web page transmission. Demonstrating a performance improvement for the web page transmission use case has much greater significance~\cite{image_video_internet}. In recent work, Lin et al.~\cite{polarcode_JSCD} suggest using polar codes for error correction in the data plane in the 5G communication system, targeted toward ultra-high definition (UHD) videos and virtual reality (VR) applications. The authors in~\cite{polarcode_IPSN} have proposed a multimedia web page transmission system that utilizes polar codes as the error correction codes for web page data. The authors evaluate their approach for transmitting web pages over a wireless link and demonstrate the feasibility and advantages of using polar codes in the data plane of the 5G communication system.

\subsection{Unequal Error Protection}

Many error correction codes have unidentical error protection for their codeword bits. Channel coding schemes that exhibit such a property are called unequal error protection (UEP) codes~\cite{uep_meaning}. Practical polar code systems have a finite codeword length. In such situations, the channel polarisation effect is not perfect. The information bits experience a range of error probabilities. The authors in~\cite{uep_polar_code} utilize the UEP property of polar codes to improve the transmission of images over a lossy channel. Image transmission is a stand-alone application. On the internet, data is transmitted in the form of web pages. Web pages have text, images, as well as video content. The work by Shreshtha et al.~\cite{polarcode_IPSN} utilizes polar codes to improve the performance of multimedia web page transmission over the internet. The authors of the paper~\cite{uep_ldpc_code} show that LDPC codes have an inherent UEP property. Building on this idea further, the authors also suggest a code construction algorithm with an improved UEP property. The authors in~\cite{uep_ldpc_code_image} utilize the inherent UEP of LDPC codes to improve the PSNR (image quality metric) of the received image in an image transmission system compared to an equal error protection (EEP) based system. In our work, we first characterize this inherent UEP property in the LDPC codes used in the data plane of the 5G NR communication system. Then, we utilize this inherent property of LDPC codes to build an LDPC code-based approximate multimedia web page transmission system.

\subsection{Approximate Communication}

The authors in~\cite{approximate_wireless_media} design a system based on approximate communication for media transmission when the channel is error-prone. We can relax the accuracy constraint on the data during transmission. The approximate communication in~\cite{approximate_wireless_media} is achieved using the differential protection of data bits caused when the data bits are modulated using QAM modulation. In our work, we utilize the differential protection of data bits due to the channel coding operation to design the approximate communication system.

\subsection{Web Page Transmission}

Web pages are transmitted using the HTTP protocol, which happens over the transmission control protocol (TCP). There has been much research to reduce the download times of web pages. SAP~\cite{SAP} is an architecture that improves web page transmission performance over wireless links using selectively approximate communication. SAP uses TCP to transmit text data, including HTML tags. However, for image data, it uses the UDP transport layer protocol. This approach decreases the web page transmission time compared to baseline TCP. However, this method is not suitable for widespread adoption. The approach requires changes in the MAC layer of the protocol stack. The approach proposed in our current work is agnostic to the choice of the transport layer protocol. The authors in~\cite{quic_tcp_evaluation} compare the performance of the TCP and the QUIC protocols under network conditions that mimic the internet. Using the page loading time as the comparison metric, the authors conclude that \emph{QUIC performs worse than TCP}. Video transmission on the internet is conventionally done using the Dynamic Adaptive Streaming over HTTP (DASH) protocol tied to the TCP transport layer protocol. The authors in~\cite{not_so_QUIC} modify the DASH protocol to run on the QUIC transport layer and compare the quality of experience (QoE) of the DASH protocol running on TCP versus QUIC. Interestingly, the authors find that using the QUIC protocol does not provide any performance boost over the standard DASH over TCP. These works show that the performance of QUIC is comparable to that of TCP~\cite{QUIC_compare_TCP, QUIC_compare_TCP_2}. 

\section{Conclusion}
\label{conclusion}

In this paper, we show that by optimally using the hardware resources (hardware accelerator cards) available on a 5G base station, we can save on the compute-node requirements of the base station. Our proposed approach reduces the cost of a 5G base station by 17\% while simultaneously improving the performance by 24\% when compared to a conventional base station. To achieve this, we perform a detailed characterization of an LDPC and a polar code-based multimedia web page transmission system. We find that for some channel conditions, using the polar code encoder in the data plane leads to a higher performance improvement than using the conventional LDPC encoder. Based on these results and the observation that the polar code encoders have a low computational resource utilization, we propose to use polar code encoders for error correction in the data plane of a 5G-NR communication system. To intelligently distribute the workloads among the multiple LDPC and polar code encoders at a base station, we use an online resource allocation algorithm. We propose a novel stochastic multi-armed bandit algorithm and two heuristics. The performance of these algorithms is compared using Monte Carlo simulations to find the one most suitable for our use case. A telecom operator needs to install more than a million base stations for a countrywide 5G deployment. Using our proposed optimizations, a 17\% reduction in the cost of a 5G base station adds up to several billion dollars of cost savings for a telecom operator. This can potentially speed up the process of deployment of 5G services especially in developing countries.

\bibliographystyle{IEEEtran}
\bibliography{references}

\begin{thebibliography}{10}
\providecommand{\url}[1]{#1}
\csname url@samestyle\endcsname
\providecommand{\newblock}{\relax}
\providecommand{\bibinfo}[2]{#2}
\providecommand{\BIBentrySTDinterwordspacing}{\spaceskip=0pt\relax}
\providecommand{\BIBentryALTinterwordstretchfactor}{4}
\providecommand{\BIBentryALTinterwordspacing}{\spaceskip=\fontdimen2\font plus
\BIBentryALTinterwordstretchfactor\fontdimen3\font minus
  \fontdimen4\font\relax}
\providecommand{\BIBforeignlanguage}[2]{{%
\expandafter\ifx\csname l@#1\endcsname\relax
\typeout{** WARNING: IEEEtran.bst: No hyphenation pattern has been}%
\typeout{** loaded for the language `#1'. Using the pattern for}%
\typeout{** the default language instead.}%
\else
\language=\csname l@#1\endcsname
\fi
#2}}
\providecommand{\BIBdecl}{\relax}
\BIBdecl

\bibitem{polarcode_IPSN}
A.~Shreshtha, P.~Singla, and S.~R. Sarangi, ``{Poster Abstract: Polar
  Code-based Approximate Communication System for Multimedia Web Pages},'' in
  \emph{2022 21st ACM/IEEE International Conference on Information Processing
  in Sensor Networks (IPSN)}, 2022.

\bibitem{rural_broadband_ericcson}
\BIBentryALTinterwordspacing
P.~Linder. (2020) Putting the spotlight on 5g in rural areas. [Online].
  Available:
  \url{https://www.ericsson.com/en/blog/2020/7/5g-in-rural-areas-spotlight}
\BIBentrySTDinterwordspacing

\bibitem{entertainment_microsoft}
\BIBentryALTinterwordspacing
J.~Cooper. (2020) 5g brings a new vision of the future for media and
  entertainment. [Online]. Available:
  \url{https://cloudblogs.microsoft.com/industry-blog/media/2020/07/20/5g-brings-a-new-vision-of-the-future-for-media-entertainment/}
\BIBentrySTDinterwordspacing

\bibitem{online_gaming_financial_express}
\BIBentryALTinterwordspacing
A.~Singh. (2022) How 5g will accelerate the growth of the online gaming
  industry. [Online]. Available:
  \url{https://www.financialexpress.com/brandwagon/how-5g-will-accelerate-the-growth-of-the-online-gaming-industry/2501639/}
\BIBentrySTDinterwordspacing

\bibitem{remote_education_qualcomm}
\BIBentryALTinterwordspacing
S.~M. Armstrong. (2020) How 5g can transform education for remote learning and
  beyond. [Online]. Available:
  \url{https://www.qualcomm.com/news/onq/2020/12/how-5g-can-transform-education-remote-learning-and-beyond}
\BIBentrySTDinterwordspacing

\bibitem{telemedicine_qualcomm}
\BIBentryALTinterwordspacing
P.~Baldwin. (2021) How 5g can transform telemedicine to tackle today’s
  toughest challenges. [Online]. Available:
  \url{https://www.qualcomm.com/news/onq/2021/01/how-5g-can-transform-telemedicine-tackle-todays-toughest-challenges}
\BIBentrySTDinterwordspacing

\bibitem{remote_surgery_business_insider}
\BIBentryALTinterwordspacing
C.~Frost. (2019) 5g is being used to perform remote surgery from thousands of
  miles away, and it could transform the healthcare industry. [Online].
  Available:
  \url{https://www.businessinsider.in/tech/5g-is-being-used-to-perform-remote-surgery-from-thousands-of-miles-away-and-it-could-transform-the-healthcare-industry/articleshow/70699125.cms}
\BIBentrySTDinterwordspacing

\bibitem{ruralembb}
S.~Henry, A.~Alsohaily, and E.~S. Sousa, ``{5G is Real: Evaluating the
  Compliance of the 3GPP 5G New Radio System With the ITU IMT-2020
  Requirements},'' \emph{IEEE Access}, 2020.

\bibitem{developing_country_perspective}
G.~Anríquez and L.~Stloukal, ``{Rural Population Change in Developing
  Countries: Lessons for Policymaking},'' \emph{European View}, 2008.

\bibitem{business_model}
A.~M. Cavalcante, M.~V. Marquezini, L.~Mendes, and C.~S. Moreno, ``{5G for
  Remote Areas: Challenges, Opportunities and Business Modeling for Brazil},''
  \emph{IEEE Access}, 2021.

\bibitem{lmlc}
\BIBentryALTinterwordspacing
R.~K. Ganti. (2020) Low mobility large cell (lmlc). [Online]. Available:
  \url{https://tsdsi.in/wp-content/uploads/2020/02/LMLC_ver1_RIT-Prof-Ganti.pdf}
\BIBentrySTDinterwordspacing

\bibitem{virtual_reality_surgery}
W.~S. Khor, B.~Baker, K.~Amin, A.~Chan, K.~Patel, and J.~Wong, ``{Augmented and
  virtual reality in surgery—the digital surgical environment: applications,
  limitations and legal pitfalls},'' \emph{Annals of translational medicine},
  2016.

\bibitem{video360degree}
G.~Lampropoulos, V.~Barkoukis, K.~Burden, and T.~Anastasiadis, ``{360-degree
  video in education: An overview and a comparative social media data analysis
  of the last decade},'' \emph{Smart Learning Environments}, 2021.

\bibitem{polarcode_JSCD}
J.~Lin, Y.~Zhang, N.~Li, and H.~Jiang, ``{Joint Source-Channel Decoding of
  Polar Codes for HEVC-Based Video Streaming},'' 2022.

\bibitem{EL8000}
\BIBentryALTinterwordspacing
H.~P. Enterprise. (2020) Vran 2.0 on hpe infrastructure. [Online]. Available:
  \url{https://h50146.www5.hpe.com/products/servers/document/pdf/edgeline/vran2.0.pdf}
\BIBentrySTDinterwordspacing

\bibitem{EL8000_datasheet}
\BIBentryALTinterwordspacing
------. (2022) Hpe edgeline el8000 converged edge system. [Online]. Available:
  \url{https://www.hpe.com/psnow/doc/a00067727enw.html}
\BIBentrySTDinterwordspacing

\bibitem{vDU}
\BIBentryALTinterwordspacing
Qualcomm. (2022) Hewlett packard enterprise and qualcomm technologies announce
  collaboration to deliver the next-generation 5g virtualized distributed unit
  solutions. [Online]. Available:
  \url{https://www.qualcomm.com/news/releases/2022/02/hewlett-packard-enterprise-and-qualcomm-technologies-announce-collaboration}
\BIBentrySTDinterwordspacing

\bibitem{hpe_telco_server}
\BIBentryALTinterwordspacing
H.~P. Enterprise. (2022) Hpe proliant dl110 gen10 plus telco server. [Online].
  Available: \url{https://www.hpe.com/psnow/doc/a50002566enw}
\BIBentrySTDinterwordspacing

\bibitem{DUX100_accelerator_card}
\BIBentryALTinterwordspacing
Qualcomm. (2021) Qualcomm introduces new 5g distributed unit accelerator card
  to drive global 5g virtualized ran growth. [Online]. Available:
  \url{https://www.qualcomm.com/news/releases/2021/06/qualcomm-introduces-new-5g-distributed-unit-accelerator-card-drive-global}
\BIBentrySTDinterwordspacing

\bibitem{T1_offload}
F.~Kaltenberger, H.~Wang, and S.~Velumani, ``{Performance evaluation of
  offloading LDPC decoding to an FPGA in 5G baseband processing},'' in
  \emph{WSA 2021; 25th International ITG Workshop on Smart Antennas}, 2021.

\bibitem{OpenAirInterface}
\BIBentryALTinterwordspacing
OAI. (2022) Open air interface. [Online]. Available:
  \url{https://openairinterface.org/}
\BIBentrySTDinterwordspacing

\bibitem{T1_product_brief}
\BIBentryALTinterwordspacing
A.~Xilinx. (2021) Xilinx t1 telco accelerator card. [Online]. Available:
  \url{https://www.xilinx.com/content/dam/xilinx/publications/product-briefs/xilinx-t1-product-brief.pdf}
\BIBentrySTDinterwordspacing

\bibitem{cost_base_station}
\BIBentryALTinterwordspacing
D.~Burstein. (2021) 5g base stations \$13,000 in china. [Online]. Available:
  \url{https://circleid.com/posts/20210713-5g-base-stations-13000-in-china}
\BIBentrySTDinterwordspacing

\bibitem{number_base_station}
\BIBentryALTinterwordspacing
D.~Khan. (2022) Number of wireless transmission points need to double to serve
  india's 5g market: Atc. [Online]. Available:
  \url{https://economictimes.indiatimes.com/industry/telecom/telecom-news/number-of-wireless-transmission-points-need-to-double-to-serve-indias-5g-market-atc/articleshow/92066549.cms}
\BIBentrySTDinterwordspacing

\bibitem{approximate_communication}
Y.~Gao, W.~Liu, and F.~Lombardi, ``{Design and Implementation of an Approximate
  Softmax Layer for Deep Neural Networks},'' \emph{2020 IEEE International
  Symposium on Circuits and Systems (ISCAS)}, 2020.

\bibitem{arikan}
E.~Arikan, ``{Channel Polarization: A Method for Constructing
  Capacity-Achieving Codes for Symmetric Binary-Input Memoryless Channels},''
  \emph{IEEE Transactions on Information Theory}, 2009.

\bibitem{frozen_bits}
R.~Wang and R.~Liu, ``A novel puncturing scheme for polar codes,'' \emph{IEEE
  Communications Letters}, 2014.

\bibitem{list_decoding}
I.~Tal and A.~Vardy, ``{List decoding of polar codes},'' in \emph{2011 IEEE
  International Symposium on Information Theory Proceedings}, 2011.

\bibitem{crc_aided_list_decoding}
K.~Niu and K.~Chen, ``{CRC-Aided Decoding of Polar Codes},'' \emph{IEEE
  Communications Letters}, 2012.

\bibitem{polar_code_matlab}
\BIBentryALTinterwordspacing
Mathworks. (2022) 5g new radio polar coding. [Online]. Available:
  \url{https://in.mathworks.com/help/5g/gs/polar-coding.html}
\BIBentrySTDinterwordspacing

\bibitem{ldpc_decoder_gbps}
R.~Zarubica, S.~G. Wilson, and E.~Hall, ``{Multi-Gbps FPGA-based low density
  parity check (LDPC) decoder design},'' in \emph{IEEE GLOBECOM 2007-IEEE
  Global Telecommunications Conference}.\hskip 1em plus 0.5em minus 0.4em\relax
  IEEE, 2007.

\bibitem{tanner_decoder_arch}
E.~Boutillon, G.~Masera, D.~Declercq, M.~Fossorier, and E.~Biglieri,
  ``{Hardware design and realization for iteratively decodable codes},'' in
  \emph{Channel coding: Theory, algorithms, and applications}.\hskip 1em plus
  0.5em minus 0.4em\relax Elsevier, 2014.

\bibitem{ber_convolution_turbo_ldpc_polar}
B.~Tahir, S.~Schwarz, and M.~Rupp, ``{BER comparison between Convolutional,
  Turbo, LDPC, and Polar codes},'' in \emph{2017 24th International Conference
  on Telecommunications (ICT)}, 2017.

\bibitem{gallager_ldpc}
R.~Gallager, ``{Low-density parity-check codes},'' \emph{IRE Transactions on
  information theory}, 1962.

\bibitem{ldpc_parallel_decoder}
J.~Nadal and A.~Baghdadi, ``{Parallel and Flexible 5G LDPC Decoder Architecture
  Targeting FPGA},'' \emph{IEEE Transactions on Very Large Scale Integration
  (VLSI) Systems}, 2021.

\bibitem{ldpc_polar_decoder_comparison}
A.~Balatsoukas-Stimming, P.~Giard, and A.~Burg, ``{Comparison of Polar Decoders
  with Existing Low-Density Parity-Check and Turbo Decoders},'' in \emph{2017
  IEEE Wireless Communications and Networking Conference Workshops (WCNCW)},
  2017.

\bibitem{multiarmed_bandit_resource_allocation}
J.~Zuo and C.~Joe-Wong, ``{Combinatorial Multi-armed Bandits for Resource
  Allocation},'' in \emph{2021 55th Annual Conference on Information Sciences
  and Systems (CISS)}, 2021.

\bibitem{contextual_mab}
T.~Lu, D.~P{\'a}l, and M.~P{\'a}l, ``{Contextual multi-armed bandits},'' in
  \emph{Proceedings of the Thirteenth international conference on Artificial
  Intelligence and Statistics}.\hskip 1em plus 0.5em minus 0.4em\relax JMLR
  Workshop and Conference Proceedings, 2010.

\bibitem{polar_util}
J.~Wang and Y.~Hu, ``{Architectural and Cost Implications of the 5G Edge NFV
  Systems},'' in \emph{2019 IEEE 37th International Conference on Computer
  Design (ICCD)}, 2019.

\bibitem{matlab_5G_toolbox}
\BIBentryALTinterwordspacing
Mathworks. (2021) 5g toolbox. [Online]. Available:
  \url{https://in.mathworks.com/products/5g.html}
\BIBentrySTDinterwordspacing

\bibitem{ldpc_accelerator}
E.~A. Papatheofanous, D.~Reisis, and K.~Nikitopoulos, ``{LDPC Hardware
  Acceleration in 5G Open Radio Access Network Platforms},'' \emph{IEEE
  Access}, 2021.

\bibitem{polar_accelerator}
P.~Giard, G.~Sarkis, A.~Balatsoukas-Stimming, Y.~Fan, C.-y. Tsui, A.~Burg,
  C.~Thibeault, and W.~J. Gross, ``{Hardware decoders for polar codes: An
  overview},'' in \emph{2016 IEEE International Symposium on Circuits and
  Systems (ISCAS)}, 2016.

\bibitem{tcp_timing_model}
J.~Padhye, V.~Firoiu, D.~Towsley, and J.~Kurose, ``{Modeling TCP Throughput: A
  Simple Model and Its Empirical Validation}.''\hskip 1em plus 0.5em minus
  0.4em\relax Association for Computing Machinery, 1998.

\bibitem{QUIC}
A.~Langley, A.~Riddoch, A.~Wilk, A.~Vicente, C.~Krasic, D.~Zhang, F.~Yang,
  F.~Kouranov, I.~Swett, J.~Iyengar, J.~Bailey, J.~Dorfman, J.~Roskind,
  J.~Kulik, P.~Westin, R.~Tenneti, R.~Shade, R.~Hamilton, V.~Vasiliev, W.-T.
  Chang, and Z.~Shi, ``{The QUIC Transport Protocol: Design and Internet-Scale
  Deployment},'' in \emph{Proceedings of the Conference of the ACM Special
  Interest Group on Data Communication}, ser. SIGCOMM '17.\hskip 1em plus 0.5em
  minus 0.4em\relax Association for Computing Machinery, 2017.

\bibitem{QUIC_similar_TCP}
\BIBentryALTinterwordspacing
R.~H. et~al. (2016) Quic: A udp-based secure and reliable transport for http/2.
  [Online]. Available:
  \url{https://datatracker.ietf.org/doc/html/draft-tsvwg-quic-protocol-02}
\BIBentrySTDinterwordspacing

\bibitem{QUIC_compare_TCP_2}
D.~Hasselquist, C.~Lindstr{\"o}m, N.~Korzhitskii, N.~Carlsson, and A.~Gurtov,
  ``{Quic throughput and fairness over dual connectivity},'' in \emph{Symposium
  on Modelling, Analysis, and Simulation of Computer and Telecommunication
  Systems}.\hskip 1em plus 0.5em minus 0.4em\relax Springer, 2020.

\bibitem{QUIC_compare_TCP}
A.~Yu and T.~A. Benson, ``{Dissecting Performance of Production QUIC},'' ser.
  WWW '21.\hskip 1em plus 0.5em minus 0.4em\relax Association for Computing
  Machinery, 2021.

\bibitem{application_spec_protocol}
W.~B. Heinzelman, ``{Application-specific protocol architectures for wireless
  networks},'' Ph.D. dissertation, Massachusetts Institute of Technology, 2000.

\bibitem{network_slicing}
J.~Ordonez-Lucena, P.~Ameigeiras, D.~Lopez, J.~J. Ramos-Munoz, J.~Lorca, and
  J.~Folgueira, ``{Network slicing for 5G with SDN/NFV: Concepts,
  architectures, and challenges},'' \emph{IEEE Communications Magazine}, 2017.

\bibitem{replacement_TCP_IP_5G}
\BIBentryALTinterwordspacing
C.~Gabriel. (2020) Etsi’s ‘non-ip’ group is its second attempt at a 5g
  replacement for tcp/ip. [Online]. Available:
  \url{https://rethinkresearch.biz/articles/etsis-non-ip-group-is-its-second-attempt-at-a-5g-replacement-for-tcp-ip/}
\BIBentrySTDinterwordspacing

\bibitem{ETSI_protocol}
\BIBentryALTinterwordspacing
A.~Sutton and R.~Li. (2016) Next generation protocols – market drivers and
  key scenarios. [Online]. Available:
  \url{https://www.etsi.org/images/files/ETSIWhitePapers/etsi_wp17_Next_Generation_Protocols_v01.pdf}
\BIBentrySTDinterwordspacing

\bibitem{metrics_algorithm}
J.~Bruno, P.~Downey, and G.~N. Frederickson, ``{Sequencing tasks with
  exponential service times to minimize the expected flow time or makespan},''
  \emph{Journal of the ACM (JACM)}, 1981.

\bibitem{weighted_flowtime}
S.~Anand, N.~Garg, and A.~Kumar, ``{Resource augmentation for weighted
  flow-time explained by dual fitting},'' in \emph{Proceedings of the
  twenty-third annual ACM-SIAM symposium on Discrete Algorithms}.\hskip 1em
  plus 0.5em minus 0.4em\relax SIAM, 2012.

\bibitem{multi_armed_bandit}
A.~Feki and V.~Capdevielle, ``{Autonomous resource allocation for dense lte
  networks: A multi armed bandit formulation},'' in \emph{2011 IEEE 22nd
  International Symposium on Personal, Indoor and Mobile Radio
  Communications}.\hskip 1em plus 0.5em minus 0.4em\relax IEEE, 2011.

\bibitem{multi_armed_bandit_ucb}
P.~Auer, N.~Cesa-Bianchi, and P.~Fischer, ``{Finite-time analysis of the
  multiarmed bandit problem},'' \emph{Machine learning}, 2002.

\bibitem{netsim}
\BIBentryALTinterwordspacing
Tetcos. (2005) Netsim. [Online]. Available: \url{https://www.tetcos.com/}
\BIBentrySTDinterwordspacing

\bibitem{ms_ssim}
Z.~Wang, E.~Simoncelli, and A.~Bovik, ``{Multiscale structural similarity for
  image quality assessment},'' in \emph{The Thrity-Seventh Asilomar Conference
  on Signals, Systems and Computers, 2003}, 2003.

\bibitem{gop}
K.~Seshadrinathan, R.~Soundararajan, A.~C. Bovik, and L.~K. Cormack, ``{A
  subjective study to evaluate video quality assessment algorithms},'' in
  \emph{Human Vision and Electronic Imaging XV}.\hskip 1em plus 0.5em minus
  0.4em\relax SPIE, 2010.

\bibitem{big_buck_bunny}
T.~Roosendaal, ``{Big Buck Bunny},'' in \emph{International Conference on
  Computer Graphics and Interactive Techniques, {SIGGRAPH} {ASIA} 2008,
  Singapore, December 10-13, 2008, Computer Animation Festival}.\hskip 1em plus
  0.5em minus 0.4em\relax {ACM}, 2008.

\bibitem{tcp_timing_literature}
B.~Liu, D.~Goeckel, and D.~Towsley, ``{TCP-cognizant adaptive forward error
  correction in wireless networks},'' in \emph{Global Telecommunications
  Conference, 2002. GLOBECOM '02. IEEE}, 2002.

\bibitem{channel_EbNo_range}
T.~Hong, T.~Tang, X.~Dong, R.~Liu, and W.~Zhao, ``{Future 5G mmWave TV Service
  With Fast List Decoding of Polar Codes},'' \emph{IEEE Transactions on
  Broadcasting}, 2020.

\bibitem{webpage_size_distribution}
\BIBentryALTinterwordspacing
T.~Everts and K.~Hempenius. (2019) Web almanac: Page weight. [Online].
  Available: \url{https://almanac.httparchive.org/en/2019/page-weight}
\BIBentrySTDinterwordspacing

\bibitem{webpage_ratio_distribution}
\BIBentryALTinterwordspacing
J.~Teague. (2021) Web almanac: Page weight. [Online]. Available:
  \url{https://almanac.httparchive.org/en/2021/page-weight}
\BIBentrySTDinterwordspacing

\bibitem{nfv_1}
Y.~Liu, J.~Pei, P.~Hong, and D.~Li, ``{Cost-Efficient Virtual Network Function
  Placement and Traffic Steering},'' in \emph{ICC 2019 - 2019 IEEE
  International Conference on Communications (ICC)}, 2019.

\bibitem{nfv_2}
A.~Mohammadkhan, S.~Ghapani, G.~Liu, W.~Zhang, K.~K. Ramakrishnan, and T.~Wood,
  ``{Virtual function placement and traffic steering in flexible and dynamic
  software defined networks},'' in \emph{The 21st IEEE International Workshop
  on Local and Metropolitan Area Networks}, 2015.

\bibitem{polar_nbiot}
R.~S. Zakariyya, K.~H. Jewel, A.~O. Fadamiro, O.~J. Famoriji, and F.~Lin, ``{An
  Efficient Polar Coding Scheme for Uplink Data Transmission in Narrowband
  Internet of Things Systems},'' \emph{IEEE Access}, 2020.

\bibitem{image_video_internet}
W.~Xin-Gong, ``{Research on Multimedia Data Stream in Network}.''\hskip 1em
  plus 0.5em minus 0.4em\relax IEEE Computer Society, 2013.

\bibitem{uep_meaning}
I.~Boyarinov and G.~Katsman, ``{Linear unequal error protection codes},''
  \emph{IEEE Transactions on Information Theory}, 1981.

\bibitem{uep_polar_code}
W.~Du, S.~Zhang, and F.~Ding, ``{Exploiting the UEP property of polar codes to
  reduce image distortions induced by transmission errors},'' in \emph{2015
  IEEE/CIC International Conference on Communications in China (ICCC)}, 2015.

\bibitem{uep_ldpc_code}
N.~Rahnavard, H.~Pishro-Nik, and F.~Fekri, ``{Unequal Error Protection Using
  Partially Regular LDPC Codes},'' \emph{IEEE Transactions on Communications},
  2007.

\bibitem{uep_ldpc_code_image}
P.~Ma and K.~Kwak, ``{Unequal Error Protection Low-Density Parity-Check Codes
  Design Based on Gaussian Approximation in Image Transmission},'' in
  \emph{2009 IEEE Wireless Communications and Networking Conference}, 2009.

\bibitem{approximate_wireless_media}
S.~Sen, T.~Zhang, S.~Gilani, S.~Srinath, S.~Banerjee, and S.~Addepalli,
  ``{Design and Implementation of an “Approximate” Communication System for
  Wireless Media Applications},'' \emph{IEEE/ACM Transactions on Networking},
  2013.

\bibitem{SAP}
B.~Ransford and L.~Ceze, ``{SAP: an architecture for selectively approximate
  wireless communication},'' \emph{arXiv preprint arXiv:1510.03955}, 2015.

\bibitem{quic_tcp_evaluation}
K.~Nepomuceno, I.~N.~d. Oliveira, R.~R. Aschoff, D.~Bezerra, M.~S. Ito,
  W.~Melo, D.~Sadok, and G.~Szabó, ``{QUIC and TCP: A Performance
  Evaluation},'' in \emph{2018 IEEE Symposium on Computers and Communications
  (ISCC)}, 2018.

\bibitem{not_so_QUIC}
D.~Bhat, A.~Rizk, and M.~Zink, ``{Not so QUIC: A performance study of DASH over
  QUIC},'' in \emph{Proceedings of the 27th workshop on network and operating
  systems support for digital audio and video}, 2017.

\end{thebibliography}

\ifCLASSOPTIONcaptionsoff
  \newpage
\fi



%

%

\vspace{-50px}

\begin{IEEEbiography}[{\includegraphics[width=1in,height=1.25in]{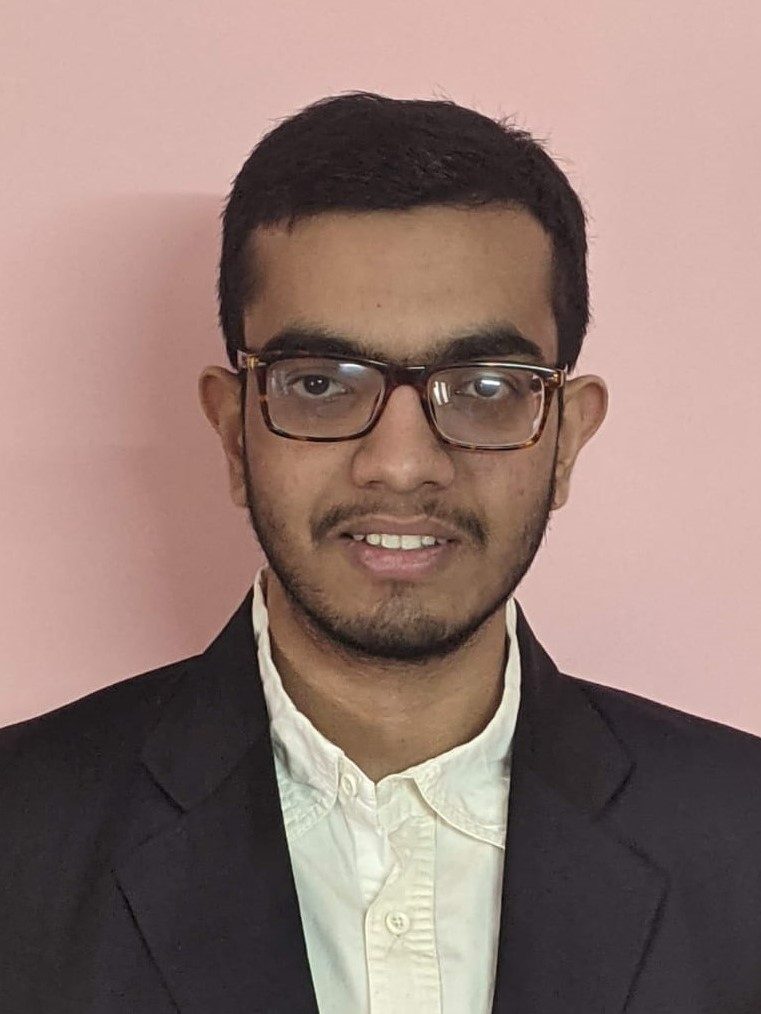}}]{Aman Shreshtha}

is currently working towards his Ph.D. degree in the Computer Science and Engineering Department, IIT Delhi. He received his B.Tech. degree in electronics engineering from IIT (BHU) Varanasi in 2020. His research interests include architectures for 5G communication, accelerators for security in high-throughput networks, and application of machine learning algorithms in commmunication systems.     

\end{IEEEbiography}

\vspace{-50px}

\begin{IEEEbiography}[{\includegraphics[width=1in,height=1.25in]{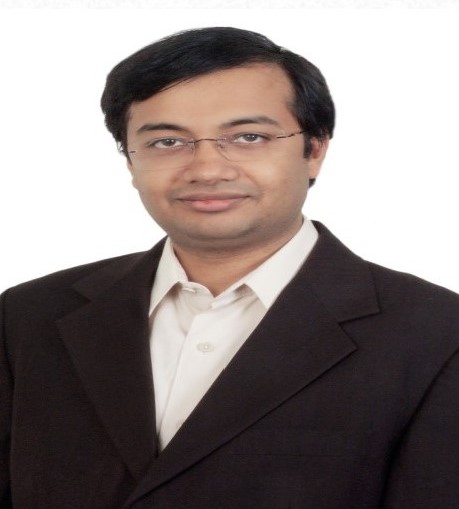}}]{Smruti R Sarangi}

is a full professor in the Computer Science and Engineering Department at IIT Delhi. He also holds a joint appointment with the Department of Electrical Engineering, the School of IT, and the Bharti School of Telecom Technology as well. He primarily works on computer architecture and operating systems. Dr. Sarangi obtained his Ph.D in computer science from the University of Illinois at Urbana Champaign(UIUC), USA  in 2006, and a B.Tech in computer science from IIT Kharagpur in 2002. After completing his Ph.D he has worked in Synopsys Research, and IBM Research Labs. He has filed five US patents, five Indian patents, and has published 117 papers in reputed international conferences and journals. Prof. Sarangi is a member of the IEEE and ACM.

\end{IEEEbiography}




\end{document}